\documentclass[11pt,a4paper]{article}
\usepackage[left=1.1in,top=1.1in,bottom=1.1in,right=1.1in,nohead]{geometry}
\usepackage[dvipdfmx]{graphicx}
\usepackage{amssymb}
\usepackage{aas_macros}
\usepackage{authblk}
\title{
Neutrino Telescope Array Letter of Intent:\\
\vspace{3mm}
{\Large
A Large Array of High Resolution Imaging Atmospheric Cherenkov and Fluorescence Detectors
for Survey of Air-showers from Cosmic Tau Neutrinos in the PeV-EeV Energy Range \\
}}
\author[1]{Makoto Sasaki\thanks{\texttt{sasakim@icrr.u-tokyo.ac.jp}}}
\author[2]{George Wei-Shu Hou\thanks{\texttt{wshou@phys.ntu.edu.tw}}}
\affil[1]{ICRR, The University of Tokyo, Kashiwa, Chiba 277-8582, Japan}
\affil[2]{Department of Physics, National Taiwan University, Taipei 10617, Taiwan}
\date{July 20, 2015}
\begin{document}
\maketitle
\begin{abstract}
This Letter of Intent (LoI) describes the outline and plan for the Neutrino
Telescope Array (NTA) project. High-energy neutrinos provide unique and
indisputable evidence for hadronic acceleration, as well as a most accurate
probe into the hidden sector of traditional astronomy or physics, such as
dark matter. However, their extremely low flux and interaction cross section
make their detection extraordinarily difficult. Recently, IceCube has reported
astronomical neutrino candidates in excess of expectation from atmospheric
secondaries, but is limited by the water Cherenkov detection method. A next
generation high-energy neutrino telescope should be capable of establishing
indisputable evidence for cosmic high-energy neutrinos. It should not only
have orders-of-magnitude larger sensitivity, but also enough pointing accuracy
to probe known or unknown astronomical objects, without suffering from
atmospheric secondaries. The proposed installation is a large array of
compound eye stations of imaging atmospheric Cherenkov and fluorescence
detectors, with wide field of view and refined observational ability of air
showers from cosmic tau neutrinos in the PeV-EeV energy range.
This advanced optical complex system is based
substantially on the development of All-sky Survey High Resolution Air-shower detector (Ashra)
and applies the tau shower Earth-skimming method to survey PeV-EeV $\nu_{\tau}$s.
It allows wide (30$^{\circ}$$\times$360$^{\circ}$) and deep ($\sim$400~Mpc)
survey observation for PeV-EeV $\nu_{\tau}$s assuming the standard GRB neutrino fluence.
In addition, it enjoys the pointing accuracy of better than 0.2$^{\circ}$
in essentially background-free conditions.
With the advanced imaging of Earth-skimming tau showers in the wide field of view,
we aim for clear discovery and identification of astronomical $\nu_{\tau}$ sources,
providing inescapable evidence of the astrophysical hadronic model
for acceleration and/or propagation of extremely high energy protons in the precisely determined direction.
In this LoI, we present main features of the NTA detector,
scientific goal and observational objects, Earth-skimming detection method, the NTA detector,
the expected detector performance, and brief summaries of time frame, organization, and funding.
\end{abstract}

\noindent {\bf Keywards:}
Astroparticle physics,
Neutrino astronomy,
Hadron acceleration,
Cosmic ray origin,
Dark matter,
Gamma ray burst,
Active galactic nuclei,
Neutrino telescope,
PeV-EeV neutrinos,
Earth-skimming detection,
Tau neutrino,
Ashra,
Neutrino Telescope Array

\maketitle

\section{Introduction}

\subsection{Background and Past-related Achievements}
High-energy neutrinos uniquely provide indisputable evidence for hadronic acceleration
in the universe.
High-energy charged cosmic rays have been observed for a long time,
but their origin is still a mystery.
The energy spectrum follows globally a broken E$^{-\alpha}$ power law,
where $\alpha = 2.7\sim3.1$, which indicates shock acceleration.
Several astronomical object classes have been proposed as potential
hadronic accelerators.
The galactic and extragalactic magnetic fields prevent us from
using the arrival direction observed on Earth to reveal the actual sources.
So far, standard astronomical observational data,
spanning the electromagnetic wavelengths from radio to $\gamma$-ray,
have not succeeded in revealing direct evidence of the non-thermal process.
On the other hand, high-energy neutrinos should be produced at the accelerators
through charged pion production in collisions with
radiation fields or the ambient matter,
in reactions such as:
$$
  p + \gamma \rightarrow \Delta^{+} \rightarrow \pi^{0} + p, \  \pi^{+} + n
$$
$$
  p + {\rm nucleus} \rightarrow \pi + X \ (\pi=\pi^{0},\pi^{\pm})~.
$$
Subsequent decay gives the approximate neutrino flavour ratio
$\nu_e:\nu_\mu:\nu_\tau=1:2:0$ at the sources, which is turned into the ratio of
$\nu_e:\nu_\mu:\nu_\tau=1:1:1$ by neutrino oscillation upon arrival at Earth.
The photopion ($p\gamma$) reaction is typically the main neutrino generation process
where extra galactic sources like jets and cores of active galactic nuclei (AGN) and $\gamma$-ray
burst (GRB) jets have been widely studied.
Some sources like starburst galaxies (SBGs) may emit the neutrino fluxes mainly through
the hadronuclear ($pp$) reaction
\cite{loeb2006cumulative}.
For many astronomical objects, ambient photons are expected to be in the UV region.
In that case,
the kinetic threshold for photopion production through delta resonance is
in the range of several PeV.

The IceCube Collaboration
claims the first observation of two PeV-Energy neutrinos,
with moderate 2.8~$\sigma$
excess over Monte Carlo expectation of background events
\cite{PhysRevLett.111.021103}.
They further extended their analysis to lower energy region
\cite{icecube2013evidence}.
Fitting to the observed photoelectron spectrum,
they estimate the diffuse neutrino flux to be
$ E^{2}\phi_{\nu_e+\nu_\mu+\nu_\tau}=3.6\times10^{-8}$~GeV~sr$^{-1}$~s$^{-1}$,
assuming an $E^{-2}$ power law flux.
The fact that no more events occur in the higher energy region
favours neutrinos from astronomical objects
but not cosmogenic neutrinos~\cite{Aartsen:2013hn}, if the events are true neutrino signals.
Assuming astronomical objects where the observed neutrinos were produced,
the estimated fluences of the neutrino beams are rather high.
It therefore becomes plausible that
a next generation high-energy neutrino telescope,
with higher sensitivity for high-energy neutrinos and wide field of view,
could make clear discovery of hadron accelerators in the Universe.

For cosmogenic neutrinos, produced by the photopion process of protons with
the cosmic microwave background,
the energy threshold is around 10$^{19.6}$~eV.
However, the Pierre Auger Observatory (Auger) claims
that heavier components dominate the highest energy region
around 10$^{19.6}$~eV \cite{PhysRevLett.104.091101}.
If true, the flux estimate of cosmogenic neutrinos is much suppressed.
Both the cosmic ray flux spectrum around threshold
and the density of cosmic microwave background are observed so well,
the detection of cosmogenic neutrinos provide a good check
of the Auger results on cosmic ray composition.
Besides high-energy neutrino detection,
the observation of PeV $\gamma$-rays could also have provided a clear proof of hadron acceleration,
from the subsequent $\pi^0\to \gamma\gamma$ decay in the above process.
However, in the PeV range, photons are absorbed by interaction with
cosmic microwave photons into electron pairs.
Therefore, especially the observation of PeV-EeV neutrinos with precise pointing accuracy
would provide unique and particularly
important identification of astronomical cosmic ray origins,
as well as examination of cosmogenic neutrino production from
extragalactic hadron propagation.

A final answer to the mystery of cosmic ray origins
requires the observation of high-energy neutrinos.
High-energy neutrinos can be
the most accurate probe into hidden sector of traditional astronomy or physics,
such as dark matter.
However, their extremely low flux and interaction cross section
make their detection extraordinarily difficult.
To discover clear, indisputable evidence of cosmic high-energy neutrinos,
a next generation detector should have orders-of-magnitude larger sensitivity.
To probe into the association with known or unknown astronomical objects
as a telescope for very high-energy (VHE) neutrinos with the energies above 1~PeV,
it should also have enough pointing accuracy,
without suffering from background events of atmospheric secondaries.

The Earth-skimming tau neutrino technique enjoys a large
target mass by detecting extensive air-showers
produced by tau lepton decays in the atmosphere.
The tau leptons, produced by VHE tau neutrinos that interact
with the Earth matter they traverse, emerge out of a mountain or the ground
facing the detector.
This method has detection sensitivity in the PeV-EeV
region, and can be used to search for neutrinos originating from hadron
acceleration in astronomical objects. Additional advantages are perfect
shielding of cosmic ray secondaries, precise arrival direction
determination, and negligible background from atmospheric neutrinos.

\begin{figure}[t!]
\hspace{2mm}
\begin{minipage}[b]{0.324\linewidth}
\includegraphics[width=\linewidth]{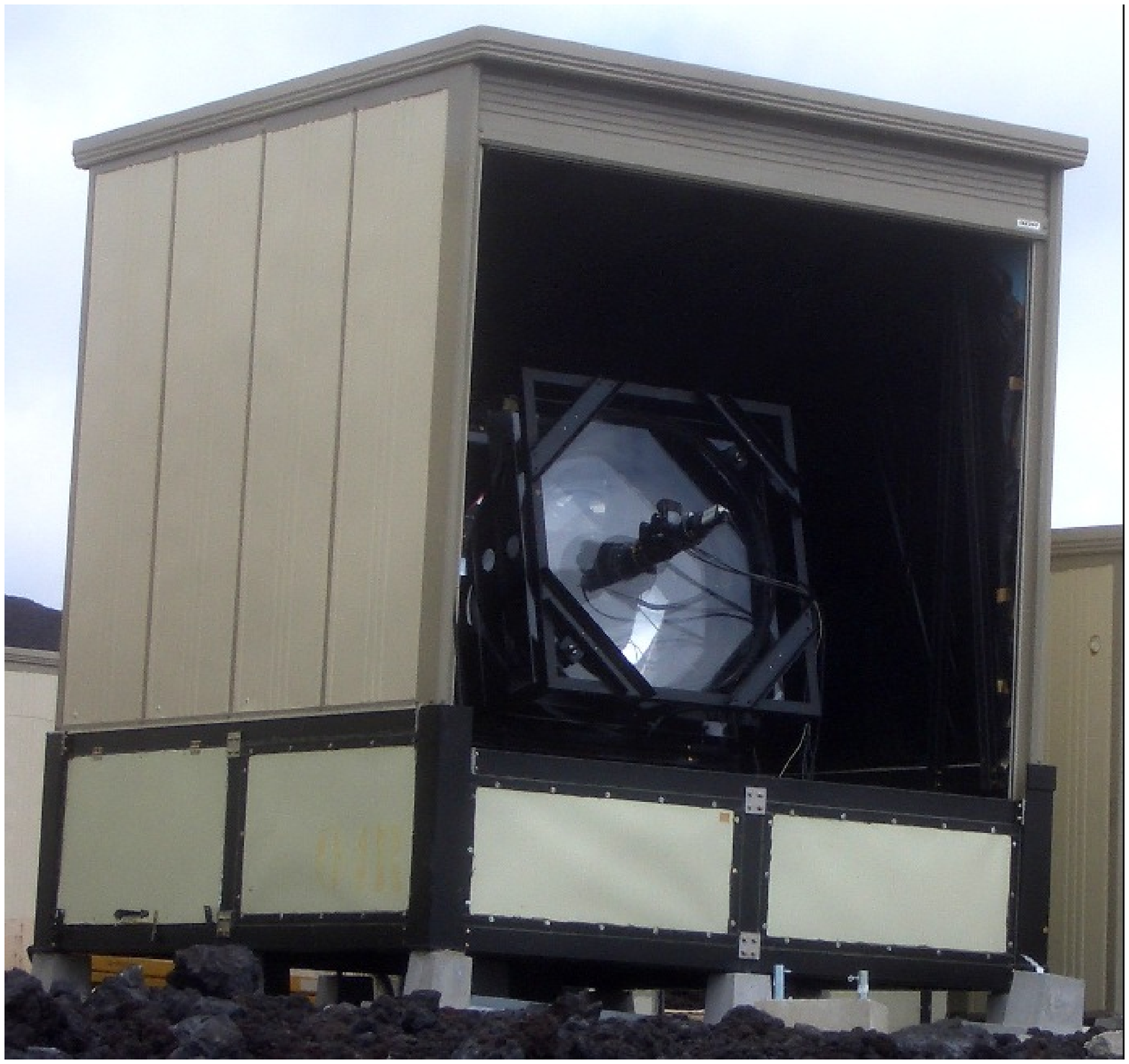}
\vskip-1mm
\caption{\label{fig:AshraLC}
An Ashra-1
light collector toward Mauna Kea. }
\end{minipage}
\hspace{2mm}
\begin{minipage}[b]{0.62\linewidth}
\includegraphics[width=\linewidth]{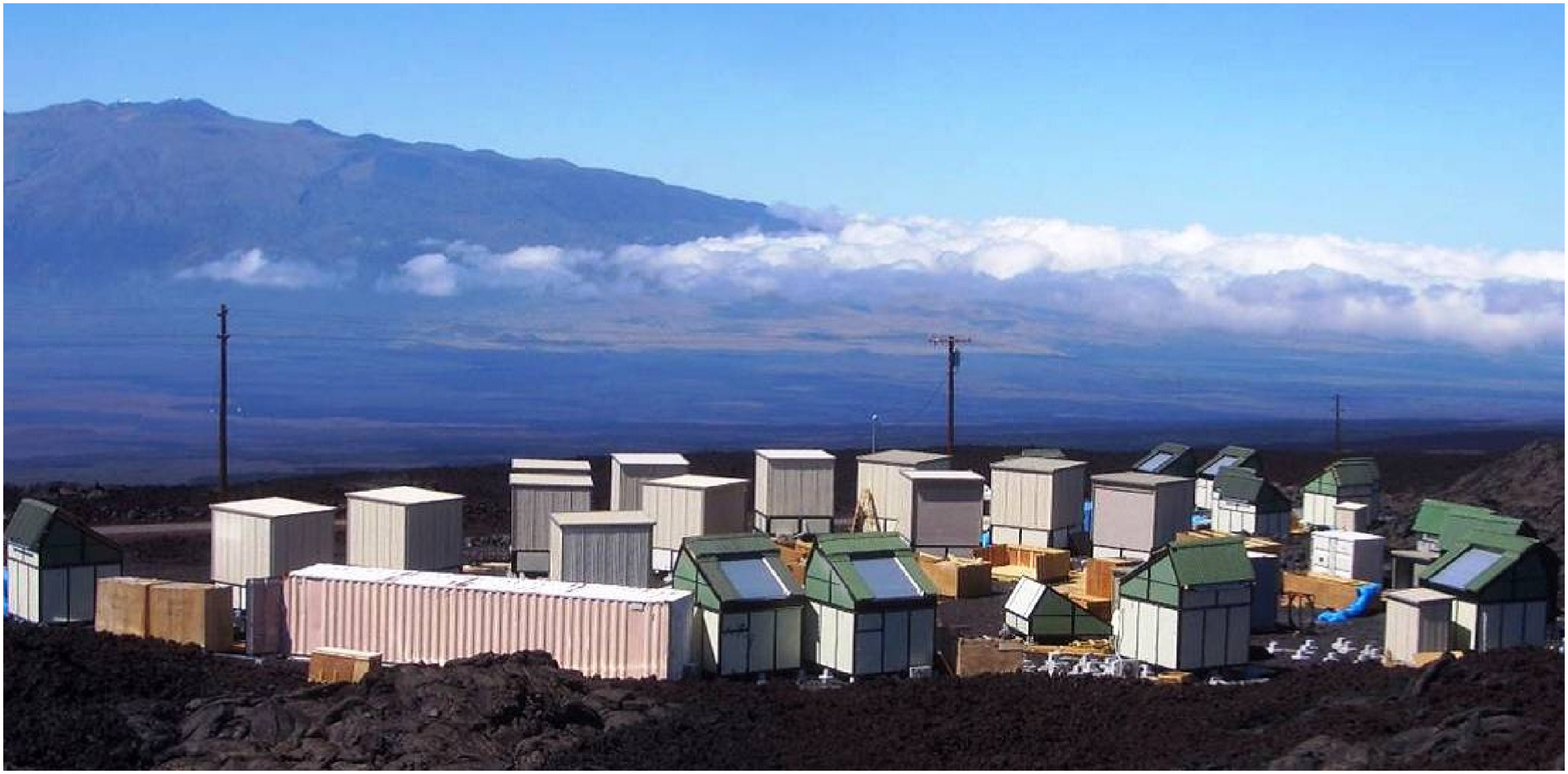}
\vskip-0.3cm
\caption{\label{fig:AshraStationMLsite}
The Ashra-1 main and sub stations at the Mauna Loa site.
}
\end{minipage}
\end{figure}

The All-sky Survey High Resolution Air-shower detector Phase I (Ashra-1) is
an optical-telescope based detector system
\cite{Sasaki08}
optimized to detect VHE particles
aiming for ``multi-particle astronomy''
\cite{Sasaki00,Barwick00}.
It is distinguished by two features:
(1) an ultra wide optical system in which
42-degree FOV (field of view) is demagnified to
1-inch diameter phospher screen on an output window
by using
photon and electron optics \cite{PLI11};
(2) high resolution imaging system with a trigger.
Ashra-1 combines these unique features, resulting in very cost-effective pixels
compared to conventional photomultiplier arrays at the focal
surface of an optical telescope
(Fig.~\ref{fig:AshraLC}).
Ashra-1 can observe the whole sky with a few arc minutes resolution,
with 12 detector units pointing at different directions,
where a detector unit consists of a few Light Collectors (LC)
pointing at the same direction.

The Ashra-1 detector system is designed so that the focal image
is split into trigger/image capture devices after amplification.
This feature enables one to simultaneously access 3 kinds of phenomena
that have different time scales, i.e., Cherenkov emission (ns), fluorescence ($\mu$s), and starlight (s),
without sacrificing the signal to noise ratio.
By fully utilizing these distinct features,
Ashra aims to undertake full-fledged astronomical observation using VHE particles,
commencing with the first detection of VHE neutrinos using Earth and mountain as
target \cite{AshraCNeu}.
It can also be used to optically observe transient objects like GRBs,
as it monitors the whole sky simultaneously
\cite{GCN8632,GCN11291}.
The principal demonstration phase, Ashra-1,
has been running at the Mauna Loa site at 3300 m above sea level
on Hawaii Island since 2008.
The deployed main and sub stations at the Mauna Loa site
are shown in Fig.~\ref{fig:AshraStationMLsite}
Ashra-1 succeeded in the first search for PeV-EeV tau neutrinos originating from
a GRB in the commissioning run \cite{AshraCNeu},
demonstrating the great sensitivity
around 100~PeV with the earth-skimming $\nu_{\tau}$ technique.
Ashra-1 has achieved the best instantaneous
sensitivity in the energy region around 100~PeV
since January 2012 after trigger upgrade.

The NuTel \cite{NuTel} project, conceived and started in 2001,
was an effort concurrent with the development of Ashra-1.
It purposed the fast construction of a limited neutrino telescope for
detecting $\nu_\tau$-originated air-showers in the energy range of 1 to 1000 PeV,
with possible sources such as AGNs, GRBs, the Galactic Center (GC), etc.
It took in the possibility of a base up the smaller Mt. Hualalai,
which provides a wide baseline view of Mauna Loa.
A multi-anode PMT-based readout electronics, aimed for Cherenkov light in the UV,
was quickly built, but the project got cut since 2004.
The group still built two 2m telescopes,
and conducted a mountain test up 2200 m in Taiwan.
An issue was the estimated event rate of $\sim 0.5$ events per year,
which was not quite convincing.
This is where the well developed Ashra-1 comes in contrast:
a mature VHE neutrino telescope, which we will call NTA, is at hand.

\subsection{Main Features of the NTA Detector}

The key technical feature of the Ashra-1 detector rests on
the use of electrostatic lenses, rather than optical lens systems,
to generate convergent beams. This enables us to realise a high resolution
over a wide field of view.
This electron optics requires:

\begin{itemize}

\item {\em wide angle high precision optics} \cite{2002NIMPA.492...49S};
a Schmidt type optical system with modified Baker-Nunn optics allows
a compromise between wide 42$^{\circ}$ field of view and 1 arc min
resolution on the focal sphere of the light collector (Fig.~\ref{fig:AshraLC}),
with pupil diameter of 1~m;

\item {\em photoelectric lens imaging-tube} \cite{PLI11};
in addition to the optical system,
the world's largest imaging-tube
uses electrostatic lens
to generate convergent beams from photo cathode of 20 inch diameter
to output phosphorus window of 1 inch diameter,
enabling a very low cost and high performance image sensor
that provides high resolution over a wide FOV;
and

\item {\em image pipeline} \cite{2003NIMPA.501..359S};
the image transportation from imaging-tube (image intensifier)
to a trigger device and image sensor of fine pixels (CCD+CMOS)
with high gain and resolution,
enables very fine images with parallel self-trigger systems
that trigger for optical flash, atmospheric Cherenkov
and fluorescence light separately.

\end{itemize}

\begin{figure}[t!]
\begin{center}
\begin{tabular}{cc}
\begin{minipage}[t]{0.4\linewidth}
\includegraphics[width=\linewidth]{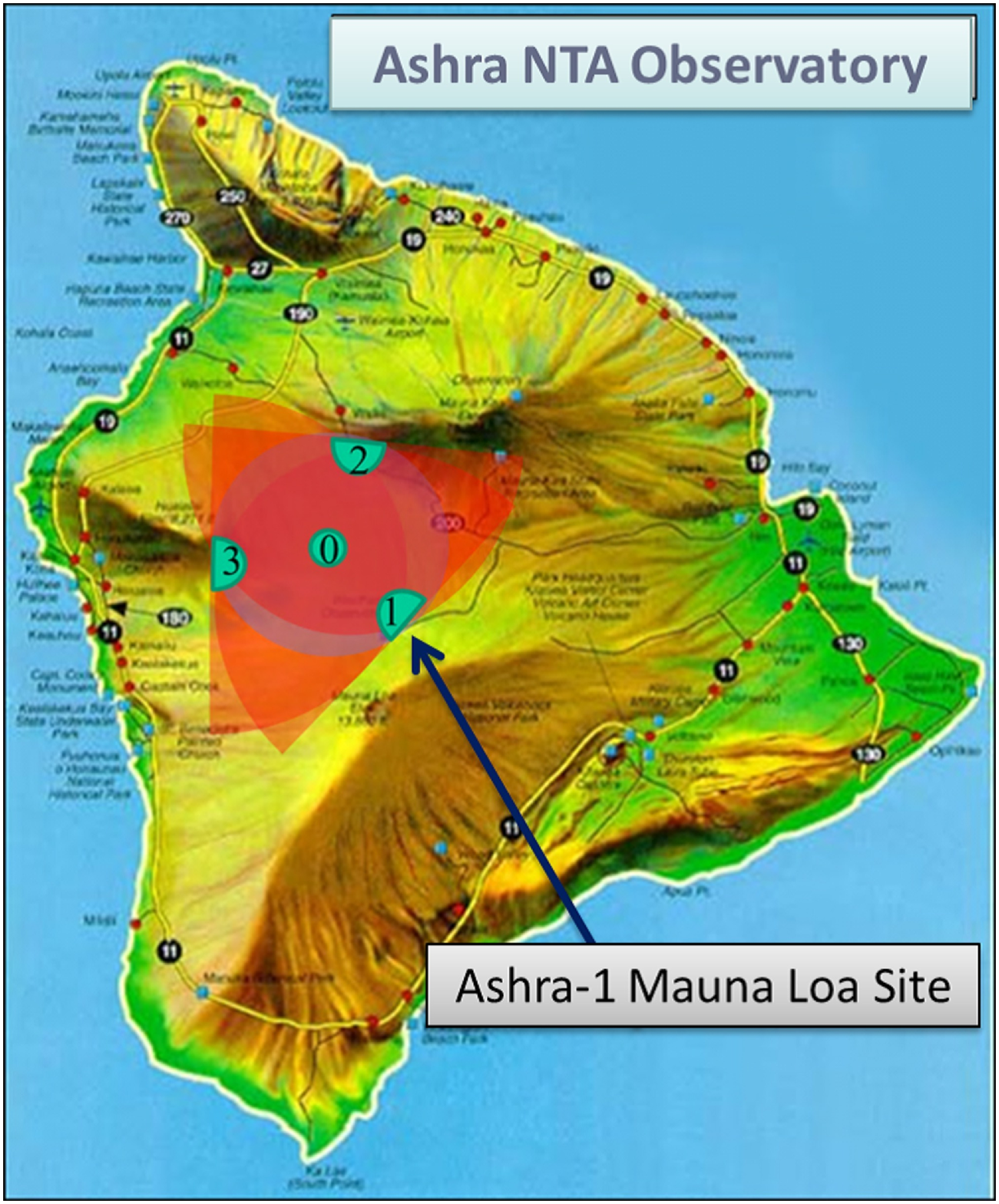}
\vskip-0.3cm
\caption{
Layout of the NTA Observatory.
Shaded region consists of
three semicircles centered
at Site1-3.
}
\label{fig:NTA-Observatory}
\end{minipage}
\hspace{2mm}
\begin{minipage}[t]{0.5\linewidth}
\includegraphics[width=\linewidth]{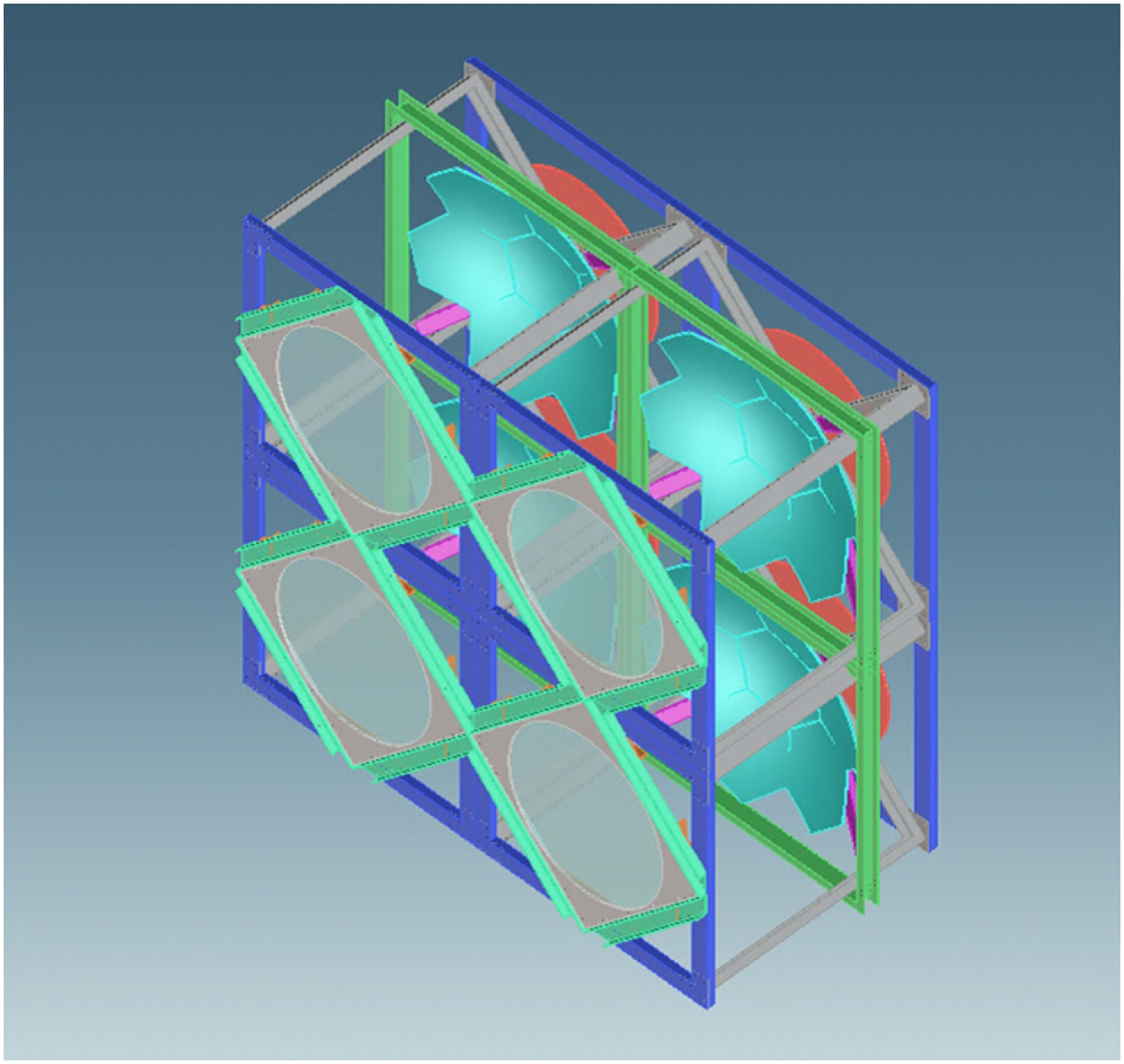}
\vskip-0.3cm
\caption{
NTA detector unit of four light collectors of the same type.
}
\label{fig:NTA-DU}
\end{minipage}
\end{tabular}
\end{center}
\end{figure}

Based on these achievements from Ashra-1,
we start to form a new collaboration for realizing
the next generation large Neutrino Telescope Array (NTA).
The conceptual layout for the NTA observatory
considers three site stations for a 25~km-side triangle,
watching the total air mass surrounded by the mountains of
Mauna Loa, Mauna Kea, and Hualalai
(Fig.~\ref{fig:NTA-Observatory}).
A single site station at the center of the triangle has half-sky coverage.
This configuration allows for tremendous
instantaneous sensitivity (equivalent to $>$100 giga ton water),
with Cherenkov-fluorescence stereoscopic observation for PeV-EeV neutrinos
in essentially background-free conditions. With the demonstrated
fine imaging of Earth-skimming tau showers
and the significant improved detection solid angle
(30$^{\circ}$ $\times$ 360$^{\circ}$)
for incoming tau neutrinos, we aim for
clear discovery and identification of astronomical tau neutrino sources.
Also interesting is the unique capability of cross observation between
optical flashes, TeV-PeV $\gamma$ rays, and PeV-EeV $\nu_{\tau}$s,
once one or more of these three kinds of self-triggers
are observed and associated with an astronomical object.

From the current baseline design of the NTA detector system,
each site has a group of detector units (DUs) for individual FOVs
(Fig.~\ref{fig:NTA-DU}).
Each DU is composed of four LCs responsible for the same FOV.
The LC has the design similar with Ashra-1
but with dimensions scaled up by 1.5 to gather more light.
Each LC is instrumented with a pupil lens, seven segmented mirrors, a set of
photoelectric imaging tube and image pipeline with a CMOS sensor.
The DU has a unified trigger system which determine if image light gathered from four LCs
through fiber-optic bundle transmission systems has enough confidence.
Detailed design studies for the NTA detector are currently underway.

\subsection{Scientific Goal and Observational Objectives}
\noindent
The main scientific goal of the NTA project is:

\vspace{2mm}
\centerline{\it the clear discovery and identification of
non-thermal hadronic process in the Universe.}
\vspace{2mm}

\begin{figure}[t]
   \begin{center}
    \includegraphics[width=0.8\hsize]{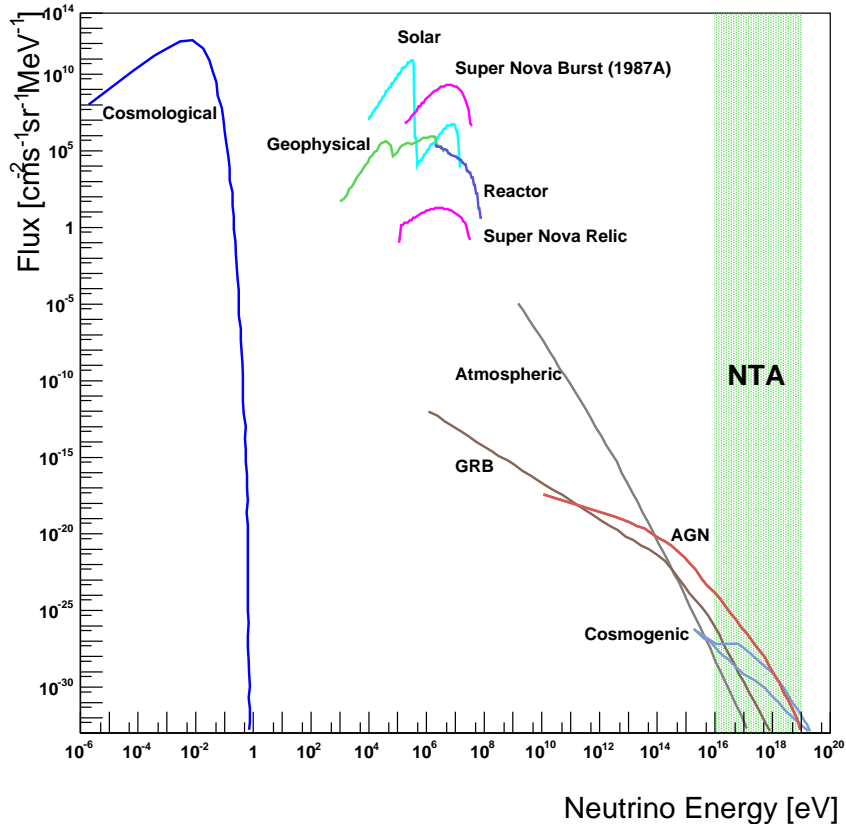}
   \end{center}
\vskip-0.3cm
 \caption{
 	Measured and expected neutrino fluxes, and sensitive energy region of NTA
 	(green band).
 }
	\label{fig:neutrinoflux}
\end{figure}

\noindent
This
has not been directly confirmed by any observation
so far and can be achieved by observing PeV-EeV neutrino emission as
direct evidence and sensitive probe for collective processes
that accelerate particles to energies many orders of
magnitude beyond thermal energies.
Fig.~\ref{fig:neutrinoflux} shows
measured and expected neutrino fluxes, and sensitive energy region of NTA
with the sensitive energy range of PeV-EeV.

\begin{figure}[t]
   \begin{center}
    \includegraphics[width=0.8\hsize]{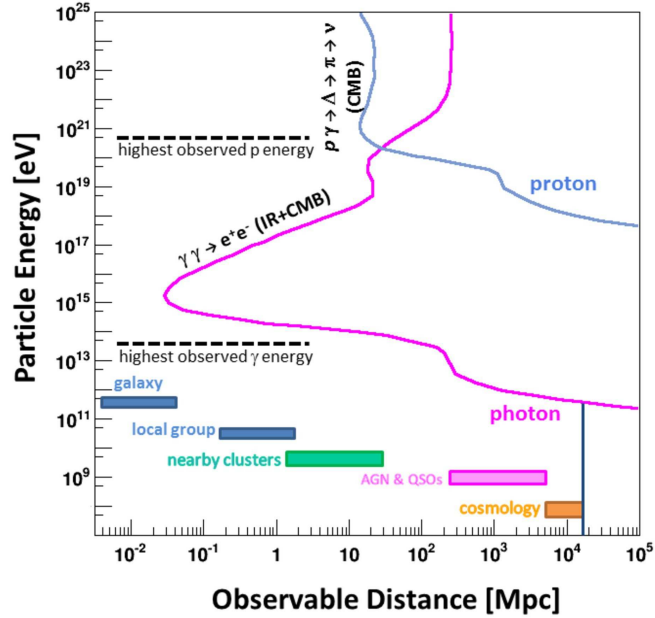}
   \end{center}
\vskip-0.3cm
 \caption{
 	Particle energies and observable distances through the interactions with background photons
 	with distance regions for sources (colored boxes), and
 	the highest observed gamma and proton energies (dashed lines).
 }
	\label{fig:EvsObservableDistance}
\end{figure}

The multimessenger connection among Cosmic Rays, photons and neutrinos of different particles is
crucial for comprehensive and deeper understandings of the fundamental non-thermal astrophysical processes.
Multimessenger is a theme to much of the recent literature
{\it e.g.}
\cite{beckerhigh-energy2008}
\cite{PhysRevD.88.121301}
\cite{0004-637X-768-2-186}.
The measured fluxes of extremely-high energy cosmic rays (EHECR) with energies above 10$^{18}$~eV (EeV)
inspire an associated flux of PeV-EeV cosmic neutrinos,
although the production mechanisms of EHECR are still unknown.
PeV-EeV neutrinos are predicted as a result of the decay of charged pions generated in
interactions of EHECRs within the source objects (astrophysical neutrinos) and
in their propagation through background photon fields (cosmogenic neutrinos)
\cite{Berezinsky-Zatsepin1969}.
Cosmic rays up to and even beyond the PeV (``knee'') are of Galactic origin.
Around EeV between 10$^{17}$ and 10$^{18.5}$~eV (``ankle''), at maximum,
known Galactic source candidates are generically
considered running out of power and extragalactic sources start dominating the spectrum.
On the other hand, from recent calculations,
the maximum energy of accelerated particles may reach 5$\times$10$^{18}$~eV for Fe ions in Type IIb Supernova Remnants (SNRs)
\cite{0004-637X-718-1-31}
\cite{PhysRevLett.105.091101}.
Adding that, both the detailed composition and galactic-extragalactic transitions in the PeV-EeV region
is still unclear and  to be understood
\cite{koterathe2011}
\cite{aloisiodisappointing2012}
\cite{giacinticosmic2012}.
Simultaneous searches for PeV gamma rays and neutrinos would be useful to distinguish between
galactic and extragalactic sources of cosmic rays
\cite{Gupta201375}.
If EHECRs are produced from Galactic point sources, then those
 point sources are also emitting PeV gamma rays.
 We note that the detection of galactocentric PeV
 gamma rays in the future would be a signature of the presence of EeV cosmic accelerators in the Milky
 Way
 \cite{guptapev2011}.
EHECR sources in our galaxy will plausibly be investigated by the multimessenger approaches with
very-high energy gamma rays and neutrinos in the PeV-EeV region, since
the galactic size is within the observable distances of gamma rays around PeV even after the propagations
in the background photons
(Fig.~\ref{fig:EvsObservableDistance}).

IceCube recently reported the observation of three neutrinos with energies at 1-2~PeV
\cite{PhysRevLett.111.021103}
and 26 additional events at lower energies
\cite{icecube2013evidence}, which are significantly inconsistent with the background.
Many authors have discussed the origins and physics scenarios
of the IceCube signals in the context of multimessenger of
EHECRs, VHE gammas, and VHE neutrinos among galactic and extragalactic sources, {\it e.g.}
\cite{PhysRevD.88.121301}
\cite{anchordoqui2014cosmic}.

\vspace{2mm}
The following potential candidates are considered
as search objectives with the NTA survey.

\begin{itemize}
\item Galactic Sources: \\
The galactic   supernova   remnants (SNRs)   are   widely
believed to be the dominant source for the cosmic rays
(CRs)   at   energies   below   the   “knee” around PeV,
most probably through the  diffusive shock acceleration mechanism
\cite{hillas2005can}.
Recent calculations show that Super Nova Remnant (SNR) acceleration
in our galaxy can describe the whole energy spectrum of observed cosmic rays
for the region from TeV up to the ankle,
using different types of SNs and transition of composition in the galaxy
\cite{0004-637X-718-1-31}.
Galactic GRBs, which are beamed away from Earth, can be the main source of
Galactic cosmic rays at all energies
\cite{1999A&A...349..259D}.
From the observational point of view,
Imaging Air Cherenkov Telescopes (IACTs)
have detected more than hundreds of TeV $\gamma$-ray sources,
including about 30 SNRs
\cite{1367-2630-11-5-055005}.
There are three classes of such objects:
shell-type supernova remnants, pulsar wind nebulae, and binary systems.
The expected neutrino fluxes from these sources and diffuse emission
from cosmic ray interaction
are calculated~\cite{0004-637X-656-2-870}.

{\it Shell-type Supernova remnants} have long been considered as the likely
acceleration site for the bulk of the galactic cosmic rays.
The morphology of $\gamma$-ray emission from RXJ1713.7-3946
was studied~\cite{2004Natur.432...75A}.
A TeV $\gamma$-ray image of
the SNR  demonstrates that VHE particles are accelerated at
the spatially resolved remnant, which has a shell morphology
similar to that seen in X-rays.
The energy spectrum
indicates efficient acceleration of charged particles to energies
beyond 100~TeV, consistent with current ideas of particle
acceleration in young SNR shocks.
Spatial correlations of the $\gamma$-ray emission with available target material
seem to be present for the SNRs W28, IC443, RCW86 and
RX~J0852.0-4622 supernova in IACT data.
The observations of $\gamma$-rays exceeding 10~TeV in the spectrum of the
RX~J0852.0-4622 supernova
\cite{2005A&A...437L...7A}
has also strengthened the hypothesis that the hadronic acceleration is
the process needed to explain the hard and intense TeV $\gamma$-ray
spectrum.
Such a correlation is also seen in the region of the
Galactic Centre (GC), where
the acceleration site of the cosmic rays is not clear
\cite{aharonian2006discovery}.
However,
the directional distribution of the 21 cascade events suggests weakly significant excess (``hotspot'')
with a trial-corrected significance of 8~\%
\cite{icecube2013evidence}.
Possible contributions from galactic neutrino sources like SNRs are consistent with the present
diffuse $\gamma$-ray limits
\cite{ahlersprobing2014}.
If the neutrino spectrum is dominated by galactic sources,
the lack of observed CR anisotropy requires a soft neutrino spectrum with index $\sim$~2.3
in the hadronuclear ($pp$) origin scenario
\cite{anchordoqui2014cosmic}.
The required index is consistent with a  spectral index 2.2 of a point-like $\gamma$-ray source
at the Galactic center, which was reported by H.E.S.S.
\cite{aharonian2004very}.
The possibility of discrimination between $pp$ and $p\gamma$ source models by combining the measured
neutrino and $\gamma$-ray fluxes, will be one example for the multimessenger approach
\cite{PhysRevD.88.121301}.

{\it Pulsar Wind Nebulae} (PWNe) are some of the brightest TeV $\gamma$-ray sources.
The central pulsar emits material into the nebulae
such as the powerful Crab and Vela pulsars
\cite{2006A&A...448L..43A}.
A significant fraction of nuclei is suggested to exist in pulsar winds
\cite{1992ApJ...390..454H}.
The decay of pions produced in the interaction of these nuclei
can dominate the  TeV $\gamma$-ray emission, which suggests
significant production of neutrinos should occur
\cite{2006A&A...451L..51H}.
These nuclei and significant production of neutrinos may occur {\it e.g.} \cite{2006A&A...451L..51H}.
Pulsars could also be a strong source of very-high energy neutrinos
\cite{link2006flux},
although there is a pessimistic estimate of the fluxes
\cite{bhadra2009tev}.

{\it Binary systems} of a compact object and a massive star are well
established galactic TeV $\gamma$-ray sources, which are classified into
binary PWN or microquasars.
In the binary pulsar scenario, the spin-down of the neutron star is the energy source.
In the microquasar scenario, accretion is the power-source, and particle acceleration
occurs in relativistic jets produced close to the compact object
(black hole or neutron star).
The PSR~1259-63 system with 3.4-year period and the Be-star SS~2883
belong to the class of binary PWN.
LS~5039 and LSI~+61~303,
are the remaining well established systems and expected
as strong neutrino sources~\cite{Aharonian-JPhysConfSer38},
of which acceleration site has not been revealed yet.
Cyg~X-1 is expected as the best $\gamma$-ray microquasar candidate, which hosts
a black hole~\cite{2006Sci...312.1759M}.

{\it Undetected bright hard-spectrum sources} beyond $\sim$1~PeV could in principle be
missed by current Cherenkov telescopes,
since they have substantially reduced energy flux sensitivities in the higher energy region
relative to their performance around 1~TeV.
Due to the rapid rise of the effective detection area
of NTA with energy,
such sources could be promising candidates for the NTA detector.
Several candidates for sources with hadron acceleration beyond 1~PeV have been
identified in the Cygnus region by Milagro
\cite{0004-637X-688-2-1078}.

{\it Our Galactic Center} (GC)
has also been proposed as neutrino sources.
An intense diffuse emission of $\gamma$-rays with higher energies
has been observed which likely implies the
presence of a source of cosmic ray protons
and thus of neutrinos~\cite{aharonian2006discovery}.
The GC region is of
particular interest because
it is in the good sky view of NTA located
on Hawaii Island in the northern hemisphere.
A general scenario of Galactic  $\gtrsim$10 PeV cosmic-ray
interactions to produce PeV-EeV events
\cite{Gupta201375},
and plausible spectra of neutrino events
as originating from Galactic cosmic rays
\cite{2013arXiv1306.5021A},
has been considered as well.
IceCube has announced
detection of 26 neutrino events with energies in the
$\sim$30-250~TeV range
\cite{icecube2013evidence},
in addition to the two events announced earlier with
$\sim$1~PeV energy each
\cite{PhysRevLett.111.021103}.
The largest concentration of 5 shower-like events detected by IceCube
is near the Galactic Center within uncertainties of their reconstructed directions.
Adding that, IceCube observed 3 shower-like events which have their arrival directions
consistent with the Fermi bubbles~\cite{0004-637X-724-2-1044}.
There is an absence
of any track-like events in this region.
Most of the track-like
events are out of the Galactic plane, and at least 4 of them
are correlated with shower-like events in those regions
\cite{PhysRevD.88.081302}.\\

\item Extragalactic Sources:\\
As the extragalactic candidates for
PeV-EeV neutrino emission,
Gamma Ray Bursts (GRB),
Active Galactic Nuclei (AGN)
and galaxy clusters
are well motivated.
PeV-EeV neutrinos are also directly linked with the physics of
proton acceleration to extremely high energy cosmic rays (EHECR)
above EeV at cosmic ray origin objects.
Recent measurements of the composition of EHECRs
by the Pierre Auger Observatory (Auger) have suggested
that the mean nuclear mass may
increase with energy between 2~EeV and 35~EeV
\cite{PhysRevLett.104.091101}.

{\it Gamma Ray Bursts} (GRBs)
eject the most energetic outflows in the observed Universe,
with jets of material expanding relativistically into the surrounding interstellar matter
with a Lorentz factor $\Gamma$ of 100 or more.
Energy dissipation processes involving nonthermal interactions between particles
are thought to play an important role in GRBs, but remain observationally unresolved.
The detection of PeV--EeV~ neutrinos ($\nu$s) from a GRB would provide direct evidence for
the acceleration of hadrons into the EeV range, and of photopion interactions in the GRB.
The GRB standard model
\cite{Meszaros06},
which is based on internal/external shock acceleration,
has been used to describe the general features of a GRB and the
observed multi-wavelength afterglow.
However, the standard model cannot reproduce well the recent observational results~\cite{Ackermann10}.
The early X-ray afterglows detected by {\it Swift} exhibit a canonical
behavior of steep-flat-steep in their light curve
\cite{Nousek06}.
In some of GRBs, precursor activities were observed
\cite{Burlon08}.
In some cases, the precursor preceded the main burst by several hundred seconds with significant energy emission.
To better understand the ambiguous mechanisms of GRBs,
observational probes of the optically thick region of the electromagnetic components,
as well as hadron acceleration processes  throughout the precursor, prompt, and afterglow phases are required.
VHE~$\nu$s can be used as direct observational probes,  which are effective even in
optically thick regions.
The discovery of nearby low-luminosity (LL) GRB060218
suggests a much higher local event rate of LL-GRBs
\cite{Murase06},
which NTA can easily search for.
NTA can check the ratio between the observed neutrino event rates
from the Earth and the sky in the field of view of the detector,
which means the measurement of the diffuse neutrino background
from all GRBs with less systematic error.

{\it Active Galactic Nuclei}
consist of super-massive black holes with 10$^6\sim 10^9$ solar masses
in their centre.
The black hole radiates huge amount of energy typically of the order of 10$^{44}$~erg/s,
which is transfered from gravitational energy after it accretes matter.
The energy is expected to induce acceleration of particles.
A special class of AGN, Blazers,
has jet aligned closely to the line of sight,
which can be strong gamma-ray sources.
Many sources are reported at GeV and TeV energies by {Fermi LAT}~\cite{Aharonian-Buckley-Kifune2008}.
Gamma ray emission from blazars is often highly variable,
e.g. PKS 2155-304,
with the most extreme variation observed
an increase by two orders of magnitude within one hour
\cite{HESS-PKS2155}.
We
should observe the
neutrino fluxes from such a source significantly
within a short time.
An observation of outburst from
the blazar 1ES~1959+650~\cite{1ES1959} suggests another type of neutrino sources,
which is TeV emission without being accompanied by X-ray emission
as synchrotron self-Compton (SSC) models typically predict.
A hadronic model does not require TeV emission accompanied by X-ray emission.
The observed flares are encouraging sites to search for high energy neutrino emission.

{\it Starburst galaxies}
have unusually high rates
of large-scale star formation processes.
A galactic-scale wind is driven by the collective effect of supernova explosions
and massive stars at the central regions of the starburst galaxies.
IACTs have detected
the gamma ray flux at several hundred GeV from
the starburst galaxies NGC253 and M82~\cite{HESS-Science326}~\cite{VERITAS-Starburst}.
They suggests
cosmic ray densities much
higher than  typical case expected in our own Galaxy
by two to three orders of magnitude.
The diffuse neutrino flux from all starburst galaxies
is expected detectable with current detectors
\cite{loeb2006cumulative}.

{\it Cosmogenic Neutrinos} are the secondary particles of the
Greisen-Zatsepin-Kuzmin (GZK) process from the interaction of
the highest energy cosmic rays with the cosmic microwave background
\cite{Berezinsky-Zatsepin1969,PhysRevLett.16.748,zatsepin1966upper}.
Various cosmogenic neutrino models
(for example \cite{ahlers2010gzk})
which assume primary cosmic ray protons predict neutrino fluxes.
They require
4$\pi$ solid angle averaged neutrino effective area
$A_{\nu}$
to be more than 10$^{-3}~$km$^2$ at 100~PeV
to detect several cosmogenic neutrinos every year
in case of full duty cycle.
NTA satisfies
this requirement well even assuming the duty of 10\%.\\
The predicted flux has large uncertainties due to dependence on
source spectrum and on spatial distribution and cosmological evolution of the sources
\cite{koterathe2011}.
If EHECRs are heavy nuclei like irons, the yield of the cosmogenic neutrino is
strongly suppressed
\cite{avecosmogenic2005}.
Therefore NTA has sufficient sensitivity for cosmogenic neutrinos to directly test
the hypothesis of
the observed highest energy cut-off of the cosmic ray spectrum
due to a suppression induced by
the GZK propagation of pure protons
taking into account of the uncertainty of flux prediction
even in the case of null result.

\item Dark Matter and New Particles \\
{\it Weakly Interacting Massive Particles (WIMPs)}
are favoured dark matter candidates, which are preferentially discussed
in the minimal supersymmetric standard model (MSSM) framework
\cite{1367-2630-11-10-105026}.
Indirect WIMP detection use secondary particles such as
$\gamma$s, $\nu$s, weak bosons, tau pairs and so on from annihilations.
Direct WIMP detection uses recoil nuclei from elastic WIMP-nucleus scattering.
There is some complementarity between direct and indirect searches for dark matter,
given the astrophysical assumptions inherent to the calculations.
Both methods are sensitive to opposite extremes of the velocity distribution
of dark matter particles in the Galaxy
(low-velocity particles are captured more efficiently in the Sun,
high-velocity particles leave clearer signals in direct detection experiments),
as well as presenting different sensitivity to the structure of the dark matter halo
(a local void or clump can deplete or enhance the possibilities for direct detection).
IceCube has evaluated these data for evidence of dark matter annihilations in the Sun,
in the Galactic Center, and in the Galactic Halo, searching for an excess neutrino flux
over the expected atmospheric neutrino background,
which provides the results of dark matter searches for WIMPs,
Kaluza-Klein modes and super heavy candidates (Simpzillas),
using the 79-string configurations of IceCube
\cite{PhysRevLett.110.131302}.
Given that the Sun is essentially a proton target and that
the muon flux at the detector can be related to the capture
rate of neutralinos in the Sun, the IceCube limits on the
spin-dependent neutralino-proton cross section are currently well
below the reach of direct search experiments,
proving that neutrino telescopes are competitive in this respect.
The simple assumption that dark matter is
a thermal relic limits the maximum mass of the
dark matter particle,
which turns out to be a few hundred TeV for a thermal WIMP,
the so called unitarity constraint.
However, dark particles might have never experienced
local chemical equilibrium during the evolution of the Universe,
and their mass may be in the range much larger than the mass of thermal WIMPs,
which have been called {\it WIMPZILLAs}
\cite{chung1998superheavy,Blasi200257}.
NTA can perform the most sensitive indirect search for WIMPZILLAs,
with much better effective detection area for tau neutrinos from annihilation in the Sun,
especially above $\sim 10$~PeV, the complementary sensitive energy region for IceCube.

{\it Super-heavy particles}
(M$\gtrsim 10^{4}$~GeV) produced during inflation may be the dark matter,
independent of their interaction strength.
Most popular ones are SIMPZILLAs,
magnetic monopoles, supersymmetric Q-balls and nuclearites.

{\it Strongly interacting super-heavy particles
(SIMPZILLAs)}
will be captured by the Sun,
and their annihilation
in the center of the Sun will produce a flux of energetic tau neutrinos
that should be detectable by neutrino telescopes
\cite{PhysRevD.64.083504}.

{\it Magnetic monopoles} turn out to be consequence of most variants of
Grand Unified Theories
\cite{1742-6596-116-1-012005}.
The electromagnetic energy losses of monopoles in  the atmosphere,
as well as neutrinos produced from monopole-antimonopole annihilations
in the Sun and Earth, induce clear signatures in optical (Cherenkov and fluorescence)
air-shower detectors like NTA
\cite{Wick2003663}.

{\it Nuclearites
(strange quark matter or strangelets)} are hypothetical aggregates of
u, d and s quarks, combined with electrons to adjust electric neutrality.
Nuclearites, like meteors, produce visible light as they traverse the atmosphere.
Their luminosity as a function of their mass
is
$L=1.5\times 10^{-3} (M/1\mu$g) watt
\cite{de1984nuclearites}.
For example, the apparent visual magnitude of a 20 g nuclearite
at a height of 10~km is $-1.4$, equal to that of the brightest star, Sirius.
Atmospheric nuclearites at galactic velocities
($v\sim250$~km/s) can easily be distinguished from ordinary meteors bounded to the Solar System,
moving no faster than 72~km/s.
It could be identified with clear evidence
with the wide FOV high resolution optical detector of NTA.

{\it Q-balls} are hypothetical coherent states of quarks, sleptons and Higgs fields
\cite{Kusenko1997108}.
Neutral Q-ball (Supersymmetric Electrically Neutral Solitons, SENS) could
catalyse proton decay along their path, similar to GUT monopoles.
Electrically charged Q-ball (Supersymmetric Electrically Charged Solitons, SECS)
would produce light in a similar way as nuclearites.

\end{itemize}

\section{Earth Skimming Tau Neutrino}\label{sec:skim}
\subsection{Neutrino detection method}
\begin{figure}[t!]
   \begin{center}
    \includegraphics[width=0.8\hsize]{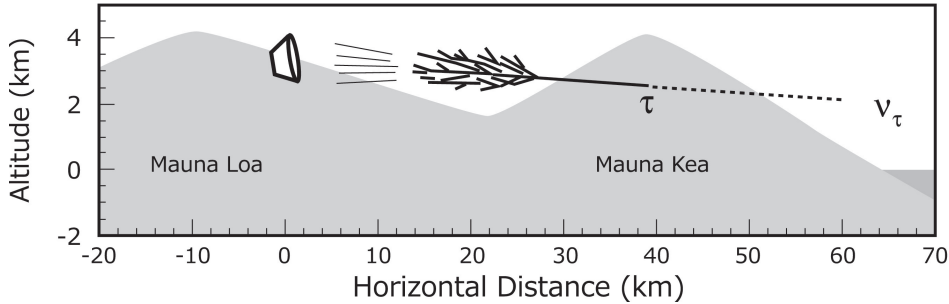}
   \end{center}
\vskip-0.3cm
 \caption{
	Schematic view of Cherenkov $\tau$ shower ES method.
	Mauna Kea is used as the target mass for neutrino charged current interaction.
	The produced air-shower is observed from Mauna Loa.
	In addition to the fact that the mountain can be viewed with
	large solid angle from the observatory, the distance of about 30~km from
	the observatory to the Mauna Kea surface is appropriate for the air-shower
	development, resulting in the huge advantage of the Ashra-1 observatory.}
	\label{fig:mtskim}
\end{figure}

\noindent
To detect VHE neutrinos, a large target volume is required in order to compensate for the
very small neutrino-nucleus cross section.
On that basis, the secondary particles produced by the first neutrino interaction
must be detected in one way or another.
The detection method using water and ice as a target detects Cherenkov light from secondary muons,
taking advantage of the fact that ice and water are optically transmissive to some extent.
This method can be categorized as the method in which the target and detection
volumes are
water and near-by rock surrounded by the water tank.
On the other hand,
the detection method using air-showers aims at the detection of higher-energy neutrinos.
This method enables us to achieve a huge detection volume as the atmosphere has
very high transmittance. However,
it is difficult to obtain a larger target mass due to low atmospheric density.
The detection method called Earth-skimming $\nu_{\tau}$ technique
\cite{Domokos98,Letessier00,Athar00,Fargion02,Feng02}
can realize a huge target mass and detection volume at the same time,
by dividing the target and detection volume
utilizing the interaction process of $\nu_{\tau}$.
The detection method is described as follows (see Fig.~\ref{fig:mtskim}).
The VHE~$\nu_{\tau}$ interacts in the Earth or mountain and produces tau lepton ($\tau$).
$\tau$ penetrates the Earth and/or mountain and appears in the atmosphere.
Subsequently, it decays and produces an air-shower.
Cherenkov photons from the air-shower are detected.
Owing to the separation of the first interaction where $\nu_{\tau}$ produces
$\tau$ and the $\tau$ decay that generates the air-shower,
air-shower observation becomes possible while preserving the huge target mass required
for the first interaction.
``Cherenkov $\tau$ shower ES method'' is defined as the detection
method which detects Cherenkov photons from tau shower appearing from the Earth or
the mountain fully utilizing this feature.
We note, for example, that Mauna Kea is over 3,200~km$^3$ in volume
and 9.3 tera tons in mass \cite{MaunaKea}.

\subsection{Deflection from parent tau neutrinos}\label{sec:tauang}
\noindent
This section describes deflection of Cherenkov $\tau$ shower compared to the
arrival direction of parent $\nu_{\tau}$,
in order to estimate the ability to trace back to the accelerator based on the direction
of the detected air-shower.
We evaluate the deflection of the propagating particle in each step of
neutrino charged current interaction, $\tau$ propagation in the Earth,
tau decay, and production of extensive air-shower.
We use PYTHIA \cite{PYTHIA6154} to evaluate neutrino charged current interaction.
Since $P_\tau < M_W$ where $P_\tau$ denotes the transverse momentum of a produced $\tau$
and $M_W$ denotes the mass of the W boson,
the deflection angle $\tau$ ($\Delta \theta_{\tau}$) with respect
to the parent $\nu_{\tau}$ should be less than 0.3~arcmin for $E_{\tau}>1$~PeV.
The simulation results with PYTHIA are consistent with this.

Second, we use GEANT4 \cite{Geant4} to evaluate the deflection of the $\tau$ due
to propagation in the Earth.
To estimate the energy loss of high energy leptons, the following parametrization
is generally adopted \cite{Dutta2001}:
\[ -\left < \frac{dE}{dX} \right > = \alpha + \beta E, \]
where $\alpha$ denotes the nearly constant ionization loss,
and $\beta$ denotes the radiative energy loss due to Bremsstrahlung, pair production
and photonuclear interaction.
Since radiative energy loss is dominant for high energy $\tau$s,
these high energy processes must be included in the ``Physics List" of GEANT4.
Thus, we apply the following processes originally defined for muons to $\tau$s,
and estimated the deflection after propagating through 10~km of rock.
\begin{itemize}
\item G4MuBremsstrahlung: Bremsstrahlung
\item G4MuPairProduction: $e^+e^-$ pair production
	\footnote{We modified the original G4MuPairProduction so that
	momentum conservation includes the produced particles, resulting
	in the inclusion of deflection.}
\item G4MuNuclearInteraction: Photonuclear Interaction
\end{itemize}

\begin{figure}[t!]
   \begin{center}
 \begin{tabular}{cc}
  \begin{minipage}{0.4\hsize}
   \begin{center}
    \includegraphics[width=\hsize]{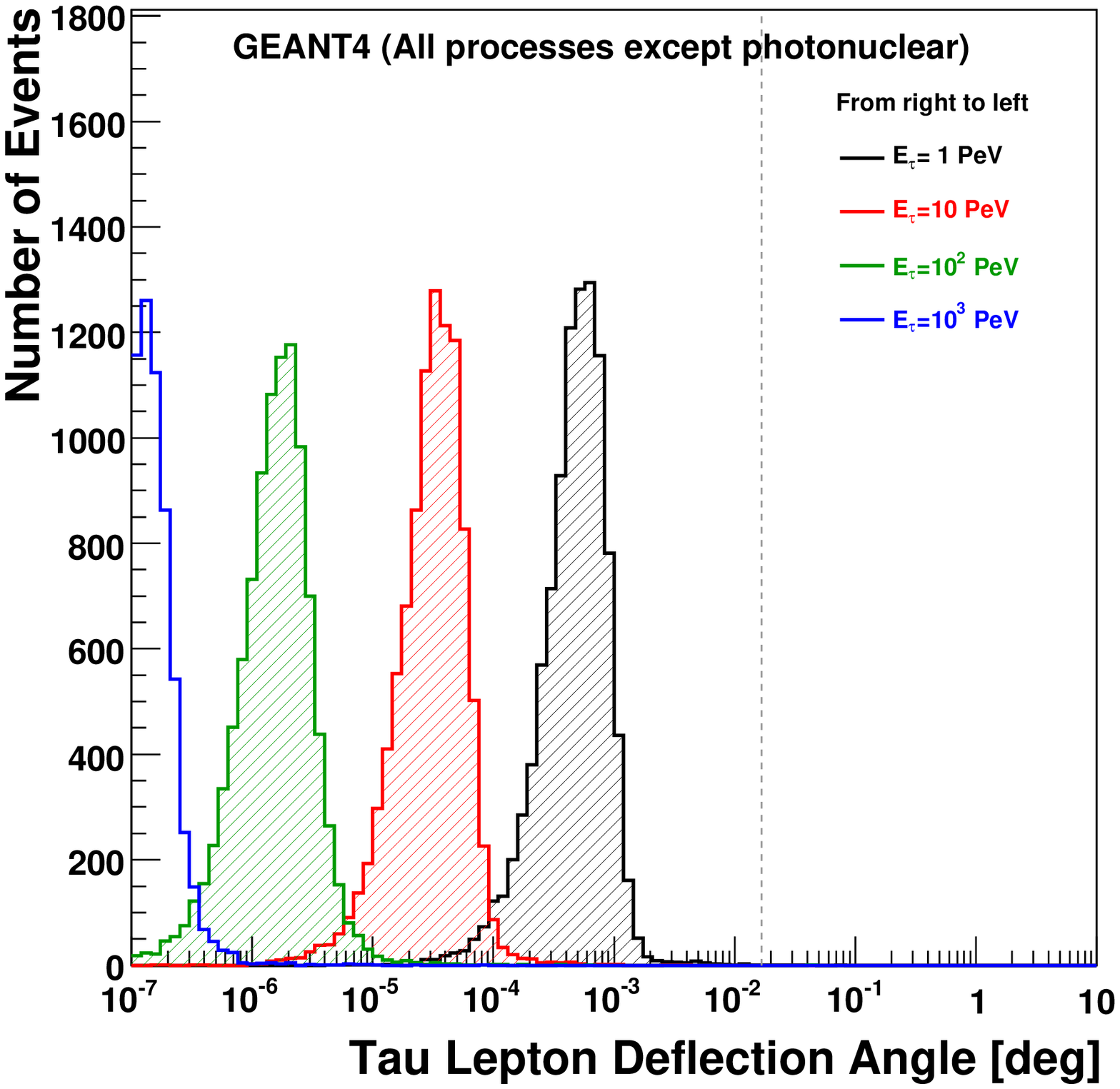}
   \end{center}
  \end{minipage}
  \begin{minipage}{0.4\hsize}
   \begin{center}
    \includegraphics[width=\hsize]{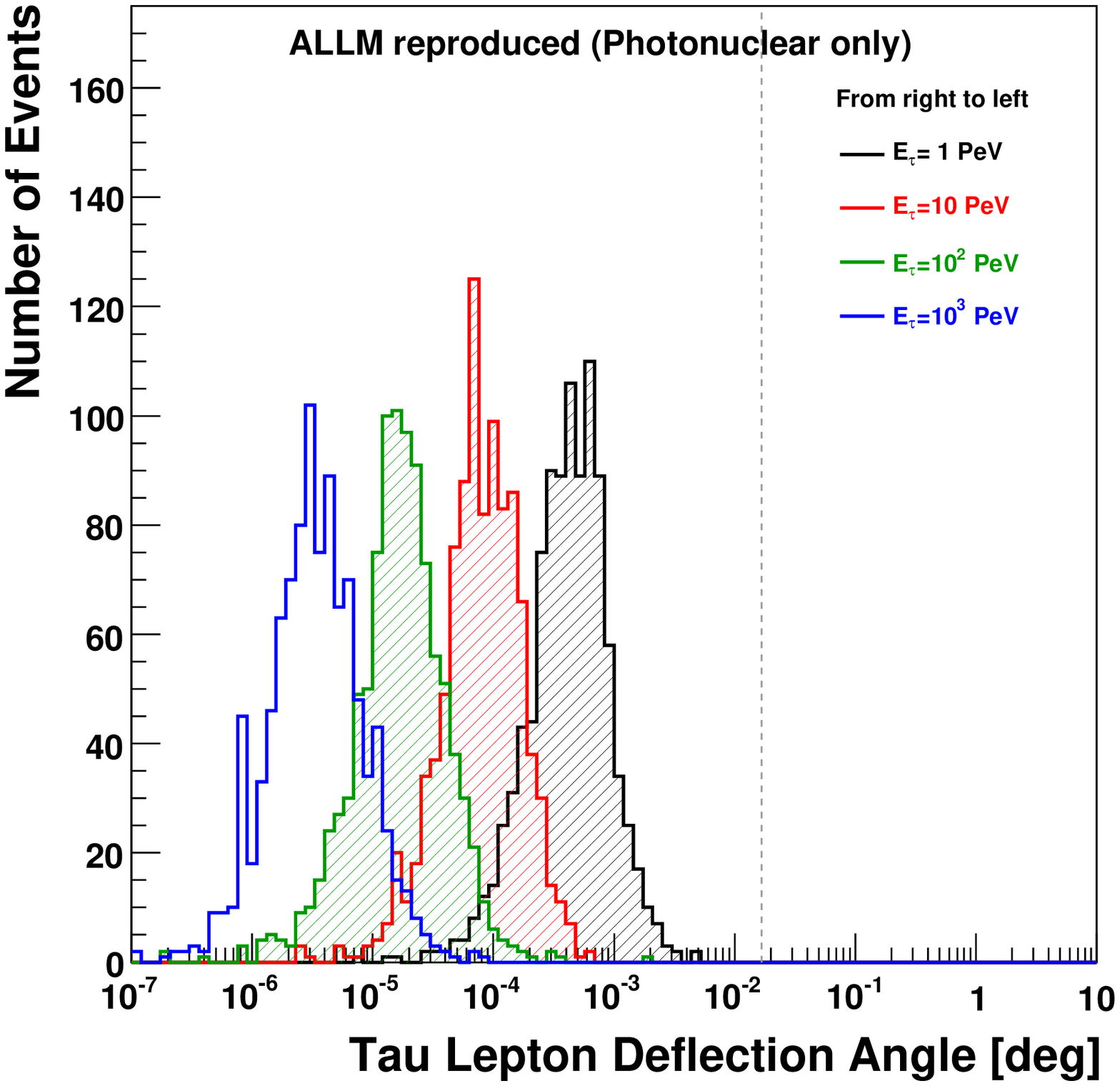}
   \end{center}
  \end{minipage}\\
 \end{tabular}
\vskip-0.3cm
 \caption{The simulation results of deflection angle
	of $\tau$s after propagating through 10~km of rock:
	({\it Left}) the GEANT4 result including all high energy processes
	except for photonuclear interaction;
	({\it Right}) the result of photonuclear interaction from custom simulation.
	Note that the decay of $\tau$ was switched off for the above simulations.
	The hatched histograms indicate that the $\tau$ range is less than 10~km.}
 \label{fig:defl}
 \end{center}
\end{figure}

To validate our GEANT4 simulation, we compare the energy dependence of $\beta$
for Bremsstrahlung, pair production, and photonuclear interaction to Ref.
\cite{Dutta2001}.
The $\beta$ energy dependence
agrees well for the former two processes, but we
find that GEANT4 produces smaller values for photonuclear interaction
at higher energy, and that the difference
is a factor of 3 at 10$^8$~GeV.
We write accordingly a toy Monte Carlo simulation for photonuclear
interaction using the formalism of Refs. \cite{ALLM, DiffCS},
reproducing the energy dependence of $\beta$
within $\pm$30~\% accuracy.

Fig.~\ref{fig:defl} shows the simulation results for $\tau$ deflection
after propagating through 10~km of rock.
The left panel shows the GEANT4 result including all high energy processes
except photonuclear interaction, while right panel shows
our ``homemade" simulation result
for the latter.
These results indicate that photonuclear interaction becomes dominant for
deflection at 1~PeV and higher.
Note that the decay of $\tau$ was switched off for the above simulations,
and the hatched histograms indicate that the $\tau$ range is less than 10~km.
For example, the $\tau$ range is 4.9~km at 100~PeV.
We conclude that the deflection angle of $\tau$s with energy greater
than 1~PeV is much less than 1~arcmin.

Next, the deflection due to $\tau$ decay is estimated by using the output of
TAUOLA \cite{TAUOLA24}, taking into account $\tau$ polarization.
From the mass $m_\tau$, the deflection angle must be less than 1~arcmin
if the energy of the secondary particle is higher than 13~TeV.
Using TAUOLA output, it was shown that the probability to have
deflection greater than 1~arcmin is quite small from the decay of PeV $\tau$s.
We conclude that the deflection angle between
decay particles which produce the air-shower and parent $\nu_{\tau}$ is less than 1~arcmin.

Finally, we evaluate the direction of the hadron air-shower using CORSIKA.
At the shower maximum, we compare the direction of the parent particle (charged pion)
to that of electrons and positrons,
the dominant producers of Cherenkov photons.
We find that the angle between the average direction of electrons and positrons and
parent particle of the air-shower is coincident within $0.1^\circ$ at 1~PeV.

In conclusion, we find that the arrival direction of PeV $\nu_{\tau}$s is
preserved within $0.1^\circ$, including the hadron air-shower generation.
The accurate reconstruction of arrival direction by means of fine imaging
will be a very powerful technique in the determination of the point sources of PeV $\nu_{\tau}$s.

\section{The NTA Detector}

\noindent
The NTA observatory will consist of four sites,
Site0, Site1, Site2, Site3, as shown in Fig.~\ref{fig:NTA-Observatory}.
The conceptual layout for the NTA observatory
considers three site stations
(Site1, Site2, and Site3)
forming a 25~km-side triangle watching the total
air mass surrounded by the mountains of Mauna Loa, Mauna Kea, and Hualalai.
A single site station, Site0, at the center of the triangle has half-sky
(extendable to full-sky) coverage.
Each site has a centralized group of Detector Units (DU).
One detector unit (Fig.~\ref{fig:NTA-DU}) has
a few Light Collecting systems (LC) with segmented mirrors.
The features of the system were
studied with the Ashra-1 station site constructed on Mauna Loa (3300~m a.s.l.).

\begin{figure}[t!]
\begin{center}
\begin{tabular}{cc}
 \begin{minipage}{0.46\hsize}
   \includegraphics[width=0.98\hsize]{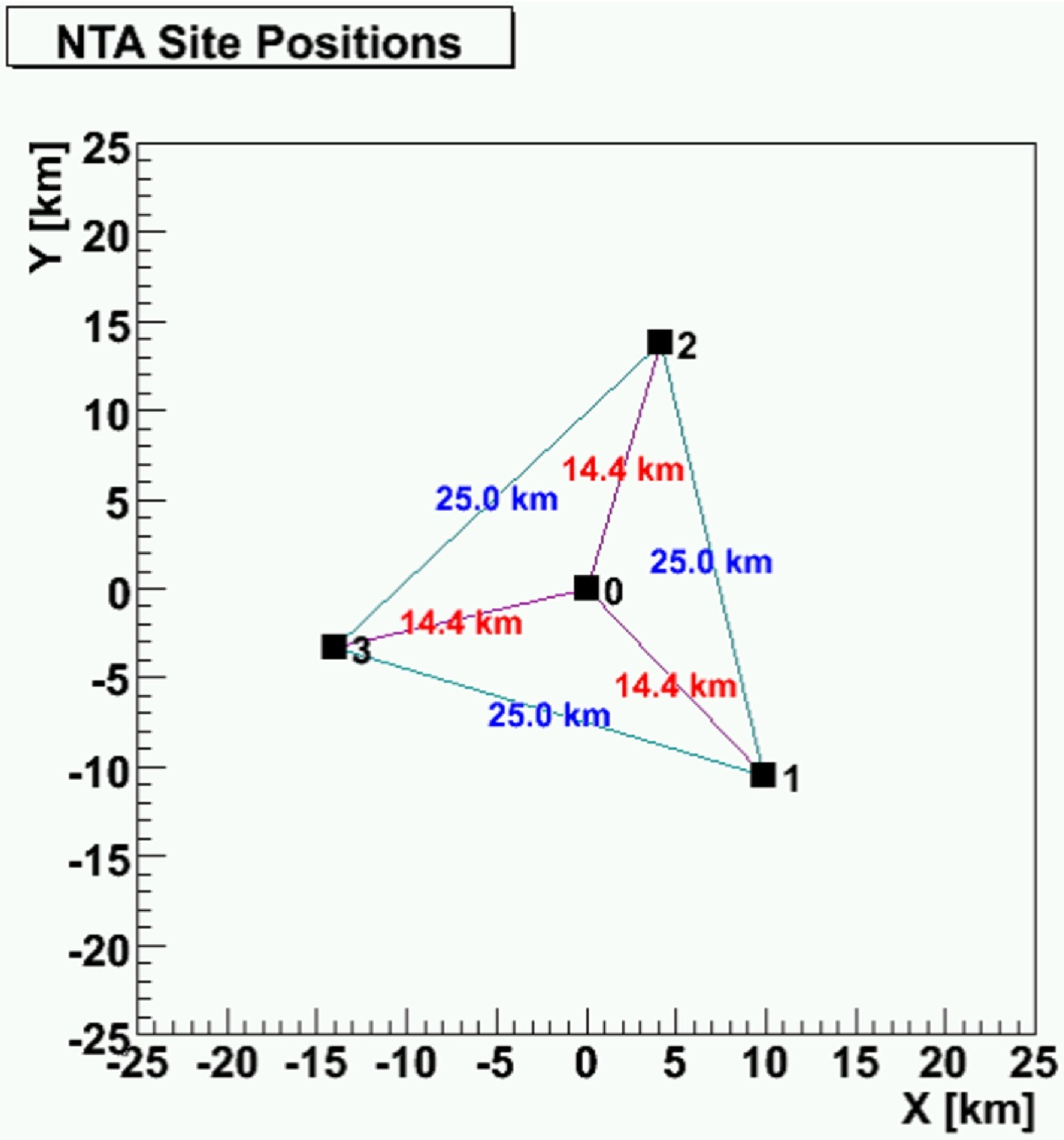}
 \end{minipage}
 \begin{minipage}{0.46\hsize}
  \begin{center}
    \includegraphics[width=0.98\hsize]{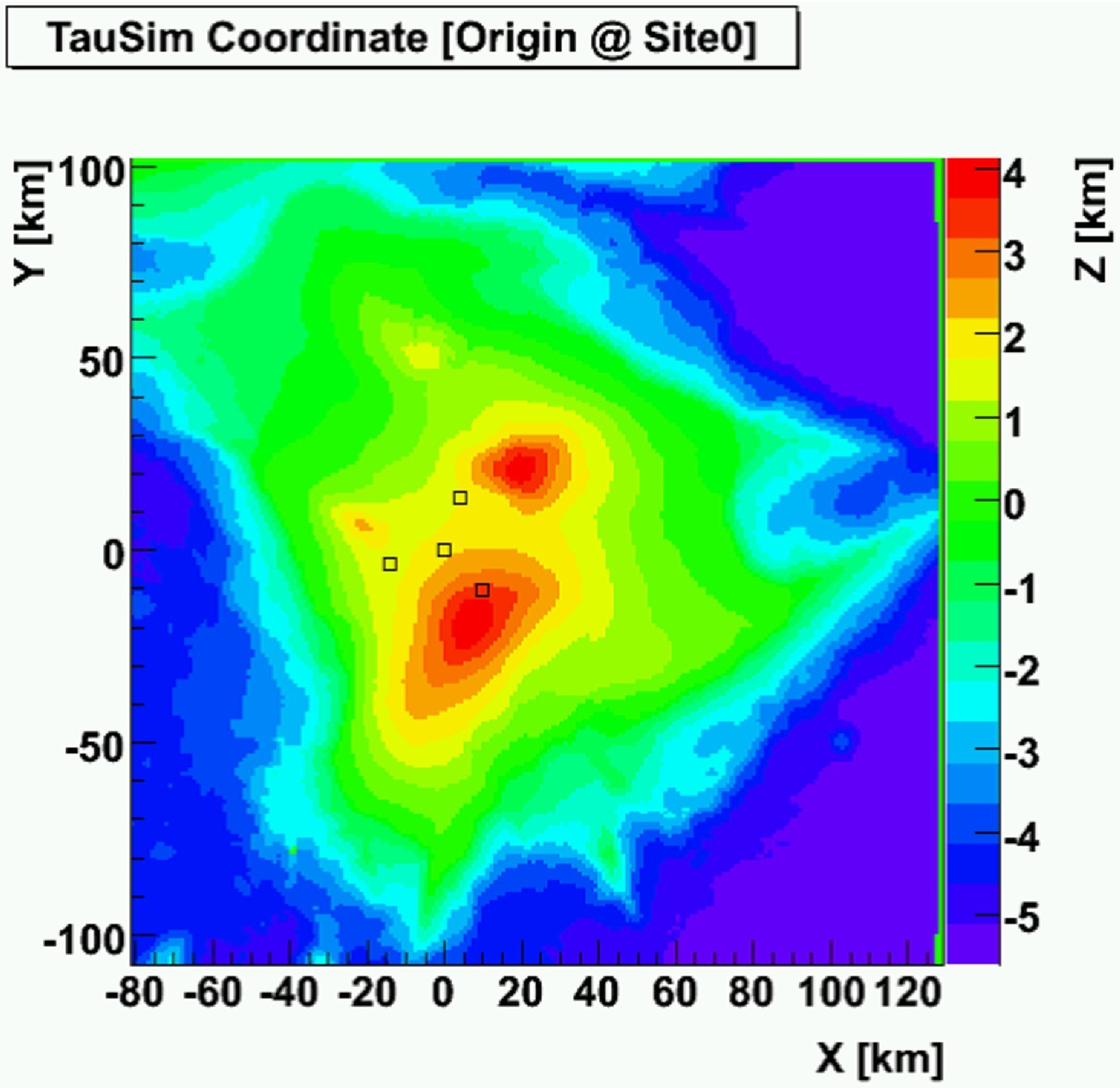}
  \end{center}
 \end{minipage}
 \end{tabular}
\vskip-0.3cm
 \caption{
 (left) The x-y coordinates of the four NTA site locations,
        with Site0 defined as the origin;
 (right) the Hawaii Island topography map superimposed with
 the four NTA site locations.
}
 \label{fig:site}
\end{center}
\end{figure}

In order to investigate the performance of the NTA detector,
we shall use the following setup conditions of
the assumed locations of
observational sites of the NTA system on Hawaii Island.
\begin{enumerate}
\item The left side of Fig.~\ref{fig:site} shows the layout of the NTA site locations.
	  The local Cartesian coordinate system is defined with
	  the origin at Site0, as denoted by ``0'' in Fig.~\ref{fig:site},
	  the positive z-axis points to the zenith,
	  and the positive y-axis points north.
	  The x-y coordinates of the site locations are from the
	  projected x-y plane, with
	  the z-coordinates from the corresponding
	  height of the topography data of Hawaii Island.
\item The three observational sites, each
    located at the vertices of a triangle
	of equal 25~km side length, are:
    Site1 at Mauna Loa Ashra-1 location, Site2 on the slope of Mauna Kea,
    and Site3 on the slope of Hualalai.
\item The central observational site, Site0, is
    at the center of gravity of the above three site locations.
    Site1--Site3 are equidistant 14.4~km from Site0.
\item The location of Site1 and Site2 are set
		at the Ashra-1 Mauna Loa Observation Site (ML-OS),
	and at 25~km distant from ML-OS in the direction of Kilohana Girl Scout Camp, respectively.
\item After above settings, the
locations of remaining
two sites
	  are automatically fixed.
\end{enumerate}

The right part of Fig.~\ref{fig:site} shows the layout of the four sites,
superimposed on the topography map image of Hawaii,
to be used as settings in the simulation program.
Table~\ref{tab:site} shows the x-y-z coordinates of the site locations
and the detection FOV coverage, as determined from the above description.
For the simulation study, given in the next section,
we assume that each LC has the total FOV of
32$^{\circ}$$\times$32$^{\circ}$,
trigger pixel FOV of
0.5$^{\circ}$$\times$0.5$^{\circ}$, and
image sensor pixel FOV of
0.125$^{\circ}$$\times$0.125$^{\circ}$.
The Site0 system consists of
12 LCs in the lower elevation angle regions,
which together cover the half-sky solid angle that is $\pi$~sr.
The other sites have only 6 LCs in the lower elevation angle region,
to cover the FOV of the half-sky solid angle which is $\pi$/2~sr.
The bottom edge of the lower elevation angle region
is defined to be $-9^{\circ}$ in elevation angle
( 9$^{\circ}$ below the horizon ).

\begin{table}[t!]
\begin{center}
\begin{tabular}{clcccc}
\hline
Site ID & Location & X [km] & Y [km] & Z [km] & FOV [sr] \\
\hline
Site0 & Center    &   0.000 &   0.00 &    2.03 &  $\pi$	\\
Site1 & Mauna Loa &   9.91  & $-$10.47 &    3.29 &  $\pi$/2 \\
Site2 & Mauna Kea &   4.12  &  13.82 &    1.70 &  $\pi$/2 \\
Site3 & Hualalai  & $-$14.02  &  $-$3.35 &    1.54 &  $\pi$/2 \\
\hline
\end{tabular}
\end{center}
\vskip-0.3cm
\caption{
The x-y-z coordinates and detection FOV coverage
of the four NTA sites, which are used in the
simulation program.
The location of Site0 is defined as the origin of the coordinate system.
}
\label{tab:site}
\end{table}

\begin{figure}[h]
\begin{center}
\begin{tabular}{cc}
 \begin{minipage}{0.48\hsize}
  \begin{center}
    \includegraphics[width=0.98\hsize]{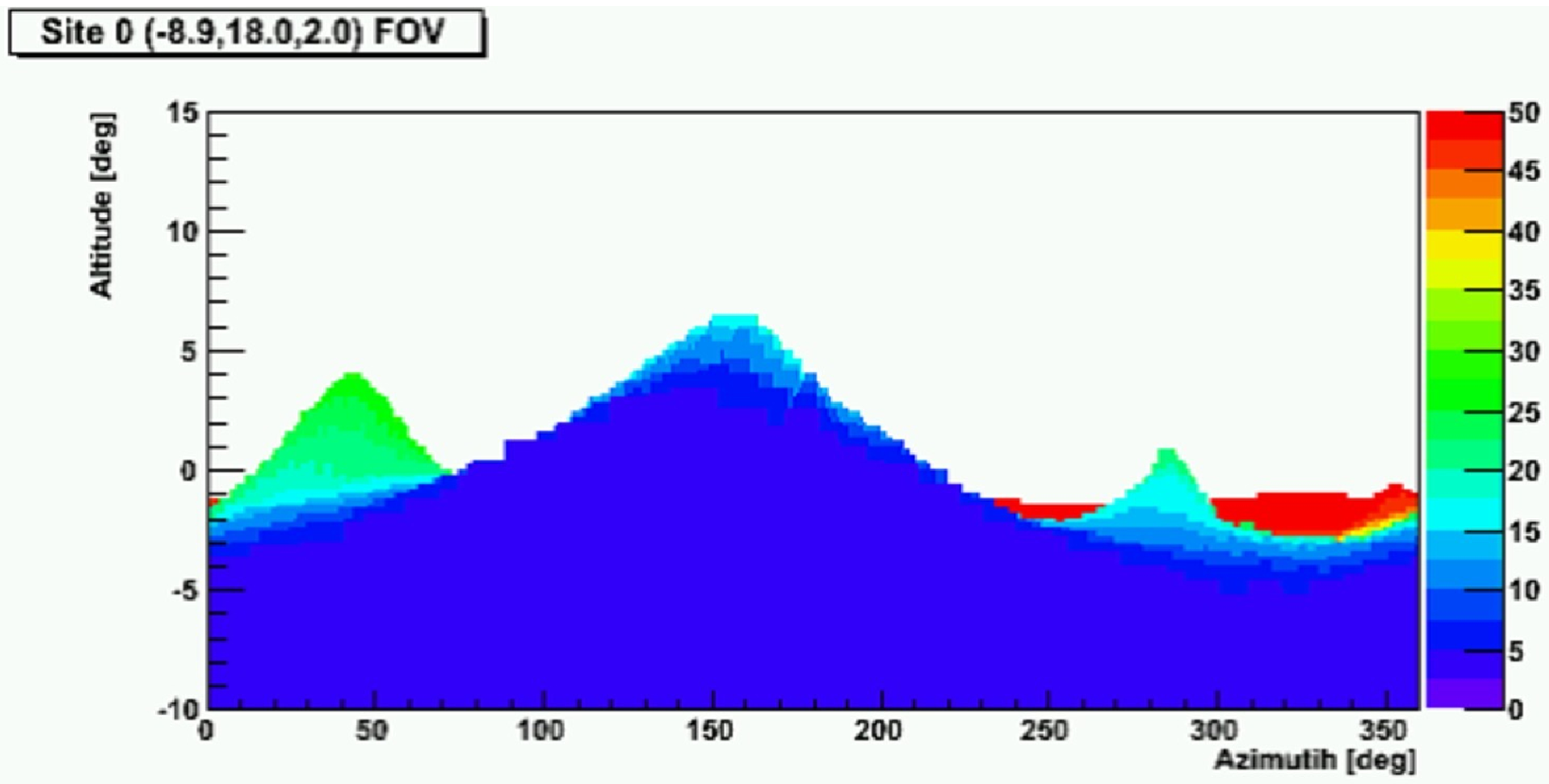}
  \end{center}
 \end{minipage} &
 \begin{minipage}{0.48\hsize}
  \begin{center}
    \includegraphics[width=0.98\hsize]{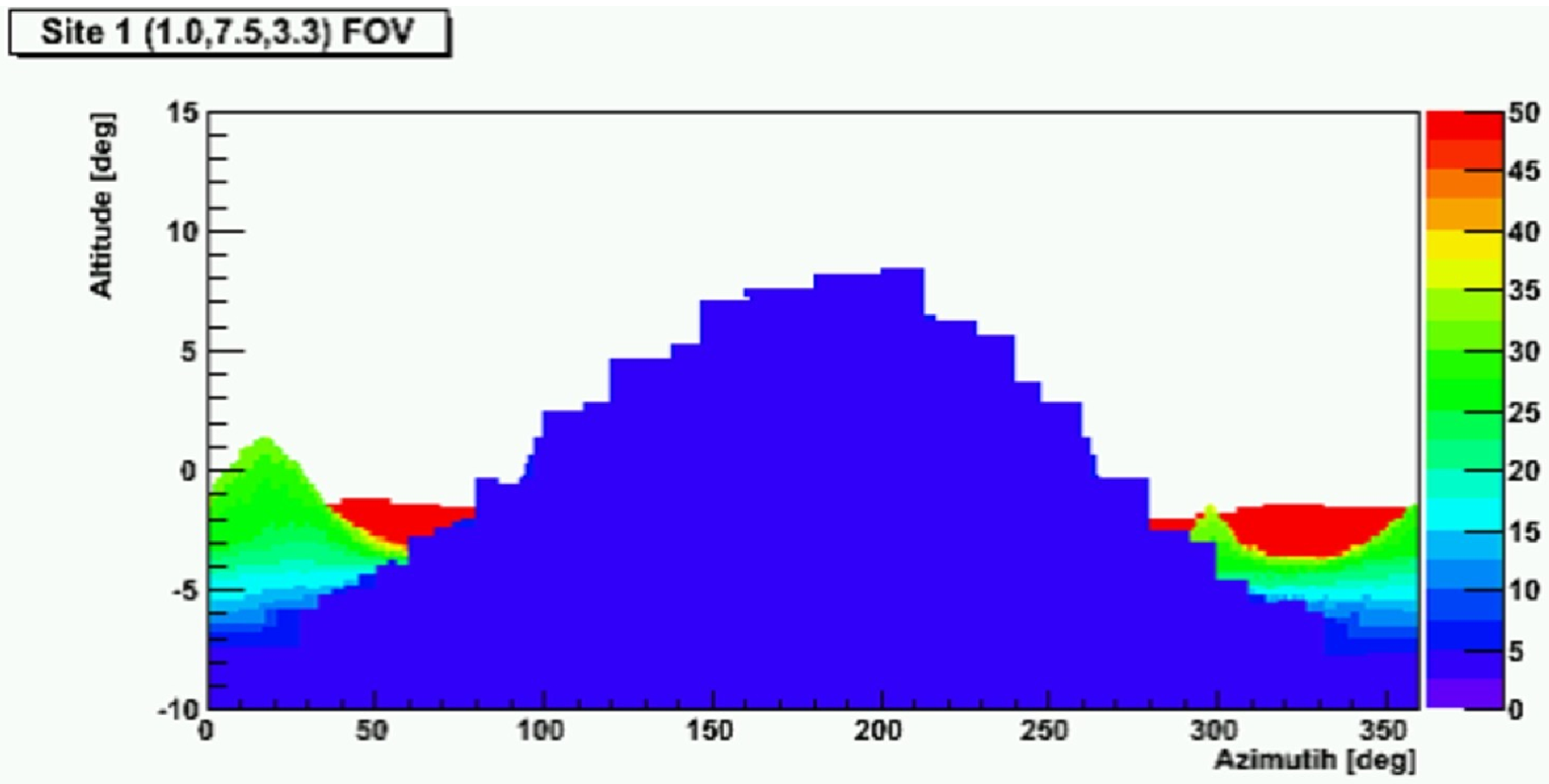}
  \end{center}
 \end{minipage} \\
 \begin{minipage}{0.48\hsize}
  \begin{center}
    \includegraphics[width=0.98\hsize]{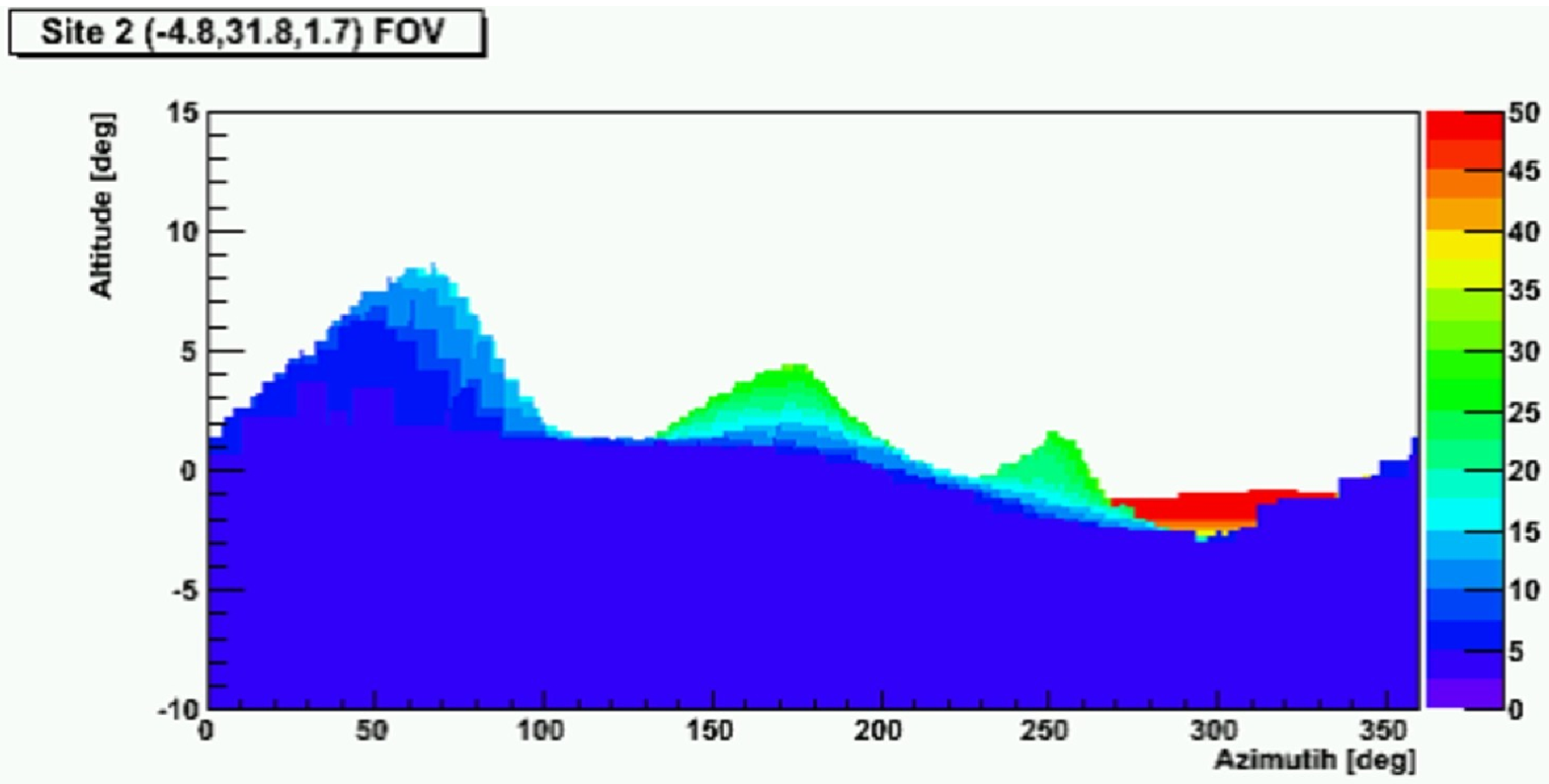}
  \end{center}
 \end{minipage} &
 \begin{minipage}{0.48\hsize}
  \begin{center}
    \includegraphics[width=0.98\hsize]{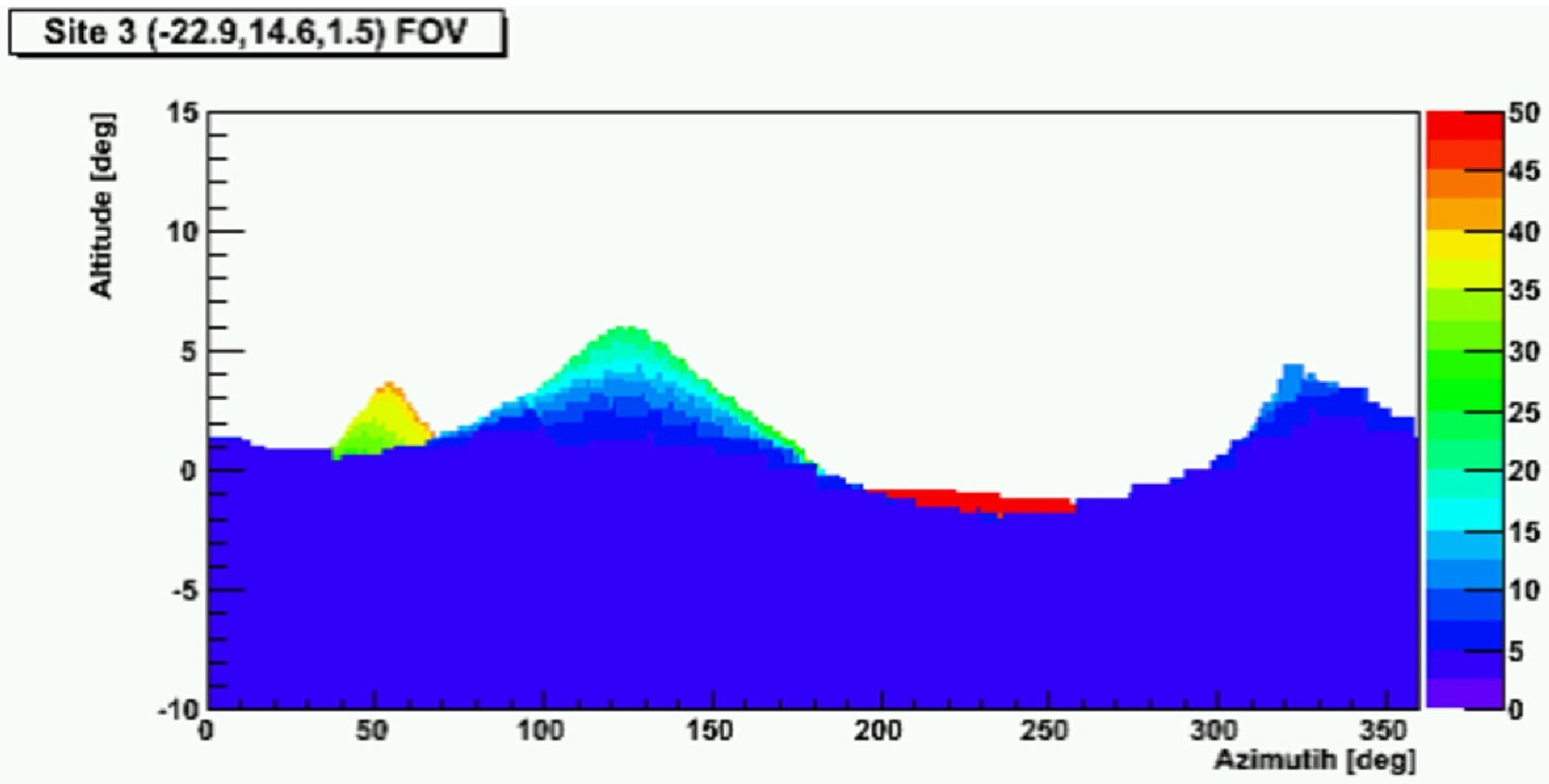}
  \end{center}
 \end{minipage} \\
 \end{tabular}
\end{center}
\vskip-0.3cm
 \caption{
 Panoramic views simulating the topographical
 image from NTA
 Site0 (top left), Site1(top right), Site2 (bottom left), Site3 (bottom right).
 Nearby obstacles with distance less than 3~km are neglected.
}
 \label{fig:fov}
\end{figure}
Fig.~\ref{fig:fov} shows the simulated panoramic views in
altitude and azimuthal directions
from the four NTA sites, with
colours indicating the distance from the corresponding site.

\begin{figure}[t!]
\begin{center}
    \includegraphics[width=0.45\hsize]{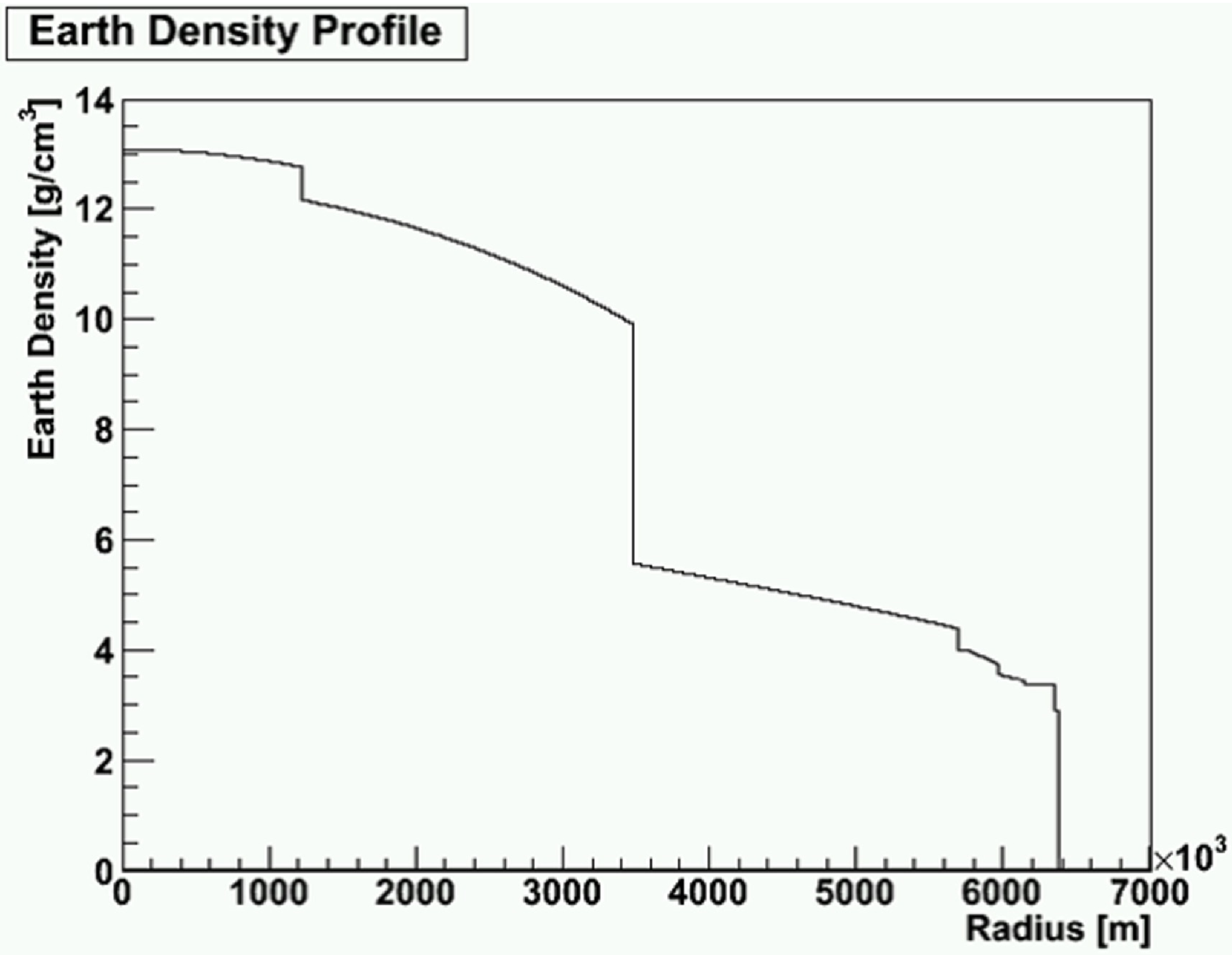}
  \end{center}
\vskip-0.3cm
 \caption{PREM (Preliminary Earth Model) density distribution of the Earth \cite{Gandhi96}}
 \label{fig:dens}
\end{figure}

\section{The NTA Detector Performance}
\noindent
We investigate NTA performance with site location setup of previous Section.
\begin{figure}[t!]
\begin{center}
\begin{tabular}{cc}
 \begin{minipage}{0.4\hsize}
  \begin{center}
     \includegraphics[width=0.98\hsize]{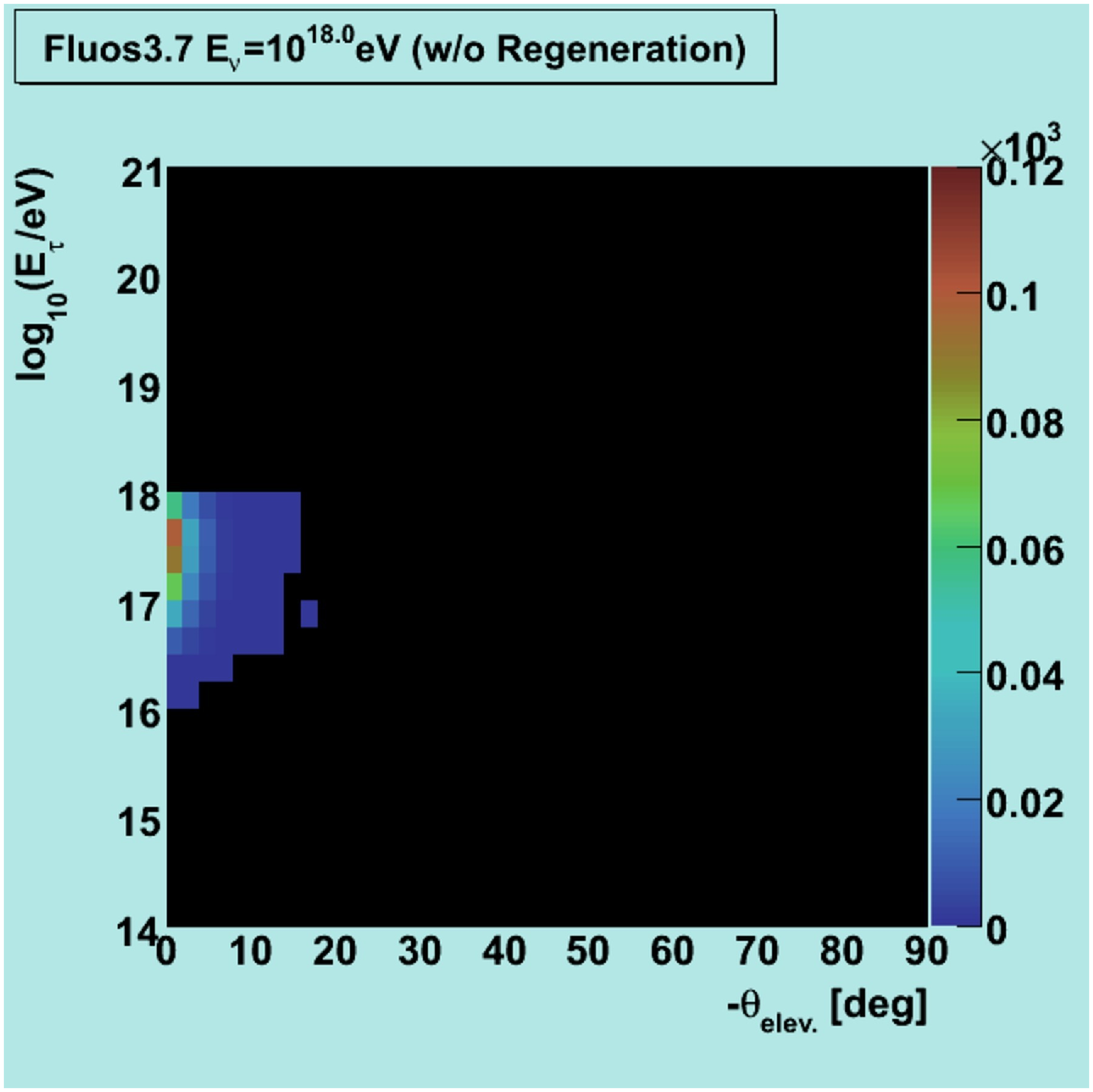}
  \end{center}
 \end{minipage} &
 \begin{minipage}{0.4\hsize}
  \begin{center}
    \includegraphics[width=0.98\hsize]{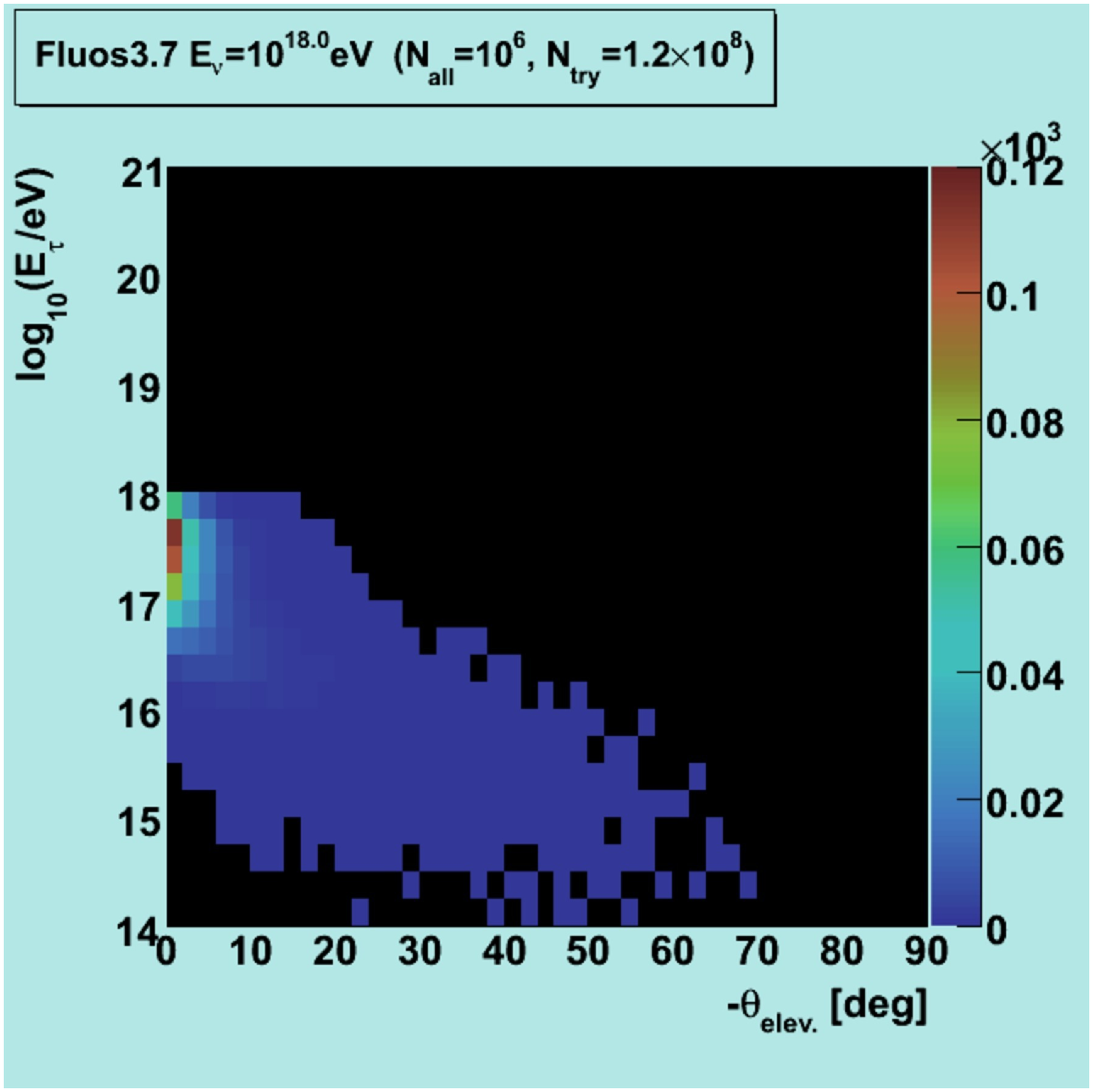}
  \end{center}
 \end{minipage} \\
  & \\
 \begin{minipage}{0.4\hsize}
  \begin{center}
     \includegraphics[width=0.98\hsize]{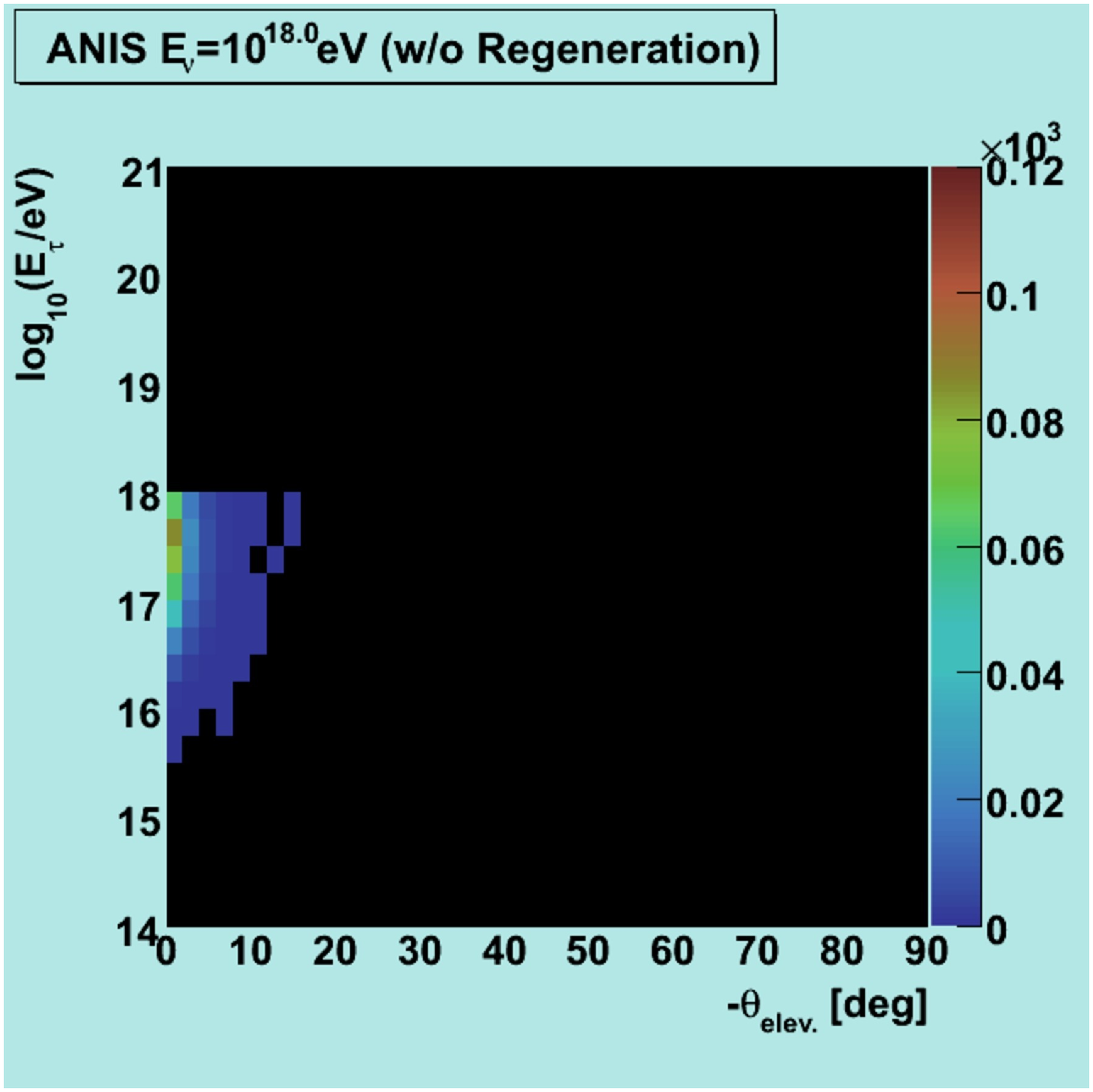}
  \end{center}
 \end{minipage} &
 \begin{minipage}{0.4\hsize}
  \begin{center}
    \includegraphics[width=0.98\hsize]{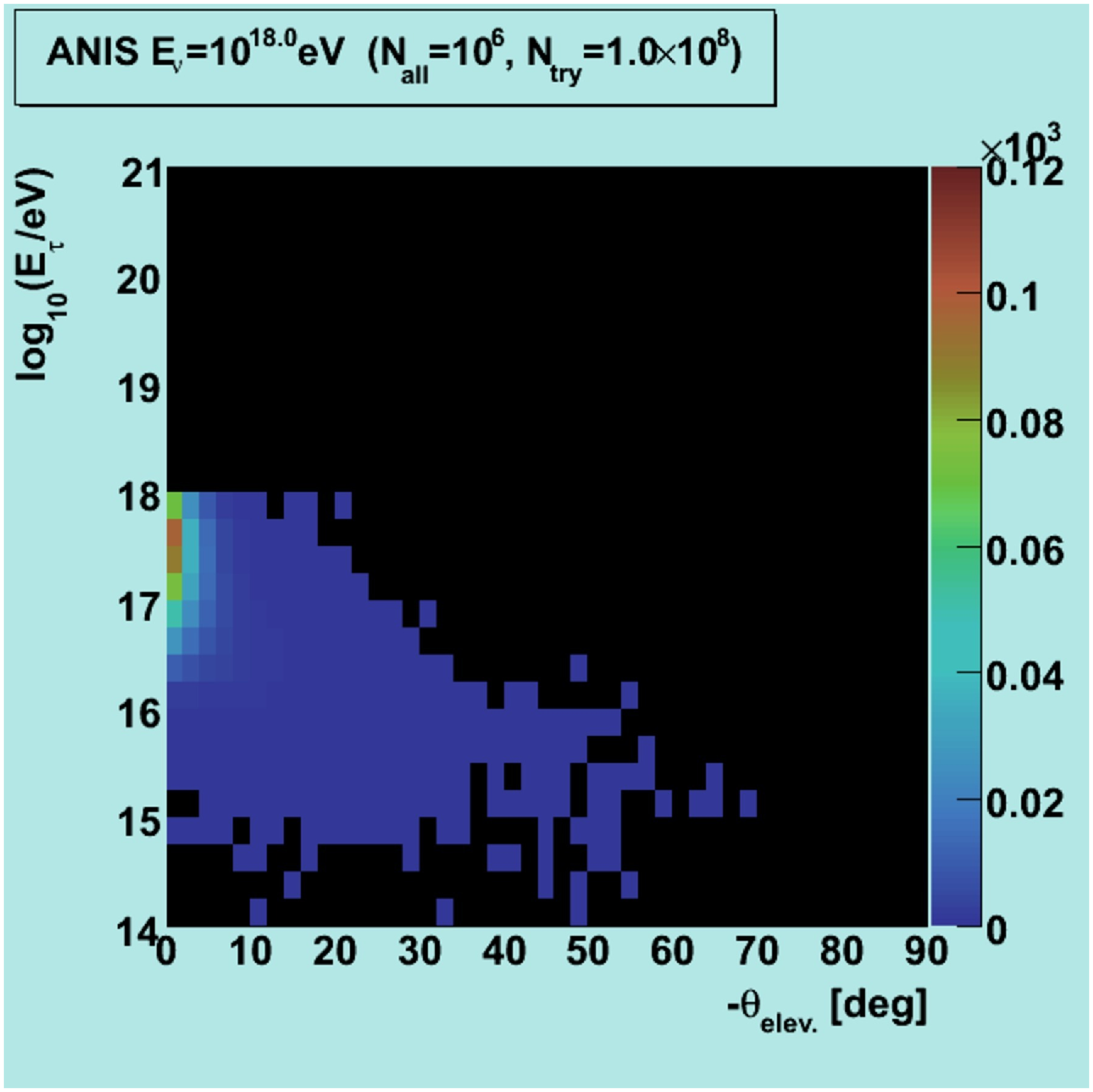}
  \end{center}
 \end{minipage} \\
 \end{tabular}
\end{center}
\vskip-0.3cm
\caption{
Distribution in the
E$_{\tau}$ and dip angle (minus elevation angle; $-\theta_{elev}$) plane
in the case of primary $\nu_{\tau}$ energy of 10$^{18}$~eV.
}
 \label{fig:regenANIS}
\end{figure}

\subsection{Propagation}\label{sec:prop}
For simulating the propagation of $\nu_{\tau}$s and $\tau$s in the Earth,
we performed the following procedure and treatment.

\begin{enumerate}
\item
The density profile of the Earth
is chosen according to the Preliminary Earth Model
\cite{Gandhi96,PREM}.
It depends strongly on the depth in the Earth as shown in Fig.~\ref{fig:dens}.
We modified the profile just for the density of the ground surface in the radius range of
$r>6356$ m from 2.6~g/cm$^{3}$ into 2.9~g/cm$^{3}$, which is suitable for Basalt rock
as the most common type of rock in the Earth's crust and most of the ocean floor
around the Island of Hawaii.

\item
We took into account both charged current interaction (CC) and neutral current interaction (NC)
of $\nu_{\tau}$s and $\tau$s in the Earth.
The energy dependence of the inelasticity parameter $y$ for CC and NC based on the CTEQ4
parametrization are shown in
\cite{Gandhi96}, and no difference is seen between those for CC and NC.

\item
We implement
$\nu_{\tau} \rightarrow \tau \rightarrow \nu_{\tau}$ regeneration in the simulation.
Because of the short lifetime of the tau,
regeneration can be an important effect as the $\nu_{\tau}$ passes
through a significant column depth through the Earth~\cite{Dutta2001}.

\item
In simulating $\tau$ propagation,
the current position of $\tau$ is evaluated
at every step of the energy loss rate of 10\%,
unless the $\tau$ comes out of the Earth or decays.

\item
In the lab frame, $\nu_{\tau}$ from $\tau$ decay on average
carries a fraction 0.4 of the $\tau$ energy
\cite{PhysRevD.66.021302}.
We used this constant average value of 0.4 as the energy of $\nu_{\tau}$ from $\tau$ decay,
without taking into account
the energy distribution.
The error from this approximation can be neglected for the moment,
because of the
good agreement
between results with our simulation and with ANIS (All Neutrino Interaction Simulation)
\cite{gazizov2005anis},
as shown in Fig.~\ref{fig:regenANIS}.
\end{enumerate}

Fig.~\ref{fig:regenANIS}
shows the distribution in the plane of
E$_{\tau}$ and dip angle (minus elevation angle; $-\theta_{elev}$)
in the case of primary $\nu_{\tau}$ energy of 10$^{18}$~eV.

The left side shows the case of neglecting
any effect from
$\nu_{\tau} \rightarrow \tau \rightarrow \nu_{\tau}$ regeneration
or NC interaction of $\nu_{\tau}$ in the Earth,
while the right side shows the case of taking into account
both effects from
$\nu_{\tau} \rightarrow \tau \rightarrow \nu_{\tau}$ regeneration
and NC interaction of $\nu_{\tau}$ in the Earth.
Each bin content in these figures is given by:
\[
\frac{d^2 N_{\tau}}{dE_{\tau} d \Omega} (E_{\tau}, \theta) \times d \log _{10} E_{\tau} \cdot 2 \pi d\theta.
\]
Fig.~\ref{fig:regenANIS}~(bottom) shows the result using ANIS
~\cite{gazizov2005anis},
which is approved for use for AMANDA and IceCube, and acknowledged well
for detailed interactions, decays, and propagation of
$\nu_{\tau}$s and $\tau$s.
In general, the results with our simulation and ANIS
agrees reasonably well.
From detailed comparison, we should consider systematic errors of
$\sim$12\% on produced $\tau$ flux in the Earth in using our simulation program.

\subsection{Simulation}\label{sec:sim}
Before evaluation of the performance of NTA using our simulation program,
we summarize our settings at the following three steps.

\begin{enumerate}
\item Simulation for the Earth-skimming： $\nu_\tau \rightarrow \tau$
	\begin{itemize}
		\item $\nu_{\tau}$ (CTEQ4) \cite{Gandhi98}
		\item Inelasticity parameter \cite{Gandhi96}
		\item Energy loss in the Earth \cite{Tseng03,Dutta2001}
	\end{itemize}
\item Air-shower simulation: $\tau \rightarrow$ Cherenkov and fluorescence light
	\begin{itemize}
		\item $\tau$ Decay (approximated; \cite{TANeu})
		\item Air-shower generation (Gaisser-Hillas $+$ NKG) \cite{TANeu}
	\end{itemize}
\item Detector simulation:
	\begin{itemize}
	\item  light collection and throughput of light
	\item  simplified triggering logic
	\item  Event reconstruction is not implemented yet.
	\end{itemize}
\end{enumerate}

We assume the following input parameters.
\begin{description}
\item{Light Collection Area:}
	$ A=7.07~{\rm m}^2 $
	(equivalent with the effective pupil diameter of $\phi$3~m)
\item{Optical filter transmittance:}
	$ \epsilon_{\rm filt} = 90\% $
\item{Quantum efficiency of photoelectric tube:}
	$ \epsilon_{\rm QE} = 24\% $
\item{LC FOV:}
	$ 32^\circ \times 32^\circ $
\item{Trigger pixel FOV:}
	$ 0.5^\circ \times 0.5^\circ$ / trigger pixel
\item{Exposure time in trigger pixel:}
	$t_{trigpix} = 50~{\rm ns} $
\item{Required trigger condition}:\\
To estimate the detection sensitivity of NTA
the event candidates must satisfy the following requirements:
\begin{itemize}
	\item total number of photoelectrons detected in one LC must satisfy:\\
			\[ N_{\rm pe}^{\rm LC} >61, \]
    \item S/N estimated in the track-associated box of the width of 4~pixels and the length of 64~pixels, which includes the candidate event air-shower track, must satisfy:\\
			  \[ S/N > 4, \]
where
the standard deviation of night sky background in the track-associated box with the exposure of 50~ns
is estimated as
$ \sigma(N_{\rm pe}^{\rm BG}) = 15.4~{\rm pe} $
from a night sky background estimate~\cite{Sasaki2001} given by
$ 2.0 \times 10^5 {\rm photons/m}^2/{\rm sr}/\mu{\rm s} $.

\end{itemize}
\end{description}

\begin{figure}[t!]
  \begin{center}
    \includegraphics[width=0.4\hsize]{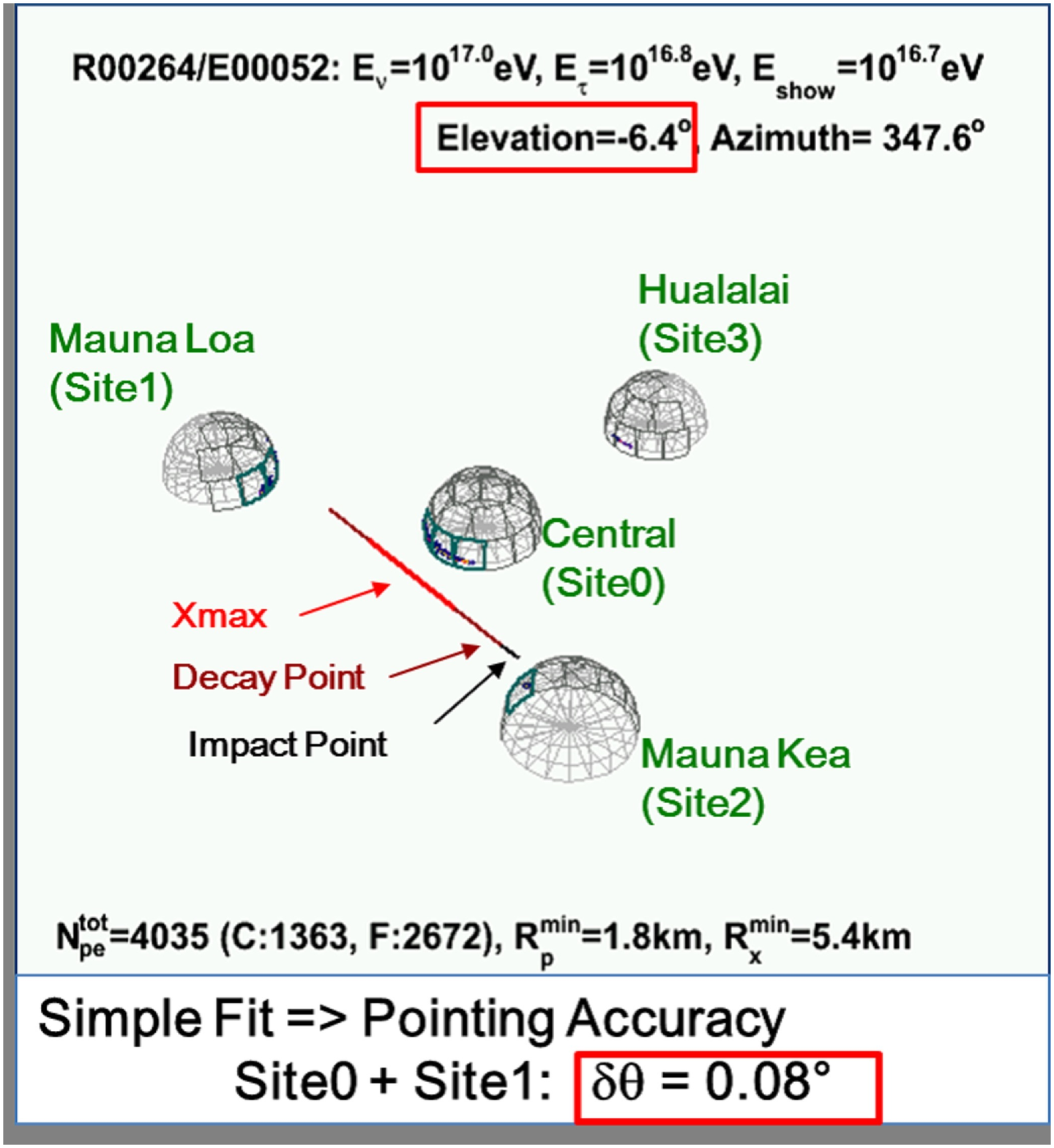}
  \end{center}
\begin{center}
\begin{tabular}{cc}
 \begin{minipage}{0.35\hsize}
  \begin{center}
    \includegraphics[width=\hsize]{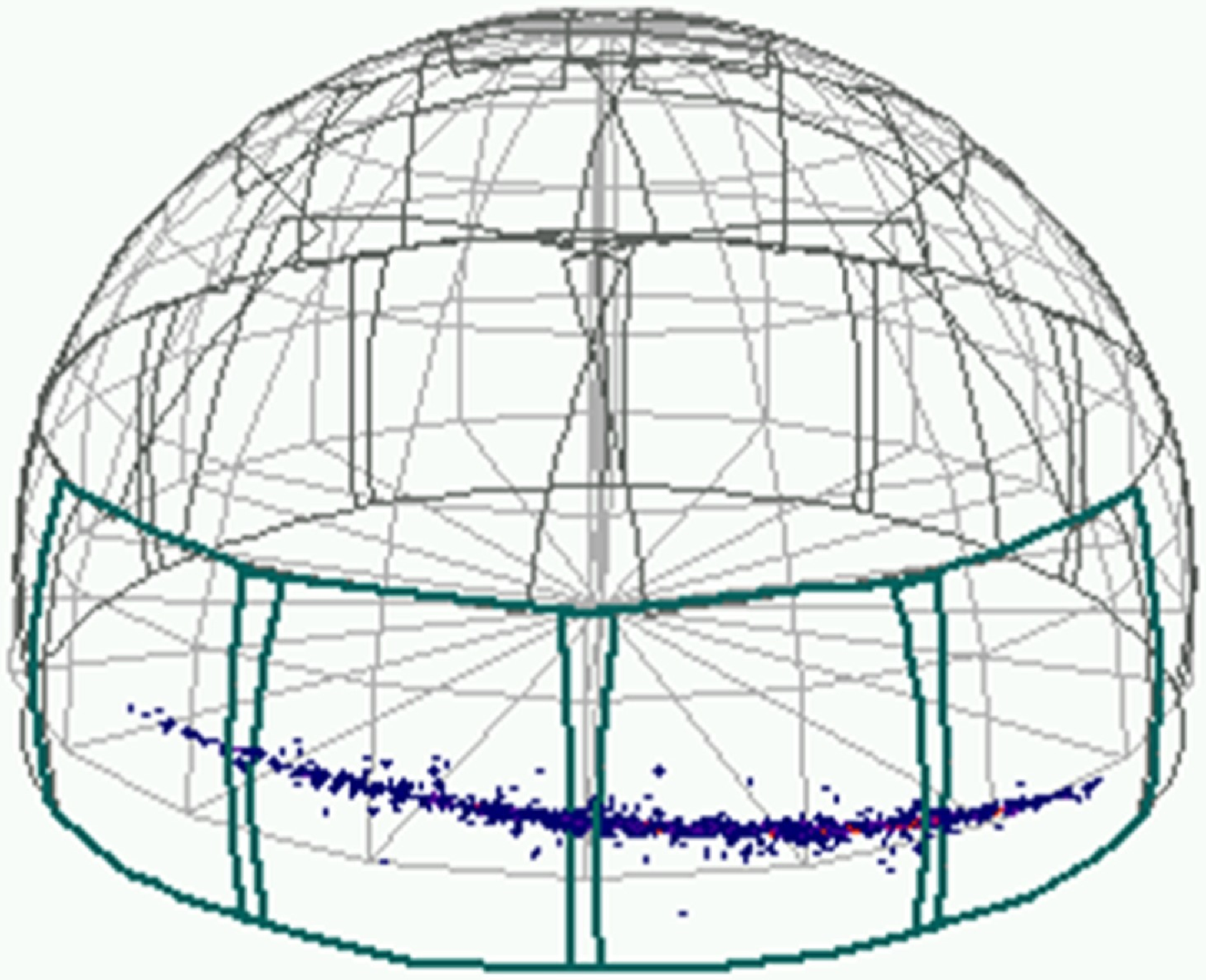}
  \end{center}
 \end{minipage} &
  \hspace{10mm}
 \begin{minipage}{0.35\hsize}
  \begin{center}
    \includegraphics[width=\hsize]{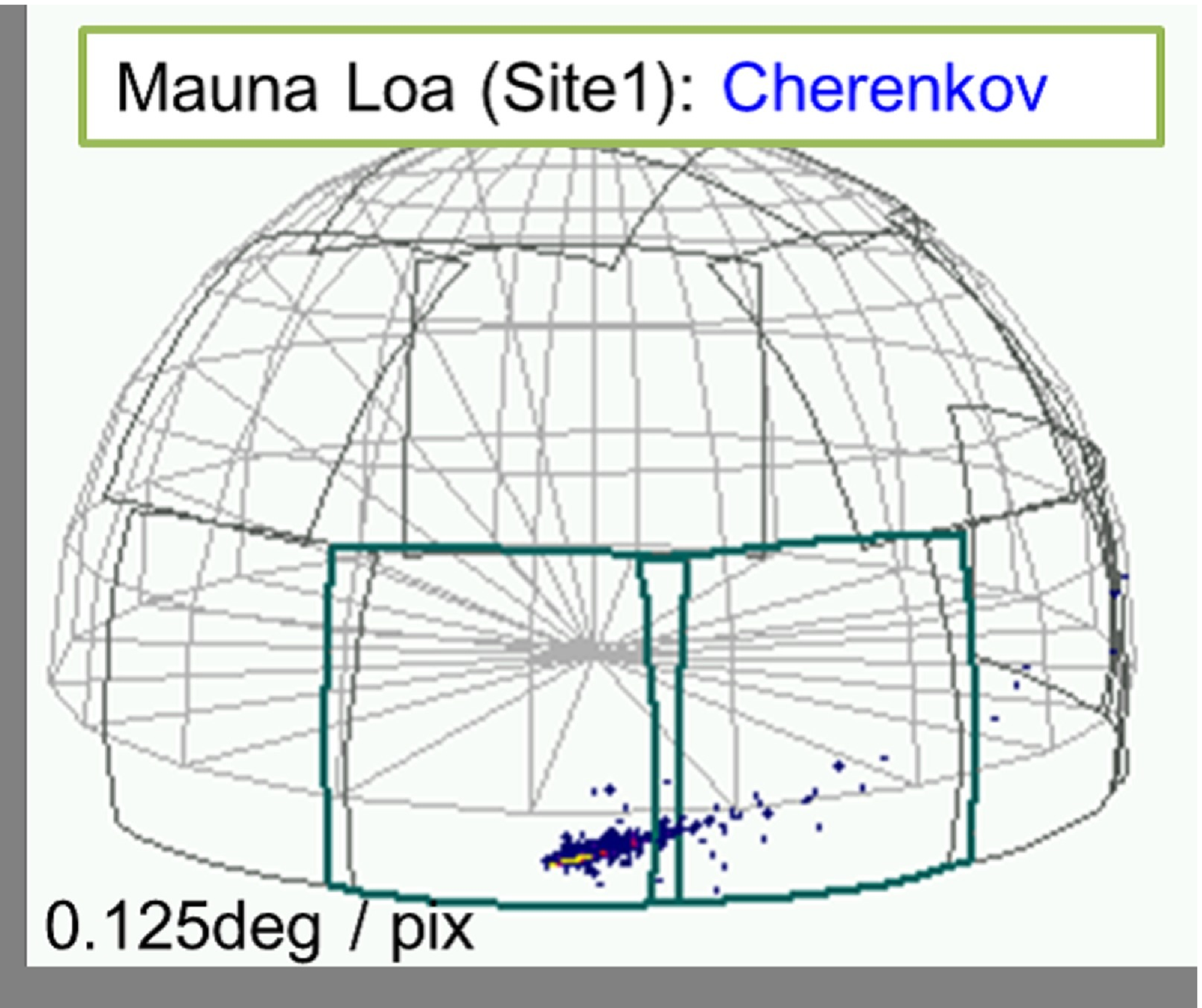}
  \end{center}
 \end{minipage} \\
 \end{tabular}
\end{center}
 \caption{
A simulated Earth-skimming $\tau$ shower event with
primary $E_{\nu_\tau}=10^{17}$ eV, which
has both fluorescence image taken by Site0 and Cherenkov by Site1.
(top) Global hit map view in the NTA system;
(bottom left) air-shower fluorescence image taken by Site0, and
(bottom right) Cherenkov image from the same event taken by Site1.
The trigger pixel and fine image FOVs are
$ 0.5^\circ \times 0.5^\circ$ and
$ 0.125^\circ \times 0.125^\circ$, respectively.
}
 \label{fig:event}
\end{figure}

A simulated Earth-skimming $\tau$ shower event with
primary $\nu_{\tau}$ energy of $E_{\nu}=10^{17}$eV
using the above settings
is shown in Fig.~\ref{fig:event}.
The reconstructed air-shower axis with simple fits to the
Cherenkov and fluorescence
hit map images taken by the two sites reproduces the primary
$\nu_{\tau}$ arrival direction with an error of 0.08$^{\circ}$.

\begin{figure}[h]
  \begin{center}
    \includegraphics[width=0.6\hsize]{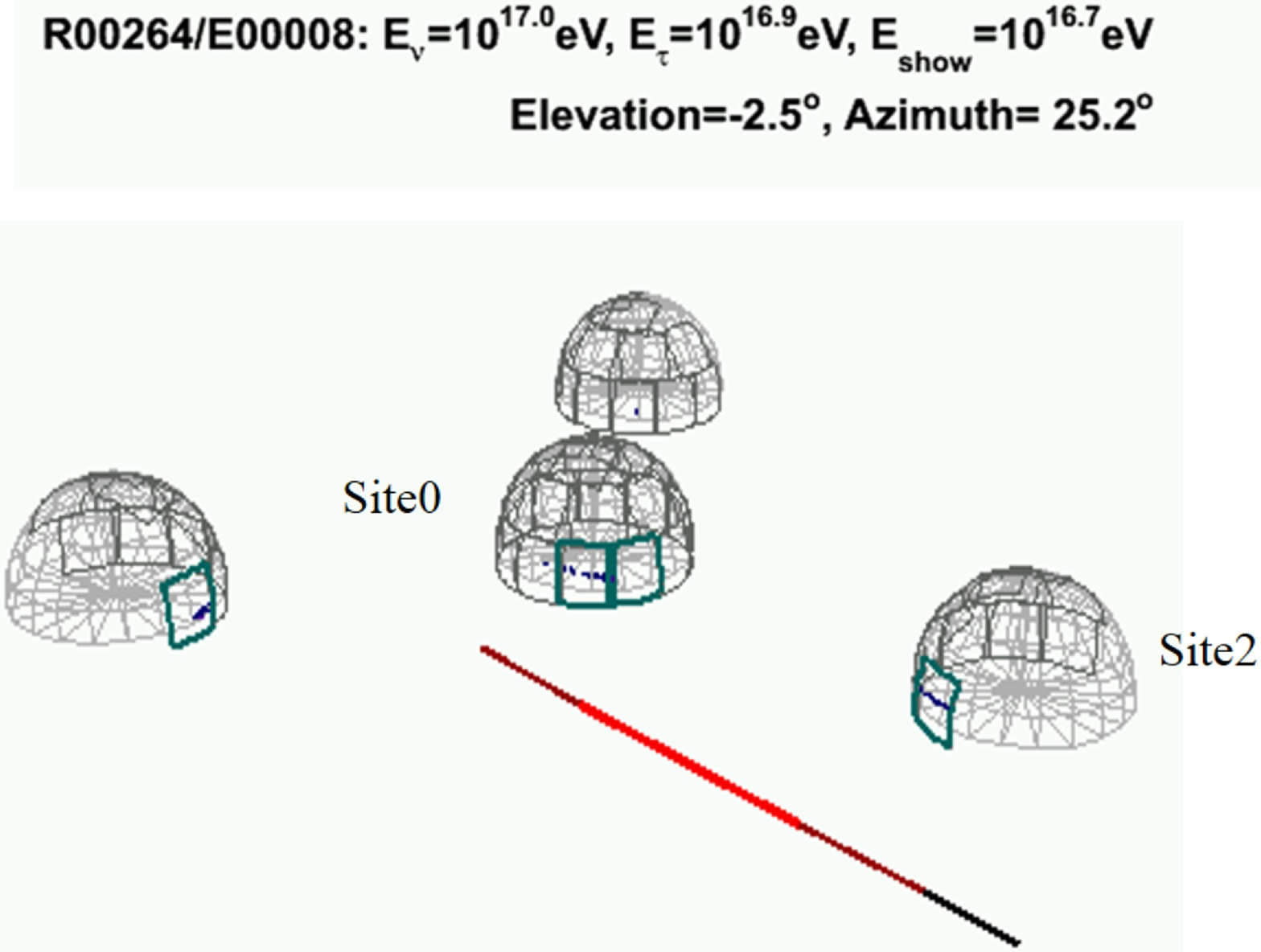}
  \end{center}
\begin{center}
\begin{tabular}{cc}
 \begin{minipage}{0.35\hsize}
  \begin{center}
    \includegraphics[width=\hsize]{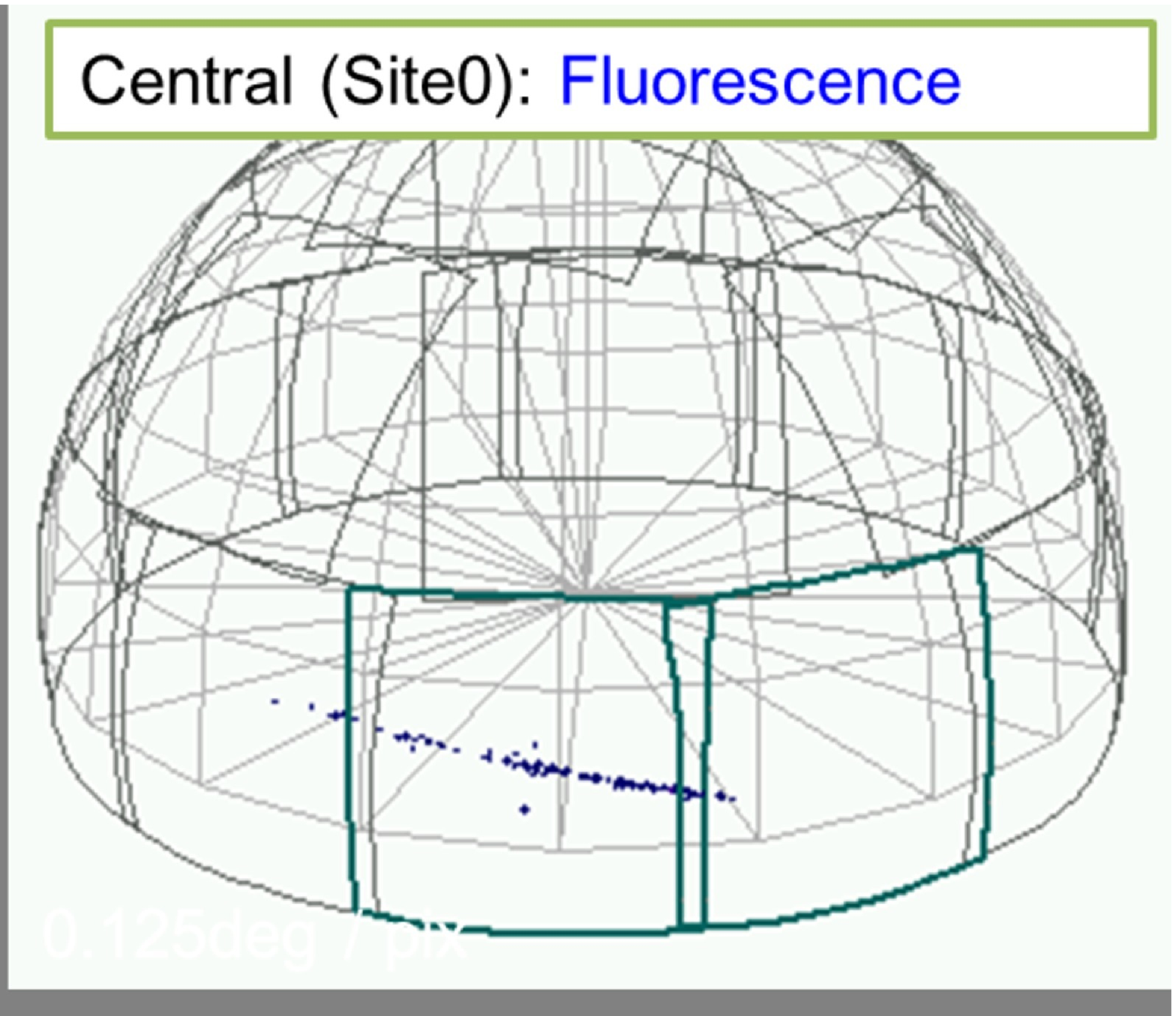}
  \end{center}
 \end{minipage} &
  \hspace{10mm}
 \begin{minipage}{0.35\hsize}
  \begin{center}
    \includegraphics[width=\hsize]{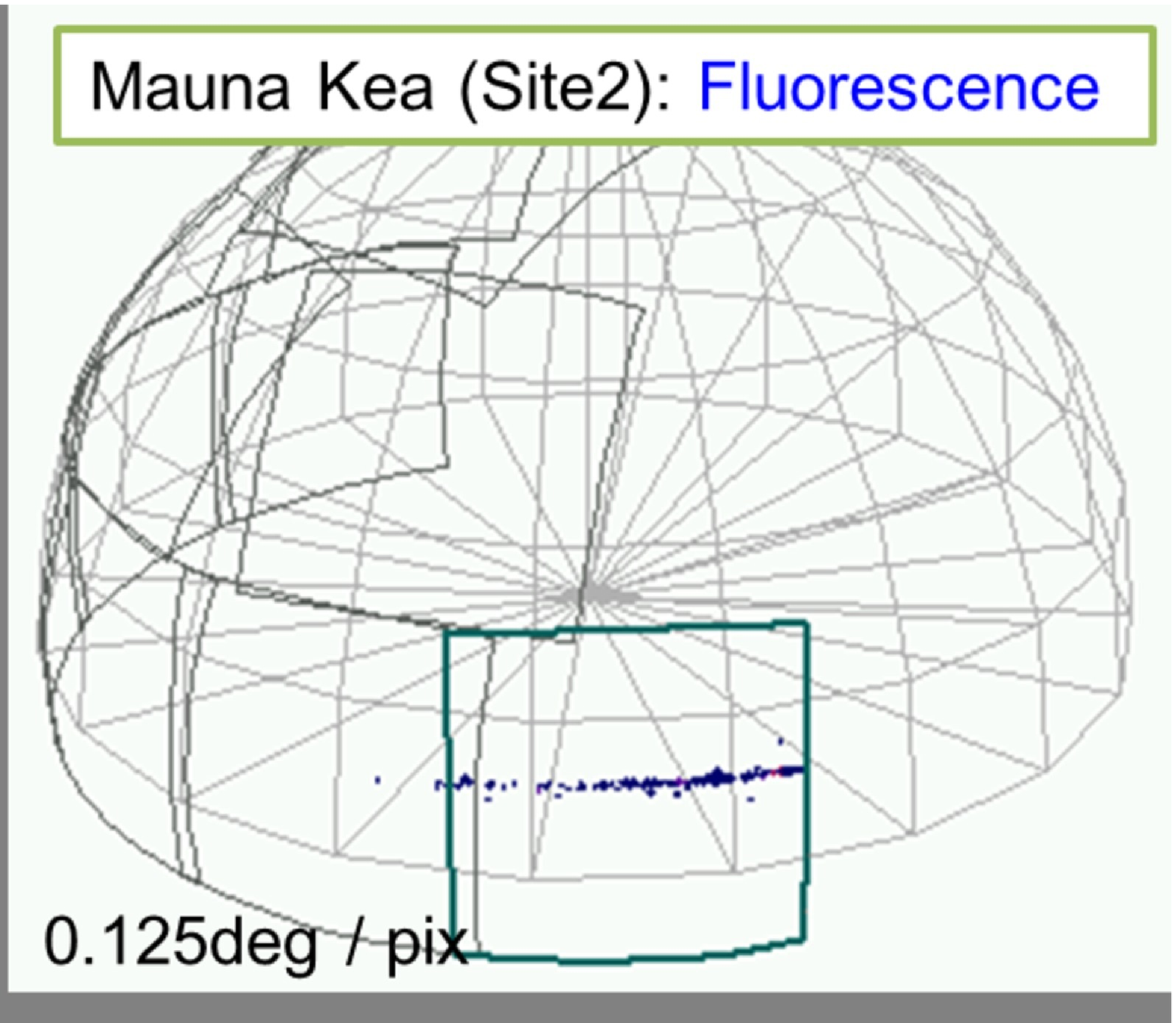}
  \end{center}
 \end{minipage} \\
 \end{tabular}
\end{center}
 \caption{
 	Similar with \ref{fig:event} but in the case of the detection of stereoscopic fluorescence signals.
}
 \label{fig:event2}
\end{figure}

\begin{table}[t!]
\begin{center}
\begin{tabular}{cccc}
\hline
$\nu_{\tau}$ Energy &  CTEQ4 $\sigma_{\rm CC}$ & $L_{\rm CC}$ & $\theta_{\rm elev}^{\rm c}$ \\
\hline
$10^{15}$~eV &   6.342$\times 10^{-34}$~cm$^2$ &  2.62$\times 10^{9}$~g/cm$^2$ & $-$30.6$^\circ$\\
$10^{16}$~eV &   1.749$\times 10^{-33}$~cm$^2$ &  9.49$\times 10^{8}$~g/cm$^2$ & $-$13.0$^\circ$\\
$10^{17}$~eV &   4.436$\times 10^{-33}$~cm$^2$ &  3.74$\times 10^{8}$~g/cm$^2$ & $-$5.71$^\circ$\\
$10^{18}$~eV &   1.049$\times 10^{-32}$~cm$^2$ &  1.58$\times 10^{8}$~g/cm$^2$ & $-$2.45$^\circ$\\
$10^{19}$~eV &   2.379$\times 10^{-32}$~cm$^2$ &  6.98$\times 10^{7}$~g/cm$^2$ & $-$1.08$^\circ$\\
\hline
\end{tabular}
\end{center}
\vskip-0.3cm
\caption{
	Based on CTEQ4~\cite{Gandhi98},
	differential $\nu_{\tau}$ CC cross section ($\sigma_{\rm CC}$),
	corresponding interaction length ($L_{\rm CC}$), and
	the critical angle ($\theta_{\rm elev}^{\rm c}$) such that
	the chord thickness at the critical angle corresponds
	to $L_{\rm CC}$.
	For the Earth density profile,
	we refer to the parametrization of PREM~\cite{gazizov2005anis,Gandhi96}.
	}
\label{tab:critangle}
\end{table}

\begin{figure}[t!]
\begin{center}
\begin{tabular}{cc}
 \begin{minipage}{0.49\hsize}
  \begin{center}
\includegraphics[width=0.98\hsize]{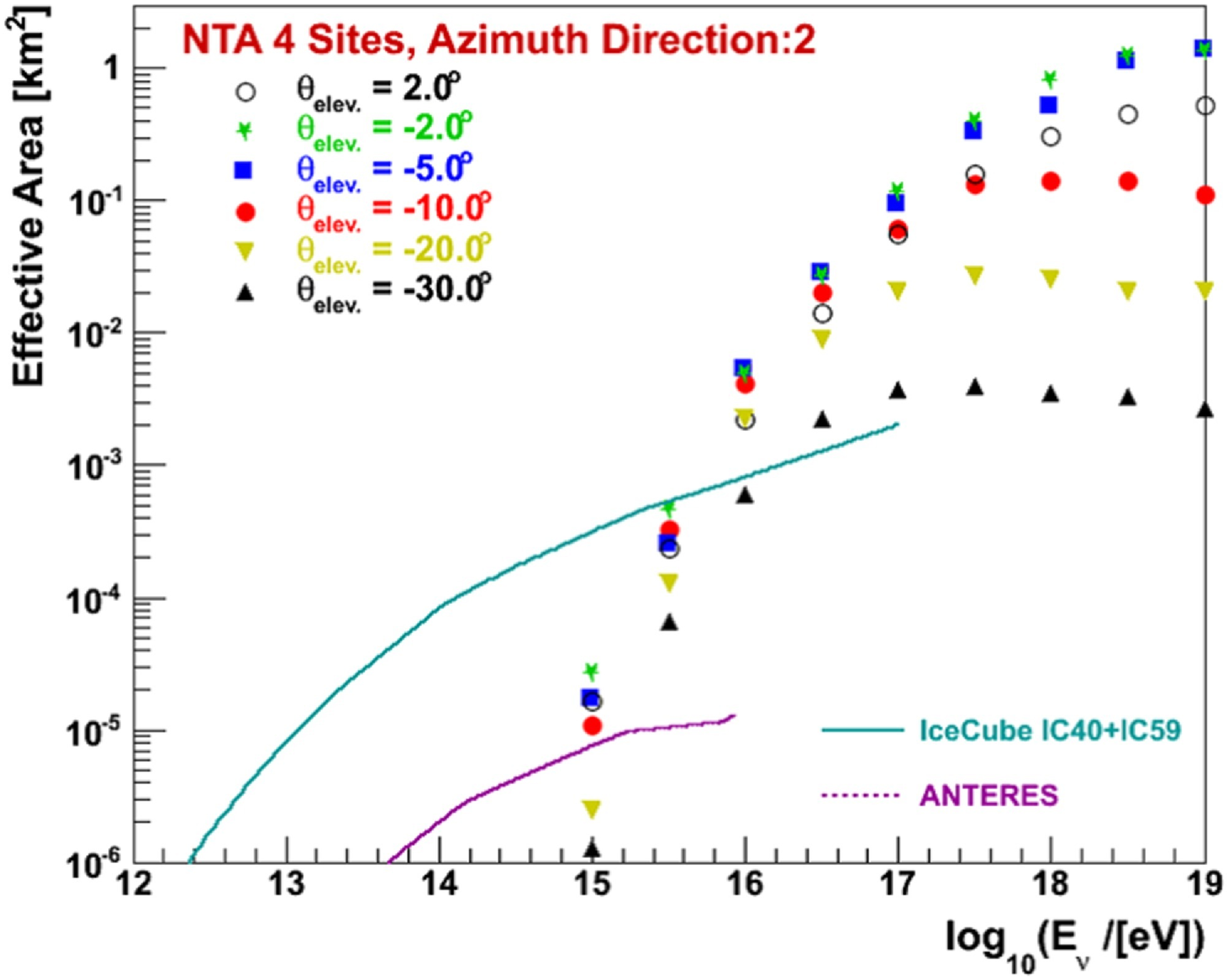}
  \end{center}
 \end{minipage} &
 \begin{minipage}{0.49\hsize}
  \begin{center}
\includegraphics[width=0.98\hsize]{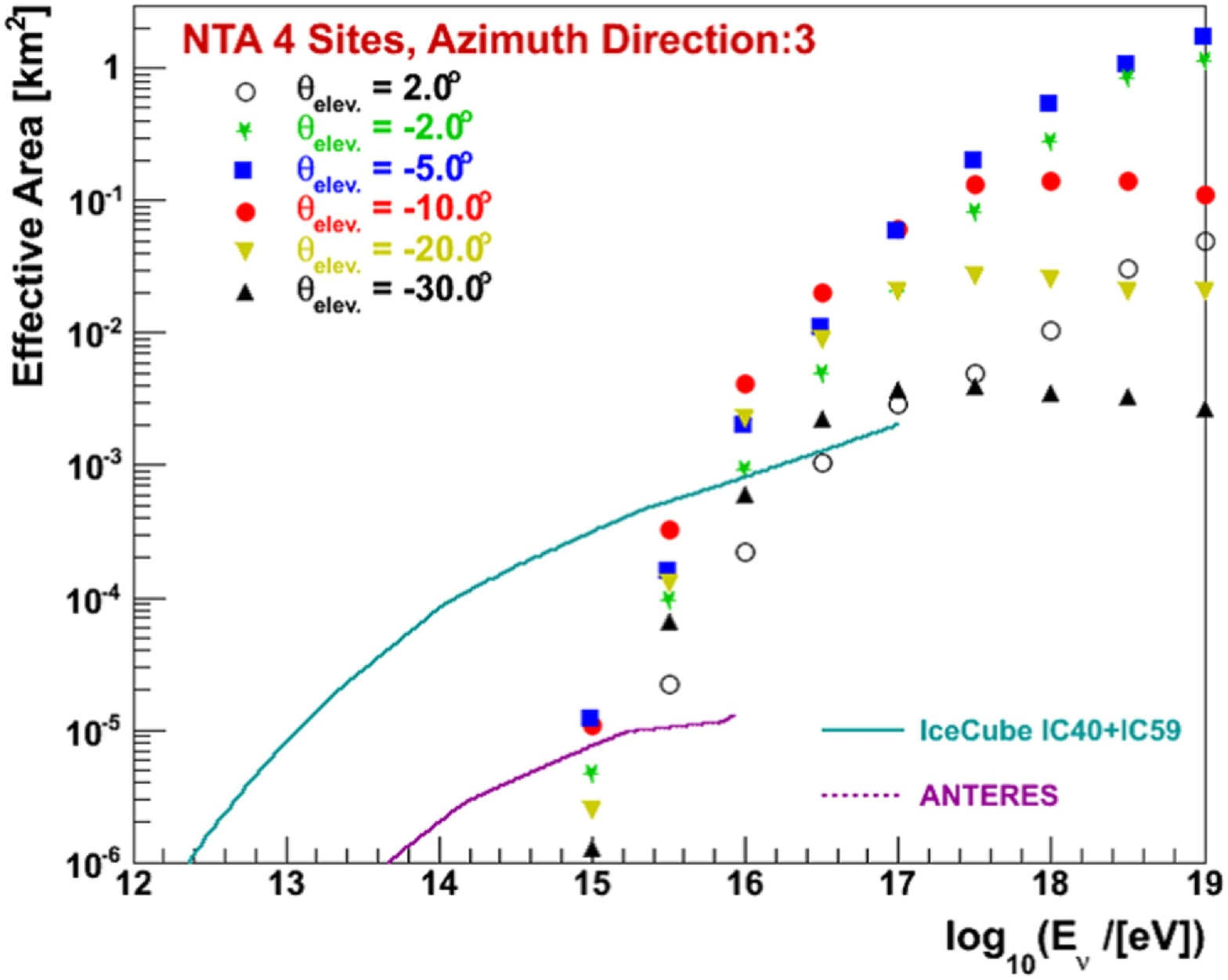}
  \end{center}
 \end{minipage} \\
 \end{tabular}
\end{center}
\vskip-0.4cm
 \caption{
Estimated effective detection area simulated for $\nu_{\tau}$
from a point source with azimuthal arrival direction
corresponding to
(left)
Mauna Loa ($\phi_2$)
and
(right)
Hualalai ($\phi_3$)
with respect to the central site of Site0,
and dip angles of
2.0$^{\circ}$ (black open circle),
$-$2.0$^{\circ}$ (green star),
$-$5.0$^{\circ}$ (blue filled box),
$-$10.0$^{\circ}$ (red filled circle),
$-$20.0$^{\circ}$ (yellow filled triangle), and
$-$30.0$^{\circ}$ (black filled triangle).
}
 \label{fig:effarea}
\end{figure}

\begin{figure}[t!]
\begin{center}
\begin{tabular}{cc}
 \begin{minipage}{0.475\hsize}
  \begin{center}
  \includegraphics[width=0.98\hsize]{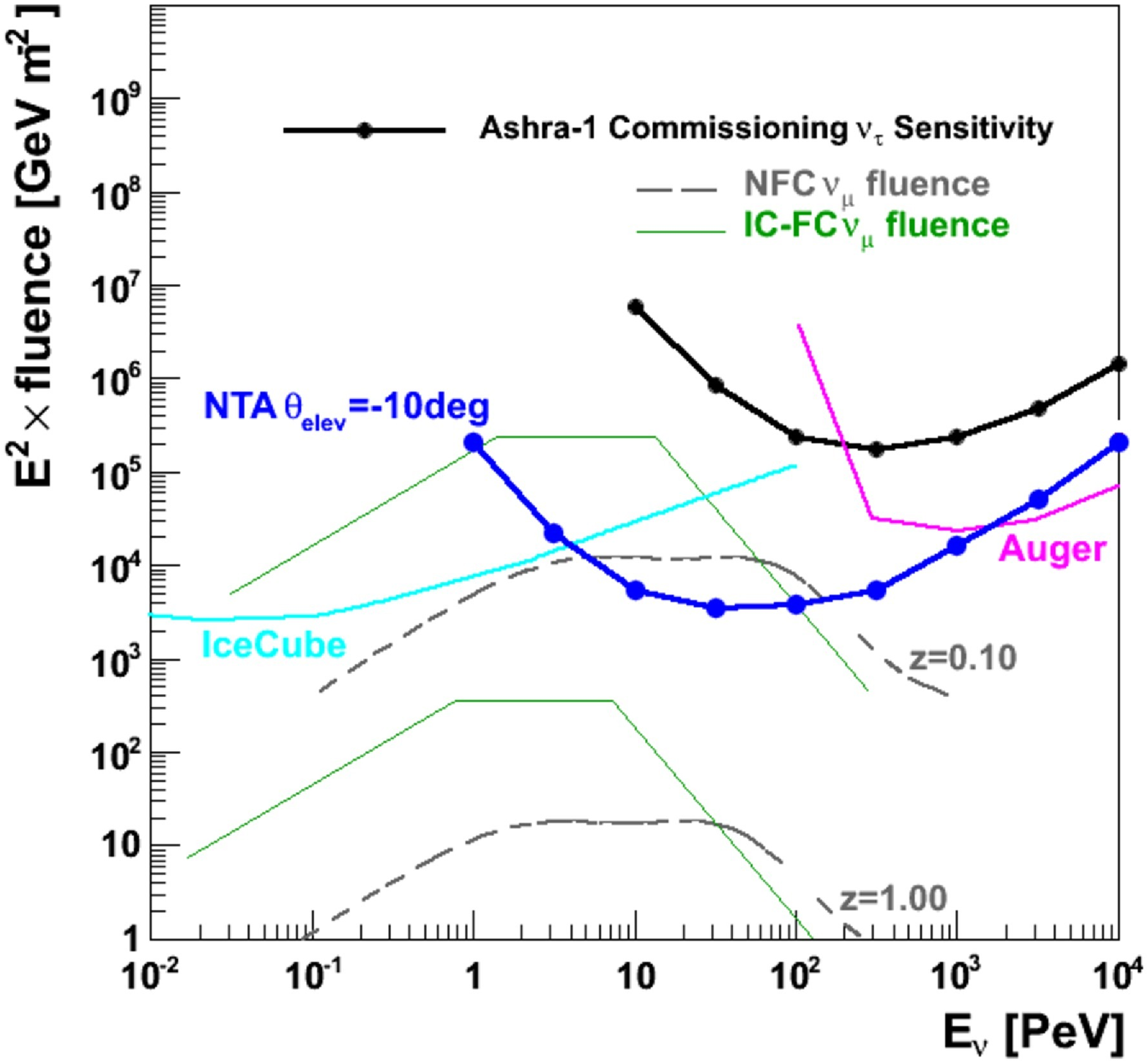}
  \end{center}
 \end{minipage} &
 \begin{minipage}{0.475\hsize}
  \begin{center}
  \includegraphics[width=0.98\hsize]{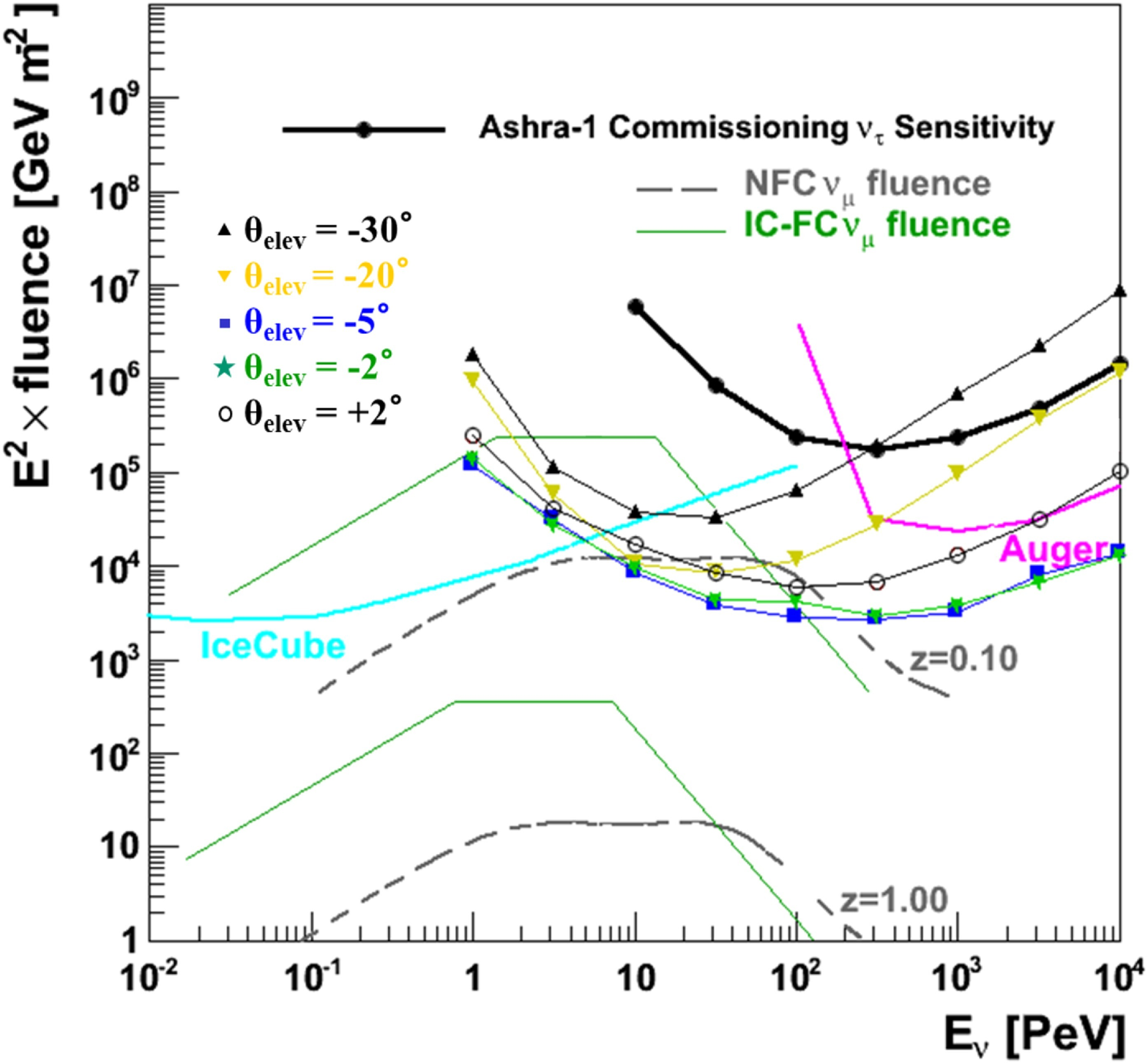}
  \end{center}
 \end{minipage} \\
 \end{tabular}
\end{center}
\vskip-0.4cm
\caption{
Comparison of
differential sensitivities
as function of $E_{\nu_{\tau}}$
for a point source of $\nu_{\tau}$,
calculated as the
Feldman-Cousins
90\% CL limit event number for null expected events,
using a light collector (LC)
from Ashra-1 commissioning~\cite{AshraCNeu}
and the NTA layout of LCs, in the cases of
(left)  $\theta_{\rm elev} = -10^{\circ}$,
(right) $\theta_{\rm elev} = +2^{\circ} ({\rm open circle}),
-2^{\circ} ({\rm green}),
-5^{\circ} ({\rm blue}),
-20^{\circ} ({\rm yellow}),
{\rm and}
-30^{\circ} ({\rm black})$.
The sensitivities published from
IceCube~\cite{IceCube2012} and
Pierre Auger Observatory~\cite{AugerES09}
are shown, as well as
theoretical estimates used for the former (solid lines)
and recalculated by H\"ummer et al. (dashed lines)~\cite{PhysRevLett.108.231101}
assuming the distance of
$z\sim 0.1$ and 1.0.
}
 \label{fig:pointsource}
\end{figure}

\subsection{Sensitivity}\label{sec:sens}
We estimate the effective detection area for $\nu_{\tau}$ fluence from a point source
with our simulation program for Earth-skimming $\tau$ showers
incorporating the appropriate Earth model~\cite{PREM},
the topography around the NTA observatory,
the interaction and propagation process of $\nu_{\tau}$ and $\tau$ in the Earth
~\cite{Gandhi96,gazizov2005anis},
the decay of $\tau$ and generation of air-shower,
with parameter choices as described before.

We define the critical dip angle (minus critical elevation angle; $-\theta_{elev}^{c}$)
as the chord thickness at the dip angle $-\theta_{\rm elev}^{c}$ that
corresponds to the CC interaction length $L_{\rm CC}(E_{\nu})$,
determined by the interaction cross-section $\sigma_{\rm CC}(E_{\rm CC})$
for a $\nu_{\tau}$ traveling with energy $E_{\nu}$.
Table~\ref{tab:critangle} shows
differential cross sections of $\nu_{\tau}$ CC interaction
$\sigma_{\rm CC}$ based on CTEQ4~\cite{Gandhi98},
the corresponding interaction length $L_{\rm CC}$, and
$\theta_{\rm elev}^{c}$
for each $E_{\nu_{\tau}}$ between 1~PeV and 10~EeV.

Taking into account the critical dip angles for the energies of $\nu_{\tau}$ in Table~\ref{tab:critangle},
we estimated the effective detection areas for $\nu_{\tau}$
from a point source with azimuthal arrival direction corresponding to
that of the Mauna Loa summit ($\phi_2$) and that of the Hualalai summit ($\phi_3$),
with respect to the central site of Site0, and dip angles of
$2.0^{\circ}$,
$-2.0^{\circ}$,
$-5.0^{\circ}$,
$-10.0^{\circ}$,
$-20.0^{\circ}$, and
$-30.0^{\circ}$,
as shown in Fig.~\ref{fig:effarea}.

Fig.~\ref{fig:pointsource} shows
the differential sensitivities,\
as a function of $E_{\nu_{\tau}}$
for a point source of $\nu_{\tau}$,
calculated as in \cite{feldman1998unified}
requiring 2.3 events in a bin size of one energy decade
$\Delta \ln E_{\nu}$.
The 2.3 events is the Feldman-Cousins 90\%
CL limit event number for null expected events.

\begin{figure}[t!]
\begin{center}
\begin{tabular}{c}
 \begin{minipage}{0.65\hsize}
  \begin{center}
  \includegraphics[width=0.98\hsize]{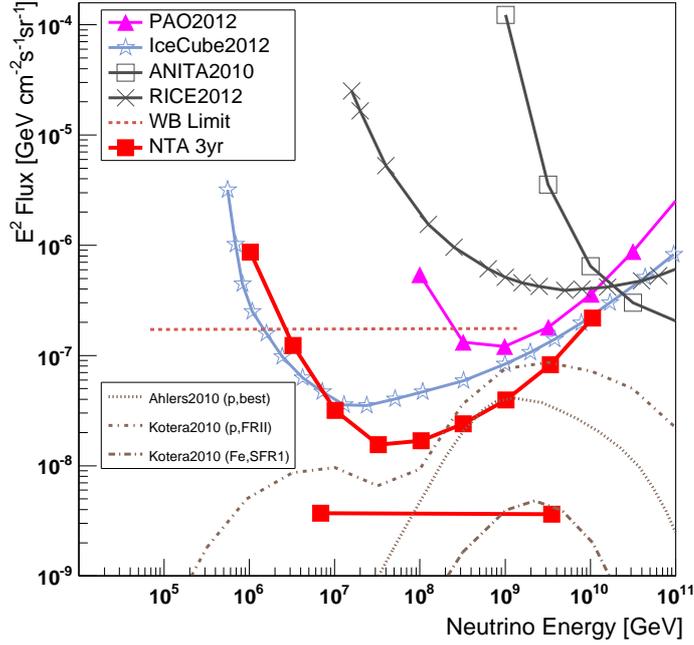}
  \end{center}
 \end{minipage} \\
 \end{tabular}
\end{center}
\vskip-0.3cm
 \caption{
 	Diffuse sensitivities for $\nu_{\tau}$ fluxes with NTA for 3 years observation time.
Both differential sensitivity (curve) and integral sensitivity
assuming the E$^{-2}$ flux spectrum (horizontal line) are shown.
The sensitivity limit is defined as $2.3E_{\nu}/(S \Omega_{\rm eff} \cdot \Delta T)$.
Comparison among NTA,
Pierre Auger Observatory~\cite{AugerES09}
and IceCube~\cite{PhysRevD.83.092003},
same as the top one but
various model predictions for cosmogenic $\nu$s,
as well as other experiments of
RICE, AMANDA, and ANITA
are superimposed
~\cite{gaisser2012high}.
For NTA,
the duty of 10\% for 3 year observation ($\sim$ 9.5$\times$10$^{6}$~s)
is assumed.
}
 \label{fig:diffuse}
\end{figure}

Fig.~\ref{fig:diffuse} (top) shows
the diffuse sensitivities for $\nu_{\tau}$ fluxes with NTA for 3 year observation time.
Both differential and integral sensitivities are given.
The sensitivity limit is defined as $2.3E_{\nu}/(S \Omega_{\rm eff} \cdot \Delta T)$.
Also shown is the comparison between NTA,
Pierre Auger Observatory~\cite{AugerES09}
and IceCube~\cite{PhysRevD.83.092003},
various model predictions for cosmogenic $\nu$s,
as well as other experiments of
RICE, AMANDA, and ANITA
are superimposed~\cite{gaisser2012high}.
For the diffuse sensitivities of NTA,
we assume the duty of 10\% for 3 years observation ($\sim 9.5 \times 10^{6}$~s)
and trigger conditions as described before.

\subsection{Exposure}\label{sec:exp}
\noindent
From Fig.~\ref{fig:effarea},
NTA can survey $\nu_{\tau}$ point sources
with best sensitivity in detection solid angle for $\nu_{\tau}$
defined as
$
-30^{\circ} < \theta_{\rm elev} < 0^{\circ} \times
 0^{\circ} < \phi_{\rm azi} < 360^{\circ}
$
in the primary $\nu_{\tau}$ energy region of 10~PeV $< E_{\nu_{\tau}} <$ 1~EeV.
From Fig.~\ref{fig:pointsource},
the survey depth can be better than redshift $z < 0.1$
corresponding to a distance of $\sim 400$~Mpc.

With the observational conditions
assumed as follows:
\begin{itemize}
\item Solar elevation angle: $< -18^{\circ}$
\item Lunar bright surface ratio: $< 0.2$
\item Ideal weather efficiency: 100\%
\end{itemize}
the total duty is estimated to be 20.5\%, which corresponds to
maximum observation time of 1800~hours per year.

\begin{figure}[t!]
\begin{center}
\begin{tabular}{cc}
 \begin{minipage}{0.49\hsize}
  \begin{center}
   \includegraphics[width=0.98\hsize]{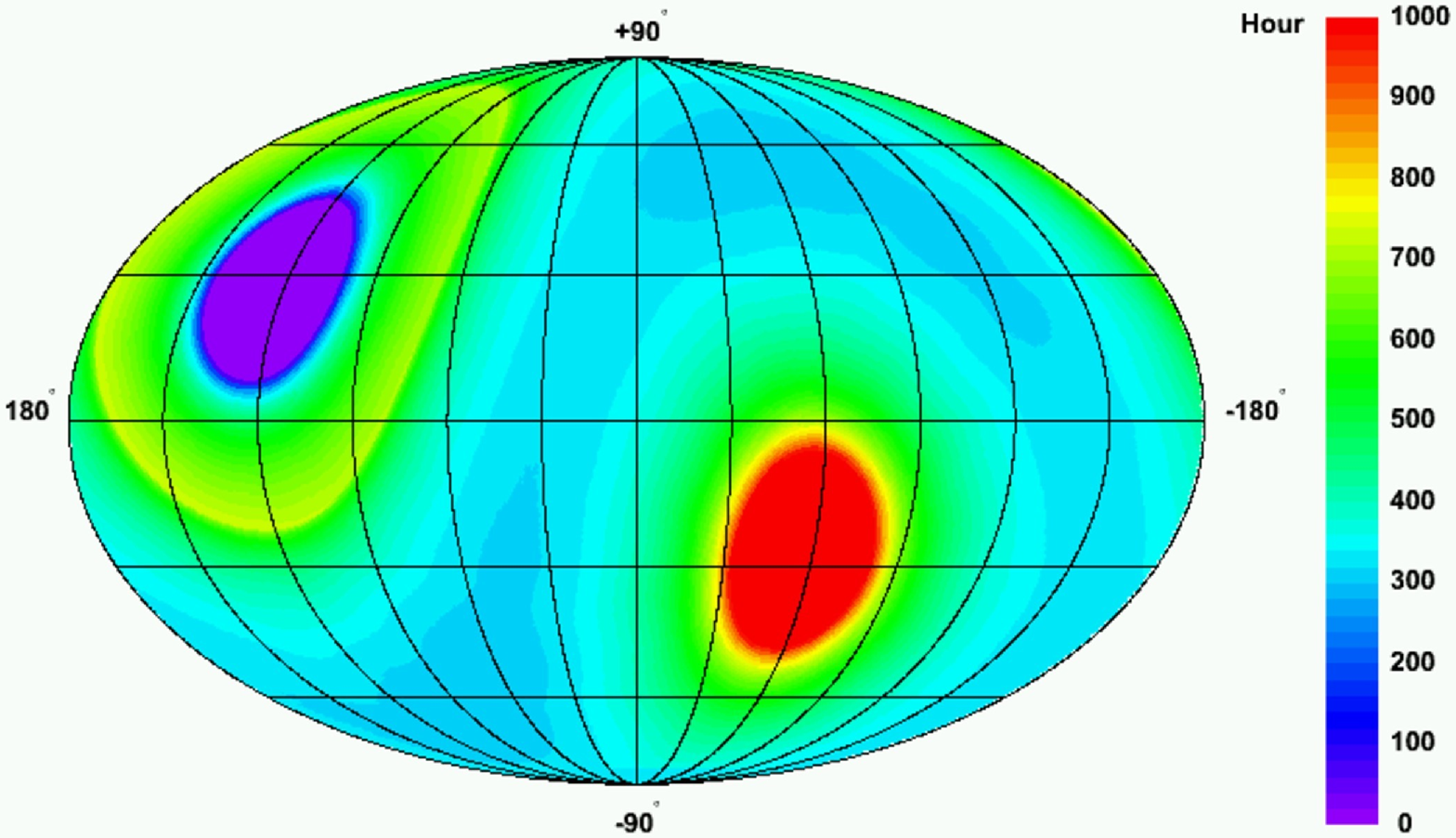}
  \end{center}
 \end{minipage} &
 \begin{minipage}{0.49\hsize}
  \begin{center}
   \includegraphics[width=0.98\hsize]{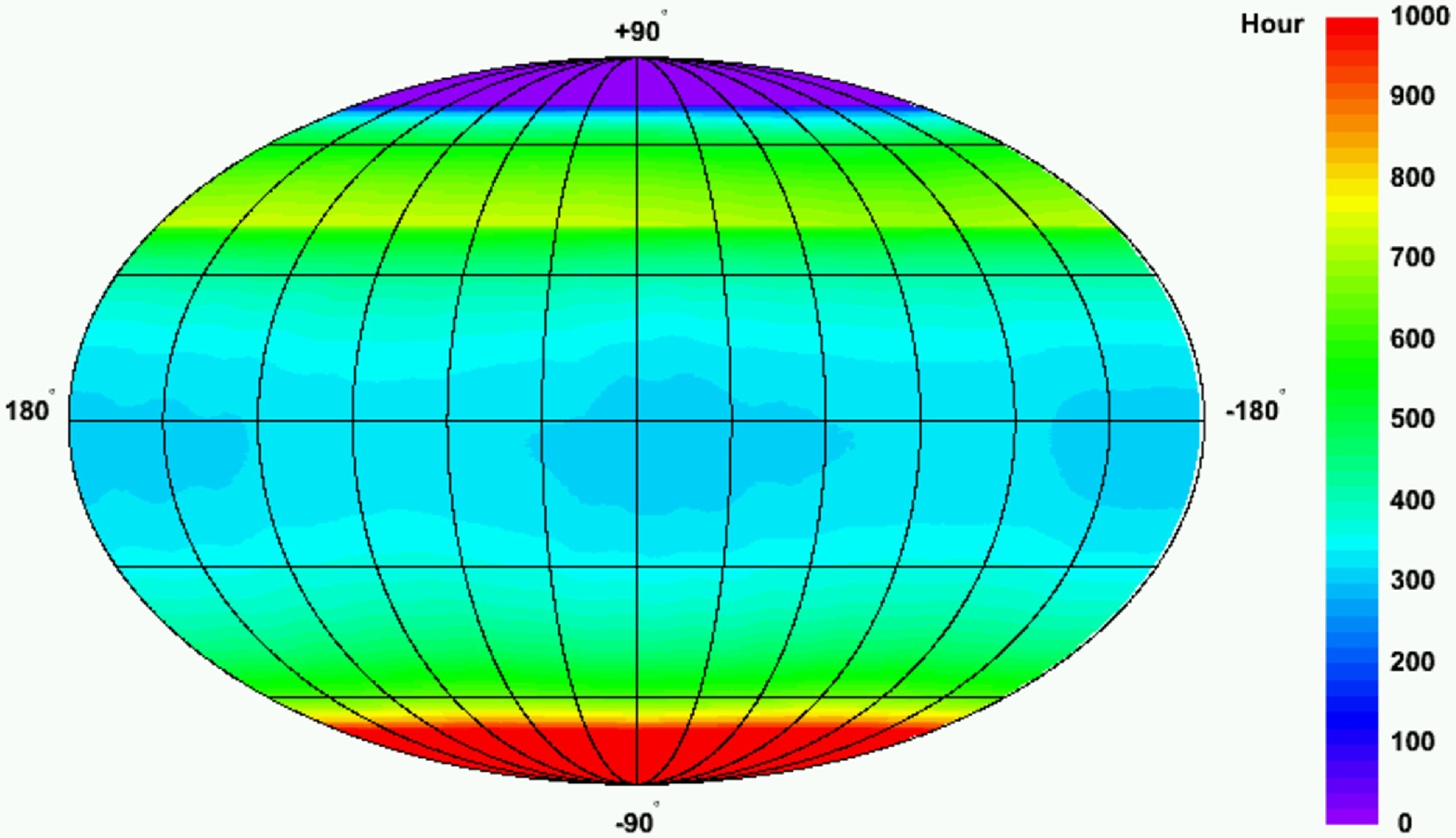}
  \end{center}
 \end{minipage} \\
 \end{tabular}
\end{center}
 \caption{
Exposure map for observation with NTA on Hawaii Island
(the Site1 position:
19$^{\circ}32'28''$ N,
155$^{\circ}34'03''$ W,
3294 m a.s.l.),
with Mollweide projection in Galactic (left) and Equatorial (right) coordinates.
Maximum observation time is normalized to 1000~hours per year (red),
where NTA can detect with maximum efficiency (total duty of 11.4\%).
 }
\label{fig:exposure}
\end{figure}

Fig.~\ref{fig:exposure} shows the exposure map
for the observation with NTA on Hawaii Island
(the Site1 position:
19$^{\circ}32'28''$ N,
155$^{\circ}34'03''$ W,
3294 m a.s.l.),
with Mollweide projection in Galactic (left) and Equatorial (right) coordinates
on the celestial sphere.
The maximum observation time is normalized to 1000~hours per year,
as shown in red in the figure where NTA can detect with maximum efficiency,
which means total duty of 11.4\%,
corresponding to about half the above ideal case.
The location of NTA on Hawaii Island allows us to enjoy
a survey of our
galactic center for more than
several hundred hours each year.

\subsection{Effect of Changing Site Layout}\label{sec:pos}
\begin{figure}[t!]
\begin{center}
\begin{tabular}{cc}
 \begin{minipage}{0.49\hsize}
  \begin{center}
    \includegraphics[width=0.98\hsize]{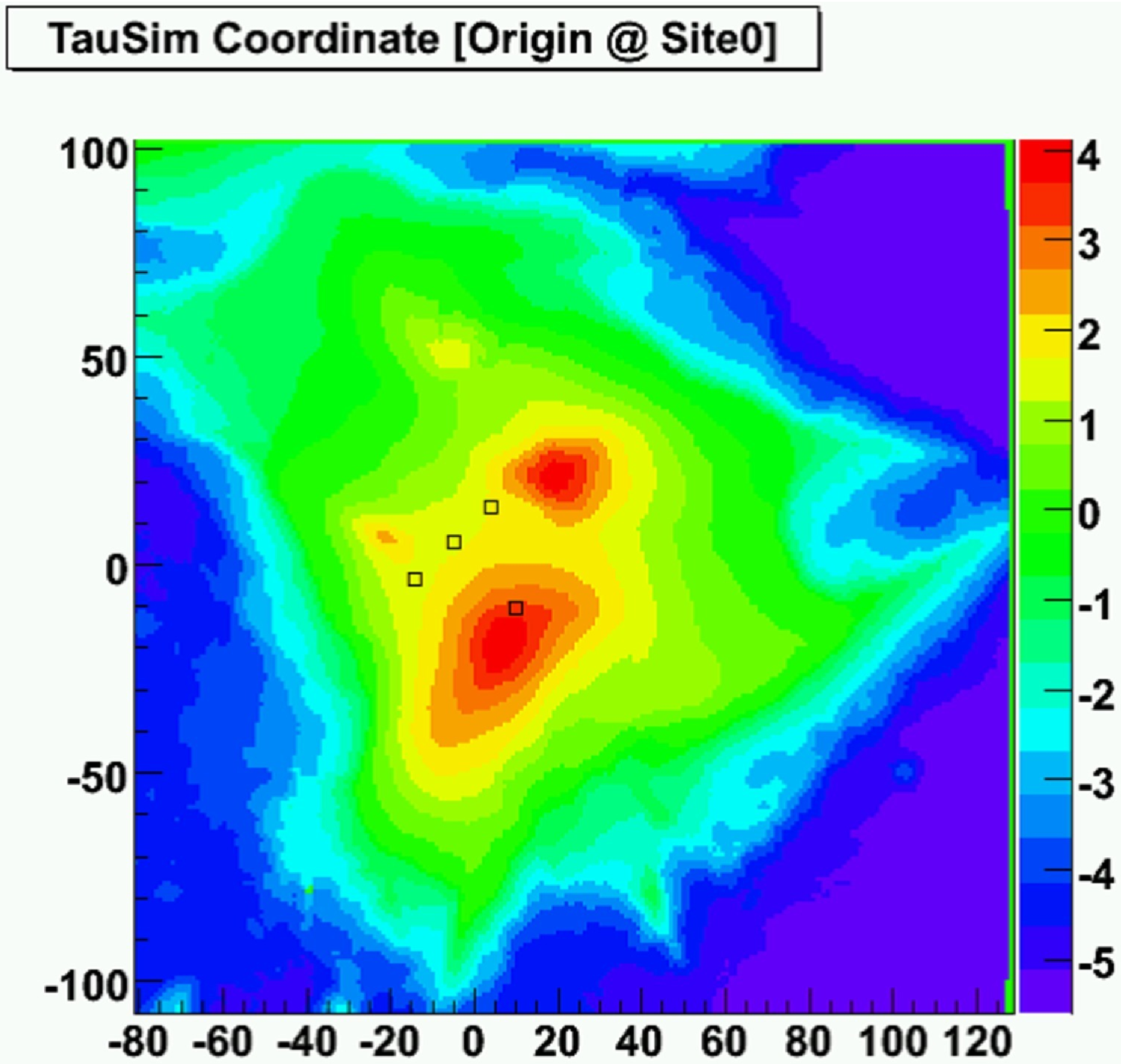}
  \end{center}
 \end{minipage}
&
 \begin{minipage}{0.49\hsize}
  \begin{center}
   \includegraphics[width=0.98\hsize]{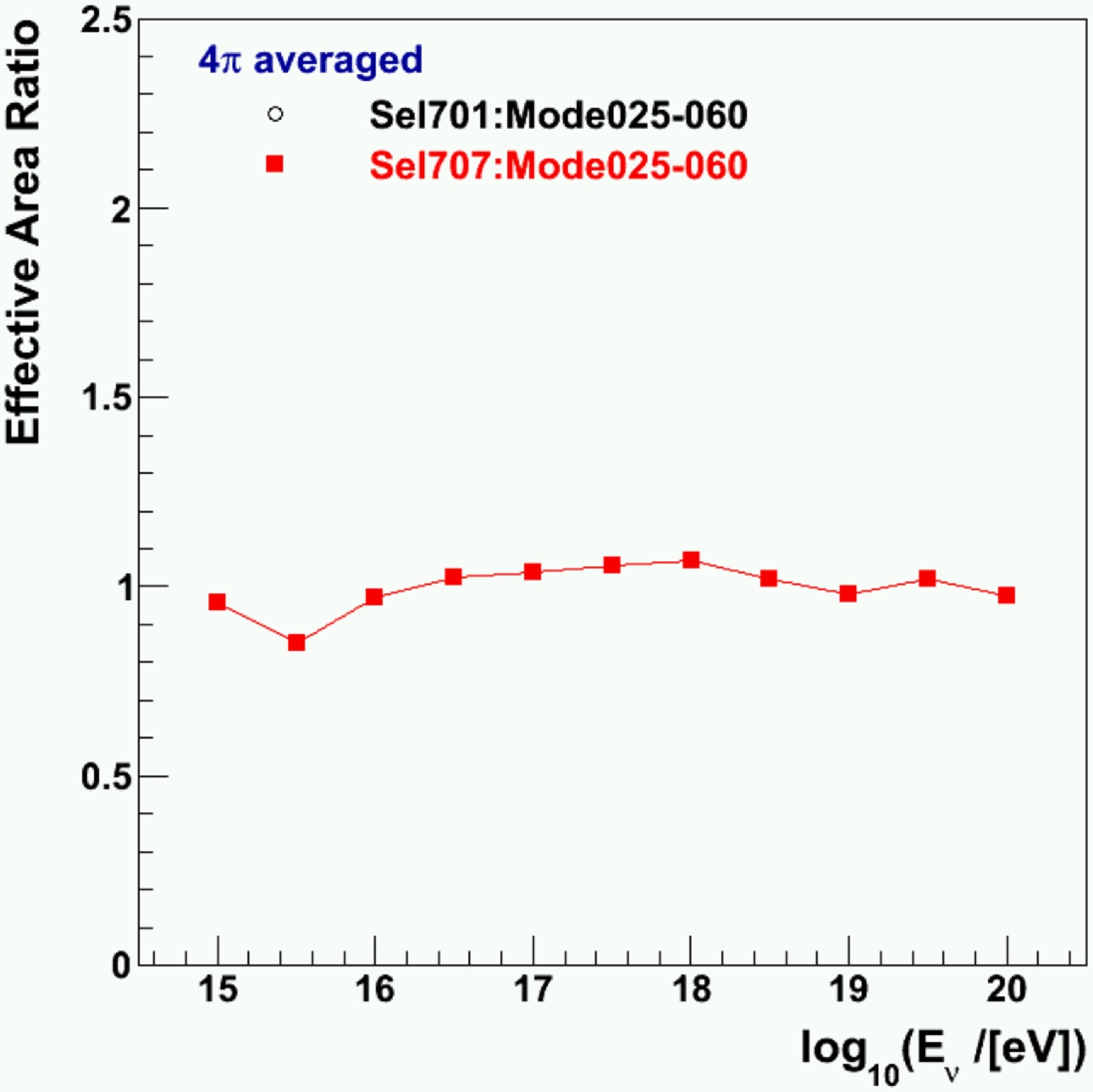}
  \end{center}
 \end{minipage} \\
 \end{tabular}
\end{center}
 \caption{
(left) Modified layout of
NTA sites, and
(right) ratio of the two sets of effective detection area
for $\nu_{\tau}$s
as a function of $E_{\nu_{\tau}}$,
obtained with modified
and regular
layouts
(Fig.~\ref{fig:site}).
The location of Site0 is changed into the midpoint between Site2 and Site3.
}
 \label{fig:sitepos}
\end{figure}

To check the effect of changing the site layout on the detection sensitivity of NTA,
we changed only the location of Site0 into the midpoint between Site2 and Site3,
as shown in Fig.~\ref{fig:sitepos}~(left),
and repeated the sequence of simulation for diffuse sources
as before.
The right side of Fig.~\ref{fig:sitepos} shows the ratio of the
two sets of effective detection area for $\nu_{\tau}$s
as a function of $E_{\nu_{\tau}}$,
which are obtained with modified layout and regular one.
We do not see any significant change over all energies in the PeV-EeV region.
The layout can therefore
be adapted to practical concerns.

\subsection{Angular resolution}
\label{sec:angreso}
\noindent
As discussed in Section \ref{sec:tauang},
a Cherenkov $\tau$ shower with $E>1$~PeV  preserves the arrival direction of
the parent $\nu_{\tau}$ to within 0.1$^\circ$ accuracy.
This means that the ability of the detector to reconstruct the arrival direction
results in the precise identification of the VHE neutrino sources and leads to the realization
of ``multi-particle astronomy''.
Owing to its high-resolution imaging capability,
the NTA detector has a huge potential to improve
the reconstruction of the arrival direction of $\nu_{\tau}$-induced air-showers.

NTA will observe quasi-horizontal air-showers with the primary energies between PeV and EeV.
Note that the location of the shower maximum in the atmosphere, i.e., the depth of maximum
development $X_{\rm max}$ is expected to be roughly
in the range of 500-1000~g/cm$^{2}$
from the first interaction in the atmosphere for tau decays of different energies~\cite{Grieder:1979ty}\cite{Hillas:1977wu}.
The depth range corresponds to the length of 6-12~km along the air-shower axis
assuming the averaged atmospheric pressure of 0.7~atm.
The shower axis is defined as the extension of the initial momentum vector of the
incident tau decay particle in the direction of cascade propagation.
The particle density in the shower core, i.e., in the central region, is very high and drops rapidly
with increasing the core distance from the shower axis.
Electron lateral distribution functions (LDFs) of air-showers are well described by
Nichimura-Kamata-Greisen (NKG) functions~\cite{Greisen:0hp}\cite{Kamata:0bi}.
The LDFs measured by KASCADE in the energy range
from 5$\times$10$^{14}$~eV up to 10$^{17}$~eV
can be reproduced accurately for the fit parameter of
electron lateral distributions $r_{\rm e}\sim$ 20-30~m with fixing $s\sim$1.6-1.8,
which are different from the original Moli\`ere radius $r_{\rm M}$ and age parameter $s$
in the NKG functions~\cite{Antoni:2001kt}; the latter are
used in the Monte Carlo simulation here.
In addition, Monte Carlo simulations of air-showers find steeper LDFs than the NKG distribution
in higher energies~\cite{Hillas:1977wu}.

Let us first we try a back-of-envelope estimate, for example,
an image of 1000 photo-electrons detected by a light collector,
which originated from an air-shower trajectory with the Gaussian LDF of
$\sigma$~=~30~m the track length of 6~km as a typical event.
From the image, we can determine
the shower detector plane (SDP) with accuracy of $\sim$~0.2~mrad (0.6~arcmin) purely
in a statistical way, neglecting the error due to image resolution,
although real LDF around the air-shower axis is exponential and steeper than the Gaussian distribution.
This means air-shower developments are fairly useful for pointing back to the original source,
and detector resolution dominantly limits
the determination of arrival direction of air-showers.
Ashra-1 has already achieved a few arcmin imaging resolution
including optical alignment errors,
and NTA will be operated with the resolution $\sim$~1.5~mrad (5~arcmin).
In that case, the SDP direction will be determined with the accuracy of 15~m,
which is finer than the measured shower LDF parameter of $r_{\rm e}$,
at the distance of 10~km as a typical case.

The simulated air-shower event shown in Fig.~\ref{fig:event2} provides a more concrete
and realistic example.
The $\nu_{\tau}$ is generated at the energy of 100~PeV with the elevation angle of $-2.5~{^\circ}$,
which is a quasi-horizontally upward event.
The converted $\tau$ emerges from the earth with energy 73~PeV
and decays into particles which induce an air-shower of 46~PeV.
The fluorescence light generated from the air-shower is
triggered by light collectors deployed at Site0 and Site2.
The closest approaches or the impact parameters ($R_{\rm P}$'s) to the air-shower axis
are 9.8~km and 12~km from the Site0 and Site2 respectively, and
the distances to the $X_{\rm max}$ from Site0 and Site2 are 16~km and 12~km respectively.
Due to the design of the layout of the four sites, i.e.
the 25~km-side triangle (Site1,2,3) with the centered site (Site0)
as shown in Fig.~\ref{fig:NTA-Observatory},
almost all air-shower axes which pass the air volume above the inner triangle area of NTA
have the closest approach of less than 12.5~km to one of the four sites.
Therefore, the simulated event shown in Fig.~\ref{fig:event2}
is an example with relatively poor signal
of all generated events of the same primary energies.
\begin{figure}[t]
\begin{center}
 \includegraphics[width=0.98\hsize]{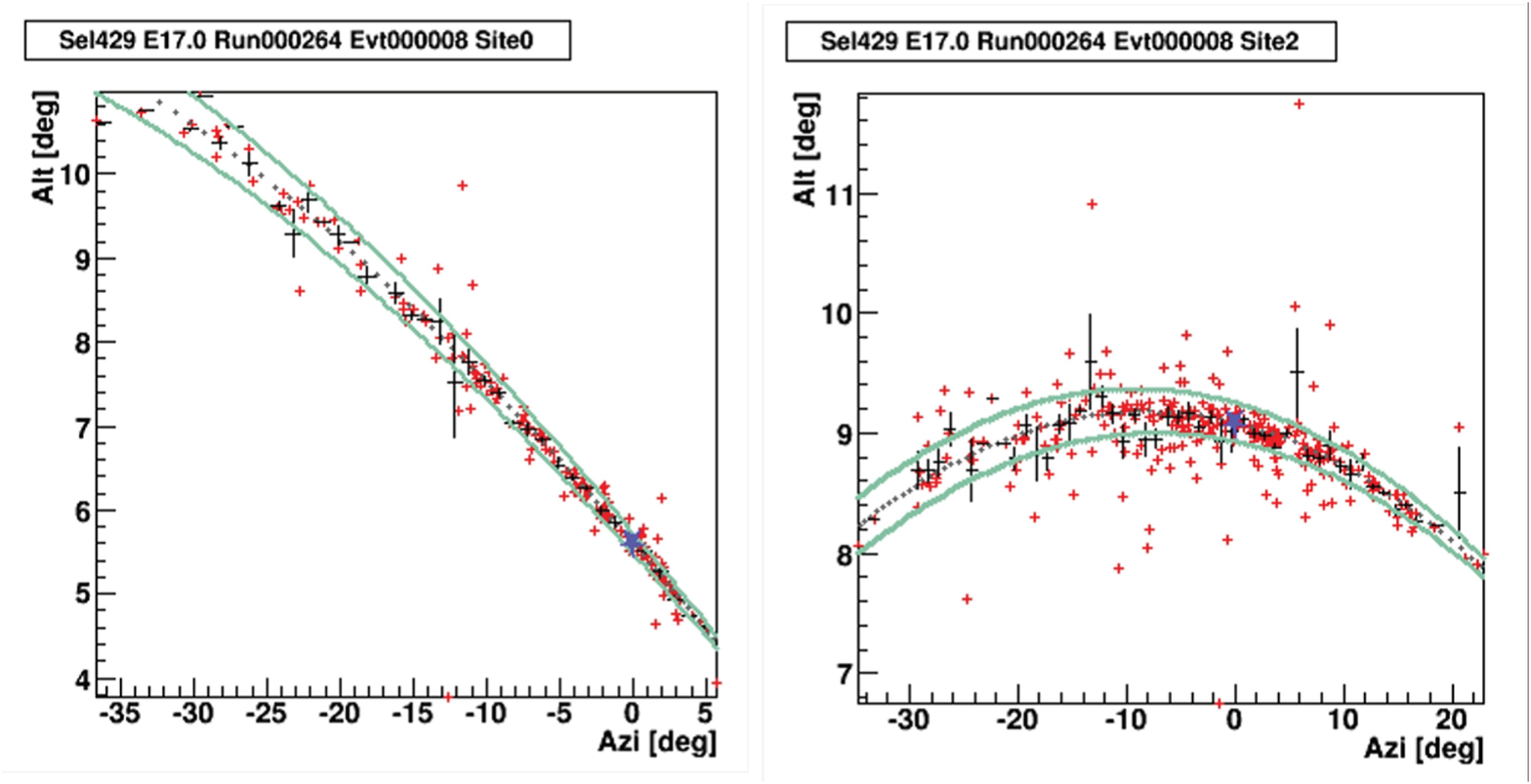}
\end{center}
\caption{
	Simulated photo-electron images
	of air-shower development (red cross marks)
	in the alt-azimuth coordinates of
	light collectors installed at Site0 and Site2.
	The shower detector plane (SDP) with
	the true air-shower axis (light green dotted line)
	and those tilted by $\pm$0.2$^{\circ}$ with respect to
	the SDP based on the true air-shower axis of which primary energy is 10$^{17}$~eV
	and altitude angle is $-2.5^{\circ}$.
}
 \label{fig:sdpfit}
\end{figure}

Fig.~\ref{fig:sdpfit} shows photo-electron images (red cross marks) on the plane of
the altitude and azimuthal angles, which are triggered and
taken from the same event shown in Fig.~\ref{fig:event2}
by light collectors at Site0 and Site2.
The SDPs that correspond to the true air-shower axis (light green dotted line)
and tilted by $\pm$0.2$^{\circ}$ (light green solid line)
are indicated on the same figure.
Note that the LDFs used in this analysis is the traditional NKG functions with the fixed
Moli\`ere radius $r_{\rm M} = $~79~m and a variable age parameter $s$ in the NKG form, not
the steeper LDF recently measured by KASCADE.
Even with the traditional NKG form the dominant components of photo-electrons of the image
is between the boundaries of the SDPs tilted by $\pm$ 0.2$^{\circ}$ with respect to
that of the true air-shower axis.
For the simple fit adopted here,
to eliminate the statistical fluctuation in the altitude angle coordinate, profile histograms
are used to display and fit the mean values of altitude angles in rebinned azimuthal
bins, which are also shown in Fig.~\ref{fig:sdpfit}.
The mean values based on the profile histograms reproduce the locations of true air-shower axis
well.
The image and trigger pixel resolutions assumed here are 0.125$^{\circ}$ and 0.5$^{\circ}$
respectively.
With the combined images taken at Site0 and Site2, the SDP can be reproduced with
fit error of 0.02$^{\circ}$.
Note that we can further improve the reconstruction accuracy here especiallyly in the higher energies
by using more sophisticated likelihood fit analysis and expected
shower developments for each shower energy with the advantage of high resolution images,
beyond our present simple treatment of profiling the lateral distribution on the alt-azimuth coordinate
plane.
In the literature \cite{Asaoka-Sasaki11},
we have described the detailed Monte Carlo study
of the angular resolution only for monostatic observation of Cherenkov images
of $\tau$ showers generated by earth-skimming PeV-EeV $\nu_{\tau}$ events with the Ashra-1 detector system.
In that work, we confirm that a likelihood analysis with fine images of shower core
structures compared with Monte Carlo expectation of air-shower development improves
significantly the arrival direction resolution.
The sophisticated and completed likelihood analysis using Monte Carlo simulated air-shower developments
is beyond the scope of this LoI, before determining detailed detector design but with only
assumed baseline concepts.

For pointing back to $\nu_{\tau}$ sources, we simultaneously fit observed data with
the four parameters of ($\phi_{SDP}$, $\theta_{SDP}$) of the normal unit vector of SDP
and the impact parameter $R_{\rm P}$ and the arrival direction angle $\chi_{0}$ of the air-shower axis
constrained on the SDP of the event.
The detailed definitions of $\chi_{0}$ and $R_{\rm P}$ on SDP can be seen in
the Fly's Eye detector paper~\cite{FlysEye}.
In the case of reconstruction of quasi-horizontal air-showers, however,
purely geometrical bistatic method is not useful, since the opening angles between two of
SDPs observed with light collectors at different sites are nearly flat, i.e. 180$^{\circ}$ and
strongly correlated with each other.
For the determination of $\nu_{\tau}$ source positions,
particularly for $\chi_{0}$ and $R_{\rm P}$ on SDP,
we should fully utilize the timing
information recorded by the trigger pixel sensor in the NTA system as well as the image data.
Our realistic baseline design of the trigger pixel sensor system has its pixel FOV of
0.5$^{\circ} \times$ 0.5$^{\circ}$
and
the least time stamp resolution of each trigger pixel of 10~ns.
An advantage of the baseline trigger design based on the developments of Ashra-1 is that
we can optimize pixel and timing resolutions of imaging system and trigger systems independently.
For each event, after determining these four parameters, they are transformed into
another set of parameters, i.e. the arrival direction of the air-shower axis
($\phi_{AS}$, $\theta_{AS}$) and the position (X$_{\tau}$,Y$_{\tau}$,Z$_{\tau}$(X$_{\tau}$, Y$_{\tau}$))
where the $\tau$ emerges from the mountain. The Z coordinate of the emerging point
Z$_{\tau}$(X$_{\tau}$, Y$_{\tau}$) is obtained  from a topographic map
as a function of X and Y coordinates.
For the aim of partially demonstrating the performance of  simultaneous fit of parameters needed
to point back to sources of observed $\nu_{\tau}$ candidates, we have prepared for
Monte Carlo data simulating $\nu_{\tau}$ events with the energies of every half decade
between 10$^{15.5}$~eV and 10$^{19}$~eV
with fixed altitude angle of $-2.0^{\circ}$ and fixed azimuthal angle toward the peak of Mauna Kea.

\begin{figure}[t!]
\begin{center}
\begin{tabular}{cc}
 \begin{minipage}{0.49\hsize}
  \begin{center}
    \includegraphics[width=0.98\hsize]{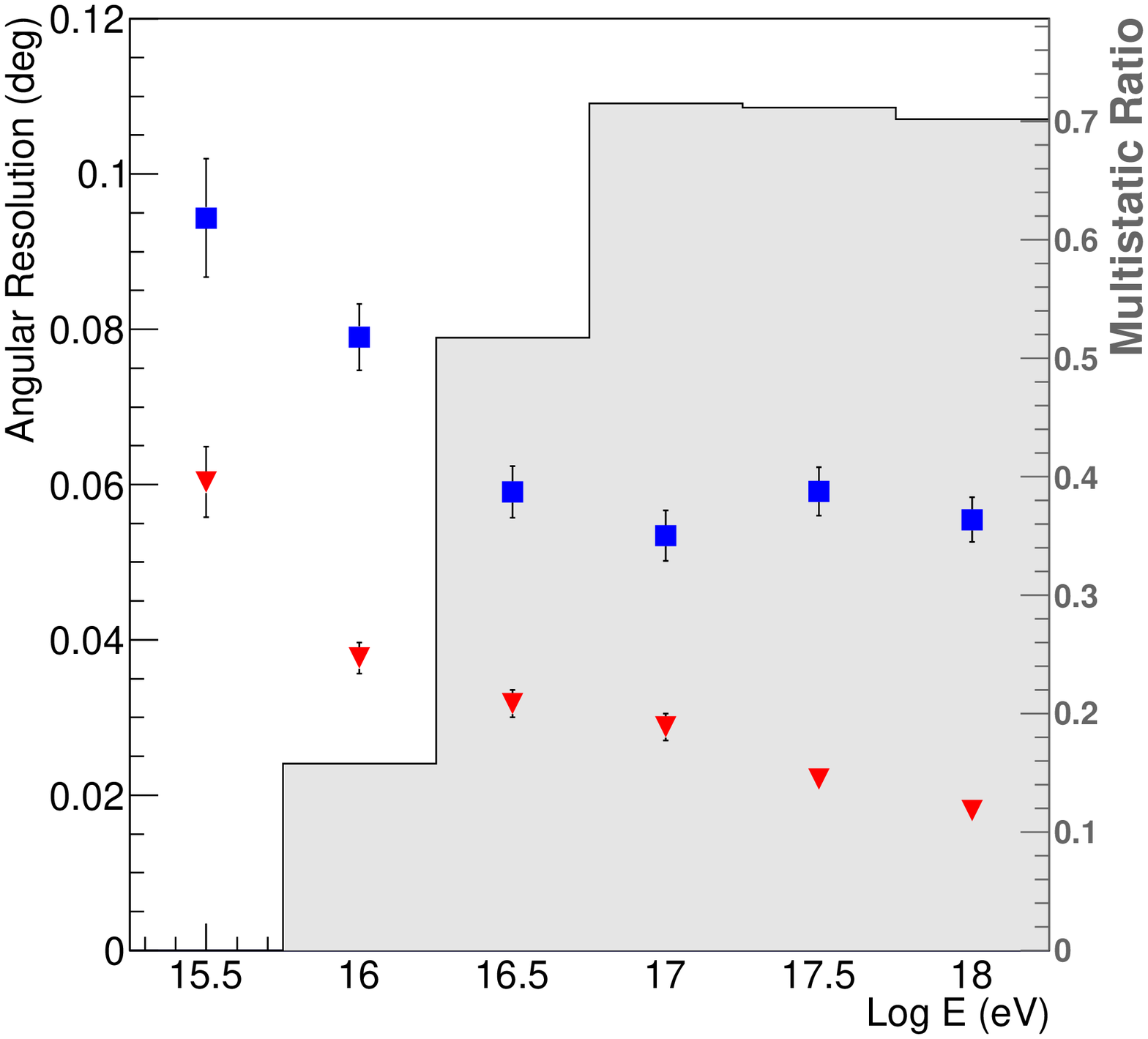}
  \end{center}
 \caption{
Angular resolutions of reconstructed arrival directions of $\tau$ from Mauna Kea.
The RMS resolutions
of the reconstructed $\tau$ arrival direction
(blue filled square marks with errors)
and the polar angle components
(red filled square marks with errors)
improves with energy.
Multistatic rate (shaded histogram), i.e. ratio of events
observed from two or more sites
saturates above 10$^{17}$~eV.
}
 \label{fig:angres_vs_logE}
 \end{minipage}
&
 \begin{minipage}{0.49\hsize}
  \begin{center}
   \includegraphics[width=0.98\hsize]{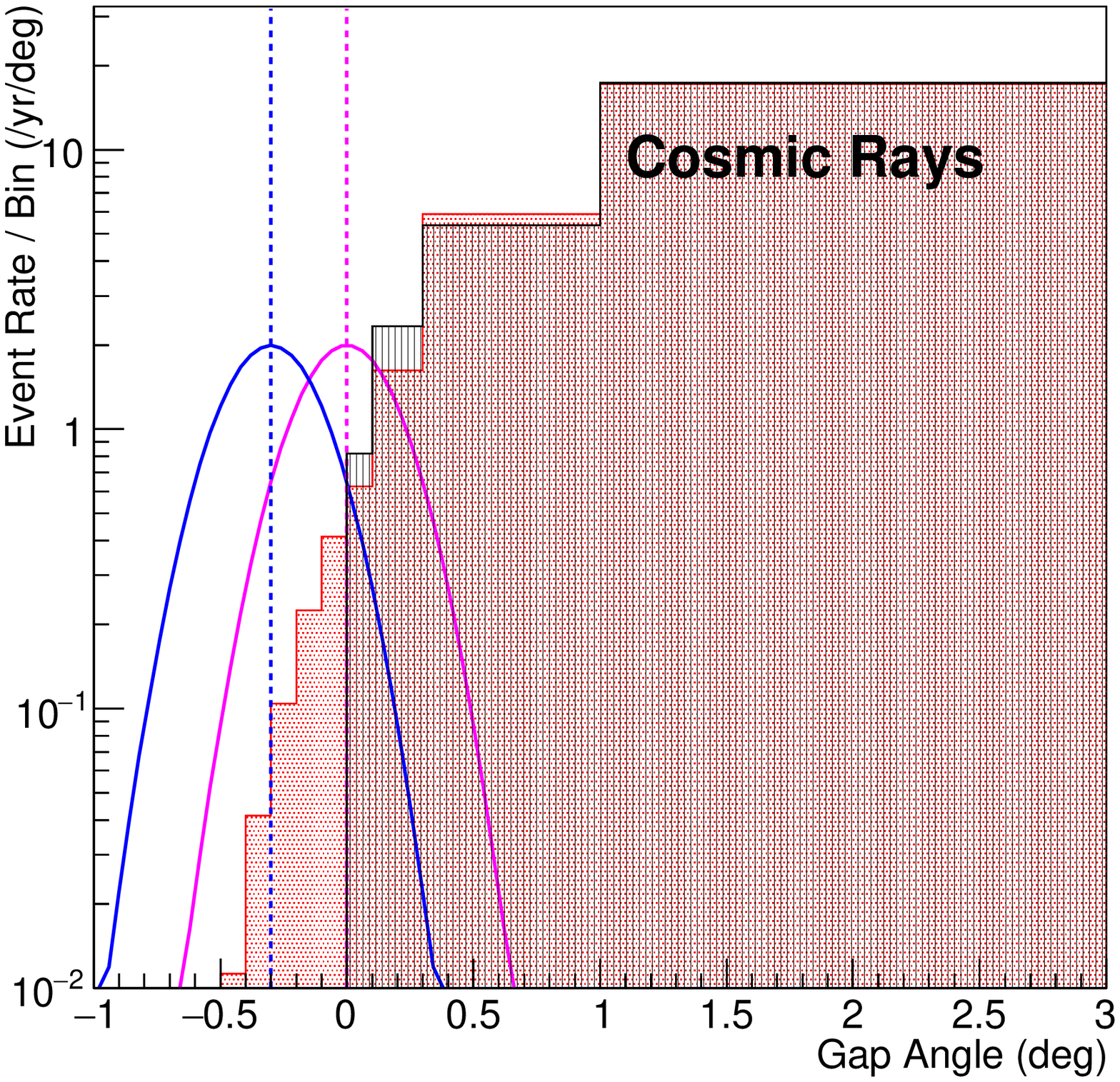}
  \end{center}
 \caption{
	 Estimate for
	 cosmic ray background contaminations
	 due to misreconstruction of their arrival directions.
	 True angle with respect of the mountain edge (Hatched histogram)
	 and observed angles (red filled histogram) assuming the angular resolution of 0.2$^{\circ}$
	 after smering out the true angular distribution.
	 Two probability density functions for signal $\tau$s
	 with the arrival directions of 0.0$^{\circ}$ and -0.3$^{\circ}$.
	 The integrated areas of these histograms are normalized to be the expected
	 event rate/year/DU.
}
 \label{fig:bg_edge}
 \end{minipage} \\
 \end{tabular}
\end{center}
\end{figure}

Fig.~\ref{fig:angres_vs_logE}
shows the results of fitting Monte Carlo data.
The blue filled square marks with error bars show the total RMS resolution of the reconstructed $\tau$ arrival direction
as function  of logarithmic energies of generated $\nu_{\tau}$s.
The red filled triangle marks with error bars
show the total RMS resolution of the polar angle component of the reconstructed $\tau$ arrival direction
as function of logarithmic energies of  generated $\nu_{\tau}$s,
which is important to eliminate cosmic ray background events due to misreconstruction of the arrival directions
as described in the next subsection.
Fig.~\ref{fig:angres_vs_logE} also shows the event rate of multistatic observation, that is ratio of
events observed with DUs at two or more sites, which is another result from this Monte Carlo simulation study.
Although the accuracy of the angular resolution increases, the multistatic rate is seen
saturated above 10$^{17}$~eV, which indicates the limitation of this simple fit method using
profiling LDS at each bin of longitudinal development along the air-shower axis.

This simple method works well when the core structure is negligible.
Although the NTA detectors with high resolution imagers resolve out the shower core structures for
high energy events, we treated the shower axis as a line without any lateral structure.
At higher energies, the effect of the shower core structure become significant.
Again likelihood analysis with fitting functions made of enough number of Monte Carlo events taking into account the
shower lateral and longitudinal development more consistently, the resolution should be recovered particularly
in the higher energy region.
Even using the simple fit method for quasi-horizontal $\tau$ shower events with Monte Carlo events with
the altitude angle of $-0.2^{\circ}$,
we have confirmed the pointing resolution of 0.1$^{\circ}$-0.06$^{\circ}$.
For the altitude angle or inclination of SDP, they can be determined within an error of 0.06-0.02$^{\circ}$.
Detailed and precise Monte Carlo studies of the NTA detector system will be performed elsewhere,
at step of the detector design report, after the publication of this LoI.

For the moment, we quote the estimate of the angular resolution to be less than $0.2^{\circ}$
as a fairly conservative estimate from simulated events of monostatic Cherenkov images
in the most pessimistic case of reconstruction of images taken with the NTA detector system.

\subsection{Background}
\label{bgd}
\noindent
In this subsection, we evaluate the background events due to air-showers.
Background events due to the detector itself are discussed in Ref. \cite{AshraCNeu}.
Air-shower background candidates are normal cosmic rays, muons, muon neutrinos,
$\tau$s, and $\nu_{\tau}$s.
From simple flux calculations, it is shown that
the neutrino components through mountain, prompt $\tau$ components and muons
are negligible \cite{Noda09, Zas93, Enberg08, Martin03, Athar03}.
Thus, we consider normal cosmic rays with large zenith angles, of which arrival directions are misreconstructed,
as the dominant background contamination in this study.
To evaluate the rate of the background contamination due to the directional misreconstruction,
we count the air-shower events which pass through the sky region above mountain edges or horizon within
assumed gap angles.
To simulate normal cosmic ray air-showers
we use CORSIKA in the same way as with the $\nu_{\tau}$ simulation
with a thinning parameter of 10$^{-5}$ as a result of confirming it to be acceptable.
Assuming the maximum weather efficiency of 100\%, 1750~hr of observation time on Hawaii Island
is expected in one year.
We use the trigger pixel threshold of 20 photoelectrons as a realistic sensitivity which is same as
the Ashra-1 DU one.
The results are shown in Table~\ref{tab:CRcontami}.
\begin{table}[t!]
\begin{center}
\begin{tabular}{|l|r|r|r|r|}
\hline
Gap angle $\delta\theta$ ($^{\circ}$) &  0.1 & 0.3 & 1.0 & 3.0 \\
\hline
CR    rate (showers/yr/DU)               & 0.082 & 0.55 & 4.3 & 39 \\
\hline
\end{tabular}
\end{center}
\vskip-0.3cm
\caption{
	The annual rate of cosmic ray showers which pass through the sky regions within assumed gap angles
	and detected with NTA with the trigger pixel threshold of 20 photoelectrons.
}
\label{tab:CRcontami}
\end{table}

From this Monte Carlo study,
we estimate the background contamination rate for the observation of NTA.
We require that the arrival direction of the candidate event
must emerge from earth or mountain.
Background events may pass through the requirement of the direction of
air-shower axis direction due to the misreconstruction of air-shower events.
However, the background rate should be very low and almost free.
From the geography of Hawaii Island and the layout of the NTA sites,
this directional misreconstruction is  potentially
caused only near the edge between earth/mountain and atmosphere.
We make Monte Carlo estimates for ordinary cosmic ray air-showers of which axis pass through
the gap spaces of the solid angle just above mountain edge as described above.

For this aim, we make a histogram of differential event rate evaluated from the result
listed in Table~\ref{tab:CRcontami} and smear out the distribution by the angular resolution
to quantitatively estimated the leakage into our $\tau$ shower candidate sample from the outside of
the fiducial volume of NTA.
Fig.~\ref{fig:bg_edge} shows the differential distribution of true air-shower axis directions
for ordinary cosmic rays (black shaded histogram) and the expected observed air-shower directions
after smearing them by very conservative angular resolution of 0.2$^\circ$ (red pearskin finish),
which is derived after detailed Monte Carlo studies on the directional reconstruction for Cherenkov air-showers
in comparison of the reconstruction with fluorescence light detection as shown in
the previous section.

As a result, the rates of cosmic rays which pass over the mountain edge (0$^{\circ}$
in Fig.~\ref{fig:bg_edge})
and constrained boundary of solid angle below the mountain edge by 0.3$^{\circ}$ are estimated to be
0.08 events/year and 0.006 events/year for one DU with the FOV of 32$^\circ \times$ 32$^\circ$.
Fig.~\ref{fig:bg_edge} also shows the probability density function for one event just on the edge
(pink curve) and that on the fiducial limit 0.3${^\circ}$ below the mountain edge, where all histograms are
normalized for the integrated area to be annual event rate for each DU respectively.

In the case of the fiducial solid angle restricted after cutting $<0.3^\circ$ with respect to the mountain edge,
total CR contamination rate in the NTA detector system with 30~DUs is estimated to be $\sim$~0.2/yr and
the efficiency of the effective aperture for Earth-skimming $\nu_{\tau}$s to be $\sim$~90~\%.
We can realize the almost background-free condition without sacrificing the superior sensitivity of NTA
for detecting Earth-skimming $\nu_{\tau}$s using good advantages of high resolution images and
the pointing accuracy.

\section{Time Frame, Organization, and Funding}
\noindent
At the present time, we are investigating various options for the site,
organization, and the design of instruments.
Also, we intend to invite other groups to either contribute
directly to this project, or to join us on the site with their complementary
instruments.
The resulting synergy effects would benefit all parties, avoid
unnecessary parallel technical developments, and
lead to cost savings for the different projects.
It is clear that collaboration forming is key to success
of the NTA scientific goal.
The time frame for the proposed project is thus
given both by considerations of budgetary and scientific aspects.
In March 19--20, 2014,
a preliminary workshop (VHEPA2014) was held
at Kashiwa campus of the University of Tokyo
to discuss the design of the project and plans with
interested colleagues.
In April 8-9, 2015
a small workshop, successive to VHEPA2014,
was held at National Taiwan University to discuss the scientific goals
and  promotion of the project.
We plan to have an informal meeting to discuss post-IceCube project
new detector project at the 34th International Cosmic Ray Conference (ICRC)
held from July 30 to August 6, 2015, in The Hague.

In January, 2016
we plan to hold
a workshop as VHEPA2016
at University of Hawaii Manoa
to discuss more detailed physics and NTA potential performance,  funding processes,
and make ready for a white paper of the project as basic documents to
use for the funding requests in each country.

We have already set up
the International Executive Board (IEB) of NTA
for decision making and steering the collaboration
since October 12, 2012.
Some IEB members have already submitted funding requests
to exchange information, detector design, meetings, and construction of the detector.
IEB selects the representative who becomes the spokesperson of the collaboration.
Each country has a Local Institutional Board (LIB), which is composed of
representatives from institutes in the country.
LIB selects the representative who becomes a member of IEB.
We will set up various Working Groups (WGs) as real working bodies.
WG leaders are nominated by LIB and decided by IEB.

Major decisions concerning the hardware implementation should
be undertaken in 2015, toward the publication of Project Proposal.
In the subsequent two years, components should be developed and tested,
and we should continue to request the Japanese government for construction funding.
We can eliminate critical developments by using experience from Ashra-1 and NuTel projects.
We aim at installing the first detector site and starting commissioning operation
in early 2017, if we succeed in the primary Japanese funding in time.
We plan to start the operation using at least a part of Site0 and Site1 of proposed four sites by 2018,
with the primary construction budget covering at least 1/4 of full operation cost.
Once we succeed in the primary funding request and start construction of the first sites,
we start the request for matching funds to the governments of various collaborating countries.
We aim at starting construction of Site2 and Site3 with the matching funds
from countries other than Japan.

Since major choices concerning implementation details are still open at this time,
it does not seem appropriate to discuss a detailed cost breakdown.
To provide a guideline, however,
we have estimated in some detail the cost of one design variant, with major components
of the detector either covered by offers from potential manufacturers
(light collector mount, mirrors, trigger-readout sensors and electronics, and so on),
or extrapolating from well-known costs of the Ashra-1 instruments.
On this basis, we estimate the production cost per light collector in the 20M~yen range,
plus total R\&D costs of about 100M to 150M~yen.
One detector unit (DU), as shown in Fig.~\ref{fig:AshraLC},
requires four light collectors and one trigger and readout unit,
so the rough cost estimate is 100M~yen per DU.
We plan to build at least 12, 6, 6, and 6 DUs
at Site0, Site1, Site2, and Site3, respectively,
for the coverage of FOV as shown in Table~\ref{tab:site},
assuming the FOV for DU
to be $32^{\circ} \times 32^{\circ}$.
30 DUs are needed for covering the total FOV.
We do not include infrastructure costs such as site access, site preparation,
light collector shelters, networking, and so on.
Roughly speaking, at least 100M~yen per DU is needed from the experience
of construction of Ashra-1 at Mauna Loa.
The current crude estimate is 5000M~yen for the construction of NTA.

The Ashra-1 collaboration has agreed to continue the observation at
the Mauna Loa site as well as explore the NTA system, at least
by the time NTA starts the construction.
In order to explore the hardware and software components, the Ashra-1
experience is recognized as an important and useful demonstration of
the challenging new detection techniques.

\section*{Acknowledgment}
\noindent
We thank P.~Binder and J.~Goldman from University of Hawaii Hilo for their useful comments and help.
Many thanks to the Ashra-1 and NuTel collaborations for their cooperation.
The Ashra Experiment is supported by the Coordination Fund for Promoting Science and Technology and
by a Grant-in-Aid for Scientific Research from the Ministry of Education, Culture, Sports, Science and Technology of Japan.
\bibliographystyle{unsrt}
\bibliography{main}

\begin{thebibliography}{100}

\bibitem{loeb2006cumulative}
A.~Loeb and E.~Waxman.
\newblock {\em Journal of Cosmology and Astroparticle Physics}, 2006(05):003,
  2006.

\bibitem{PhysRevLett.111.021103}
M.~G. Aartsen et~al.
\newblock {\em Phys. Rev. Lett.}, 111:021103, 2013.

\bibitem{icecube2013evidence}
IceCube Collaboration et~al.
\newblock {\em Science}, 342(6161):1242856, 2013.

\bibitem{Aartsen:2013hn}
M~G Aartsen, R~Abbasi, M~Ackermann, and J~Adams.
\newblock {Probing the origin of cosmic rays with extremely high energy
  neutrinos using the IceCube Observatory}.
\newblock {\em Physical Review D}, 2013.

\bibitem{PhysRevLett.104.091101}
J.~Abraham et~al.
\newblock {\em Phys. Rev. Lett.}, 104:091101, 2010.

\bibitem{Sasaki08}
M.~Sasaki.
\newblock {\em J. Phys. Soc. Jpn.}, 77SB:83, 2008.

\bibitem{Sasaki00}
M.~Sasaki.
\newblock {\em Proc. of ICRR2000 Satellite Symposium: Workshop of Comprehensive
  Study of the High Energy Universe}, pages 109--124, 2000.

\bibitem{Barwick00}
S.~Barwick.
\newblock {\em Physica Scripta.}, T85:106, 2000.

\bibitem{PLI11}
Y.~Asaoka and M.~Sasaki.
\newblock {\em Nucl. Instrum. Methods Phys. Res. A}, 647:34, 2011.

\bibitem{AshraCNeu}
Y.~Aita et~al.
\newblock {\em The Astrophysical Journal Letters}, 736(1):L12, 2011.

\bibitem{GCN8632}
Y.~Aita et~al.
\newblock {\em GCN Circular, 8632}, 2008.

\bibitem{GCN11291}
Y.~Asaoka et~al.
\newblock {\em GCN Circular, 11291}, 2010.

\bibitem{NuTel}
George~WS Hou and MA~Huang.
\newblock {\em arXiv preprint astro-ph/0204145}, 2002.

\bibitem{2002NIMPA.492...49S}
M.~{Sasaki}, A.~{Kusaka}, and Y.~{Asaoka}.
\newblock {\em Nuclear Instruments and Methods in Physics Research A},
  492:49--56, 2002.

\bibitem{2003NIMPA.501..359S}
M.~{Sasaki}, Y.~{Asaoka}, and M.~{Jobashi}.
\newblock {\em Nuclear Instruments and Methods in Physics Research A},
  501:359--366, 2003.

\bibitem{beckerhigh-energy2008}
J.~K. Becker.
\newblock {\em Physics Reports}, 2008.

\bibitem{PhysRevD.88.121301}
K.~Murase, M.~Ahlers, and B.~C. Lacki.
\newblock {\em Phys. Rev. D}, 88:121301, 2013.

\bibitem{0004-637X-768-2-186}
P.~Baerwald, M.~Bustamante, and W.~Winter.
\newblock {\em The Astrophysical Journal}, 768(2):186, 2013.

\bibitem{Berezinsky-Zatsepin1969}
V.~S. Berezinsky and G.~T. Zatsepin.
\newblock {\em Physics Letter B}, 28:423, 1969.

\bibitem{0004-637X-718-1-31}
V.~Ptuskin, V.~Zirakashvili, and E.-S. Seo.
\newblock {\em The Astrophysical Journal}, 718(1):31, 2010.

\bibitem{PhysRevLett.105.091101}
A.~Calvez, A.~Kusenko, and S.~Nagataki.
\newblock {\em Phys. Rev. Lett.}, 105:091101, 2010.

\bibitem{koterathe2011}
K.~Kotera and A.~V. Olinto.
\newblock {\em Astronomy and Astrophysics}, 2011.

\bibitem{aloisiodisappointing2012}
R.~Aloisio, V.~Berezinsky, and A.~Gazizov.
\newblock {\em Journal of Physics: Conference Series}, 2012.

\bibitem{giacinticosmic2012}
G.~Giacinti, M.~Kachelrie{\ss}, D.~V. Semikoz, and G.~Sigl.
\newblock {\em Journal of Cosmology and Astroparticle Physics}, 2012(07), 2012.

\bibitem{Gupta201375}
N.~Gupta.
\newblock {\em Astroparticle Physics}, 48(0):75 -- 77, 2013.

\bibitem{guptapev2011}
N.~Gupta.
\newblock {\em Astroparticle Physics}, 2011.

\bibitem{anchordoqui2014cosmic}
L.~A. Anchordoqui et~al.
\newblock {\em Journal of High Energy Astrophysics}, 1:1--30, 2014.

\bibitem{hillas2005can}
A.~M. Hillas.
\newblock {\em Journal of Physics G: Nuclear and Particle Physics}, 31(5):R95,
  2005.

\bibitem{1999A&A...349..259D}
A.~{Dar} and R.~{Plaga}.
\newblock {\em \aap}, 349:259--266, 1999.

\bibitem{1367-2630-11-5-055005}
J.~Hinton.
\newblock {\em New Journal of Physics}, 11(5):055005, 2009.

\bibitem{0004-637X-656-2-870}
A.~Kappes et~al.
\newblock {\em The Astrophysical Journal}, 656(2):870, 2007.

\bibitem{2004Natur.432...75A}
F.~A. {Aharonian} et~al.
\newblock {\em \nat}, 432:75--77, 2004.

\bibitem{2005A&A...437L...7A}
F.~{Aharonian} et~al.
\newblock {\em \aap}, 437:L7--L10, 2005.

\bibitem{aharonian2006discovery}
F~Aharonian et~al.
\newblock {\em Nature}, 439(7077):695--698, 2006.

\bibitem{ahlersprobing2014}
M.~Ahlers and K.~Murase.
\newblock {\em Physical Review D}, 2014.

\bibitem{aharonian2004very}
F.~Aharonian et~al.
\newblock {\em Astronomy \& Astrophysics}, 425(1):L13--L17, 2004.

\bibitem{2006A&A...448L..43A}
F.~{Aharonian} et~al.
\newblock {\em \aap}, 448:L43--L47, 2006.

\bibitem{1992ApJ...390..454H}
M.~{Hoshino}, J.~{Arons}, Y.~A. {Gallant}, and A.~B. {Langdon}.
\newblock {\em \apj}, 390:454--479, 1992.

\bibitem{2006A&A...451L..51H}
D.~{Horns} et~al.
\newblock {\em \aap}, 451:L51--L54, 2006.

\bibitem{link2006flux}
B.~Link and F.~Burgio.
\newblock {\em Monthly Notices of the Royal Astronomical Society},
  371(1):375--379, 2006.

\bibitem{bhadra2009tev}
A.~Bhadra and R.~K. Dey.
\newblock {\em Monthly Notices of the Royal Astronomical Society},
  395(3):1371--1375, 2009.

\bibitem{Aharonian-JPhysConfSer38}
F.A. Aharonian et~al.
\newblock {\em J.~Phys.~Conf.~Ser.}, 39:408, 2007.

\bibitem{2006Sci...312.1759M}
I.~F. {Mirabel}.
\newblock {\em Science}, 312:1759, 2006.

\bibitem{0004-637X-688-2-1078}
A.~A. Abdo et~al.
\newblock {\em The Astrophysical Journal}, 688(2):1078, 2008.

\bibitem{2013arXiv1306.5021A}
L.~A. {Anchordoqui} et~al.
\newblock {\em ArXiv e-prints}, 2013.

\bibitem{0004-637X-724-2-1044}
M.~Su, T.~R. Slatyer, and D.~P. Finkbeiner.
\newblock {\em The Astrophysical Journal}, 724(2):1044, 2010.

\bibitem{PhysRevD.88.081302}
S.~Razzaque.
\newblock {\em Phys. Rev. D}, 88:081302, 2013.

\bibitem{Meszaros06}
P.~M\'es\'zaros.
\newblock {\em Rep. Prog. Phys.}, 69:2259, 2006.
\newblock and references therein.

\bibitem{Ackermann10}
M.~Ackermann et~al.
\newblock {\em \apj}, 716(2):1178, 2010.

\bibitem{Nousek06}
J.~A. Nousek et~al.
\newblock {\em \apj}, 642(1):389, 2006.

\bibitem{Burlon08}
D.~Burlon et~al.
\newblock {\em The Astrophysical Journal Letters}, 685:L19, 2008.

\bibitem{Murase06}
K.~Murase et~al.
\newblock {\em The Astrophysical Journal Letters}, 651:L5, 2006.

\bibitem{Aharonian-Buckley-Kifune2008}
T.~Kifune G.~Sinnis F.A.~Aharonian, J.~Buckley.
\newblock {\em Rep. Progr. Phys.}, 71:096901, 2008.

\bibitem{HESS-PKS2155}
W.~Benbow et~al.
\newblock {\em Proc. 30th Int. Cosmic Ray Conf., Merida, Mexico}, 3:1081, 2007.

\bibitem{1ES1959}
H.~Krawczynski et~al.
\newblock {\em Astrophys. J.}, 601:151, 2004.

\bibitem{HESS-Science326}
F.~Acero and others (H.E.E.S.~Collaboration).
\newblock {\em Science}, 326:1080, 2009.

\bibitem{VERITAS-Starburst}
N.~Karlsson et~al.
\newblock {\em arXiv preprint arXiv:0912.3807}, 2009.

\bibitem{PhysRevLett.16.748}
K.~Greisen.
\newblock {\em Phys. Rev. Lett.}, 16:748--750, 1966.

\bibitem{zatsepin1966upper}
G.T. Zatsepin and V.A. Kuz'min.
\newblock {\em JETP Lett.(USSR)(Engl. Transl.)}, 4, 1966.

\bibitem{ahlers2010gzk}
M.~Ahlers et~al.
\newblock {\em Astroparticle Physics}, 34(2):106--115, 2010.

\bibitem{avecosmogenic2005}
M.~Ave et~al.
\newblock {\em Astroparticle Physics}, 23(1):19--29, 2005.

\bibitem{1367-2630-11-10-105026}
R.C. Cotta, J.S. Gainer, J.L. Hewett, and T.G. Rizzo.
\newblock {\em New Journal of Physics}, 11(10):105026, 2009.

\bibitem{PhysRevLett.110.131302}
M.~G. Aartsen et~al.
\newblock {\em Phys. Rev. Lett.}, 110:131302, 2013.

\bibitem{chung1998superheavy}
D.J. Chung, E.W. Kolb, and A.~Riotto.
\newblock {\em Physical Review D}, 59(2):023501, 1998.

\bibitem{Blasi200257}
P.~Blasi, R.~Dick, and E.W. Kolb.
\newblock {\em Astroparticle Physics}, 18(1):57 -- 66, 2002.

\bibitem{PhysRevD.64.083504}
I.F.M. Albuquerque, L.~Hui, and E.W. Kolb.
\newblock {\em Phys. Rev. D}, 64:083504, 2001.

\bibitem{1742-6596-116-1-012005}
G.~Giacomelli, S.~Manzoor, E.~Medinaceli, and L.~Patrizii.
\newblock {\em Journal of Physics: Conference Series}, 116(1):012005, 2008.

\bibitem{Wick2003663}
S.D. Wick, T.W. Kephart, T.J. Weiler, and P.L. Biermann.
\newblock {\em Astroparticle Physics}, 18(6):663 -- 687, 2003.

\bibitem{de1984nuclearites}
A.~De~R{a\'u}jula and S.L. Glashow.
\newblock {\em Nature}, 312(5996):734--737, 1984.

\bibitem{Kusenko1997108}
A.~Kusenko.
\newblock {\em Physics Letters B}, 405(1-2):108 -- 113, 1997.

\bibitem{Domokos98}
G.~Domokos and S.~Kovesi-Domokos.
\newblock 433:390, 1998.

\bibitem{Letessier00}
A.~Letessier-Selvon.
\newblock {\em AIP Conf. Proc.}, 566:157, 2000.

\bibitem{Athar00}
E.~Zas H.~Athar, G.~Parente.
\newblock {\em Physical Review D}, 62:093010, 2000.

\bibitem{Fargion02}
D.~Fargion.
\newblock {\em \apj}, 570:909, 2002.

\bibitem{Feng02}
F.~Wilczek T.~Yu J.~Feng, P.~Fisher.
\newblock {\em \prl}, 88:161102, 2002.

\bibitem{MaunaKea}
E.~Wolfe, W.~Wise, and G.~Dalrymple.
\newblock {\em U.S. Geological Survey Professional Paper}, 1557:129, 1997.

\bibitem{PYTHIA6154}
T.~Sj{\" o}strand et~al.
\newblock {\em Comput. Phys. Commun.}, 135:238, 2001.

\bibitem{Geant4}
S.~Agostinelli et~al.
\newblock {\em Nucl. Instrum. Methods A}, 506:250, 2003.

\bibitem{Dutta2001}
S.~Iyer~Dutta et~al.
\newblock {\em Physical Review D}, 63:094020, 2001.

\bibitem{ALLM}
H.~Abramowicz and A.~Levy.
\newblock {\em arXiv:hep-ph/9712415v2}, 1997.

\bibitem{DiffCS}
B.~Badelek and J.~Kwiecinski.
\newblock {\em Rev. Mod. Phys.}, 68:445, 1996.

\bibitem{TAUOLA24}
S.~Jadach et~al.
\newblock {\em Comput. Phys. Commun.}, 76:361, 1993.

\bibitem{Gandhi96}
R.~Gandhi et~al.
\newblock {\em Astropart. Phys.}, 5:81, 1996.

\bibitem{PREM}
A.~M. Dziewonski and D.~L. Anderson.
\newblock {\em Physics of the Earth and Planetary Interiors}, 25:297, 1981.

\bibitem{PhysRevD.66.021302}
J.~F. Beacom, P.~Crotty, and E.~W. Kolb.
\newblock {\em Phys. Rev. D}, 66:021302, 2002.

\bibitem{gazizov2005anis}
A.~Gazizov and M.~Kowalski.
\newblock {\em Computer physics communications}, 172(3):203--213, 2005.

\bibitem{Gandhi98}
R.~Gandhi et~al.
\newblock {\em Physical Review D}, 58:093009, 1998.

\bibitem{Tseng03}
J.~Tseng et~al.
\newblock {\em Physical Review D}, 68:063003, 2003.

\bibitem{TANeu}
M.~Sasaki, Y.~Asaoka, and M.~Jobashi.
\newblock {\em Astropart. Phys.}, 19:37, 2003.

\bibitem{Sasaki2001}
M.~Sasaki.
\newblock {\em J. Phys. Soc. Jpn.}, 70(Suppl. B):129, 2001.

\bibitem{IceCube2012}
R.~Abbasi et~al.
\newblock {\em Nature}, 484(Supplemental information):351, 2012.

\bibitem{AugerES09}
J.~Abraham et~al.
\newblock {\em Physical Review D}, 79:102001, 2009.

\bibitem{PhysRevLett.108.231101}
S.~H\"ummer, P.~Baerwald, and W.~Winter.
\newblock {\em Phys. Rev. Lett.}, 108:231101, 2012.

\bibitem{feldman1998unified}
G.~J. Feldman and R.~D. Cousins.
\newblock {\em Physical Review D}, 57(7):3873, 1998.

\bibitem{PhysRevD.83.092003}
R.~Abbasi et~al.
\newblock {\em Phys. Rev. D}, 83:092003, 2011.

\bibitem{gaisser2012high}
T.K. Gaisser.
\newblock High energy neutrinos from space.
\newblock {\em arXiv preprint arXiv:1201.6651}, 2012.

\bibitem{Grieder:1979ty}
PKF Grieder.
\newblock {Average Development and Properties of the Hadronic Moonic and
  Electromagnetic Components in Showers of 10**4 to 10**7 GEV Derived from AN
  All- {\ldots}}.
\newblock {\em International Cosmic Ray Conference}, 1979.

\bibitem{Hillas:1977wu}
A~M Hillas and J~Lapikens.
\newblock {Electron-photon cascades in the atmosphere and in detectors}.
\newblock {\em International Cosmic Ray Conference}, 1977.

\bibitem{Greisen:0hp}
K~Greisen.
\newblock {Cosmic Ray Showers}.
\newblock {\em Annual Review of Nuclear and Particle Science}, 10:63--108,
  1960.

\bibitem{Kamata:0bi}
K~Kamata and J~Nishimura.
\newblock {The Lateral and the Angular Structure Functions of Electron
  Showers}.
\newblock {\em Progress of Theoretical Physics Supplement}, 6:93--155, 1958.

\bibitem{Antoni:2001kt}
T~Antoni et~al.
\newblock {Electron, muon, and hadron lateral distributions measured in air
  showers by the KASCADE experiment}.
\newblock {\em Astroparticle Physics}, 14(4), 2001.

\bibitem{Asaoka-Sasaki11}
Y.~Asaoka and M.~Sasaki.
\newblock {\em Astroparticle Physics}, 41:7--16, 2013.

\bibitem{FlysEye}
R.~M. Baltrusaitis et~al.
\newblock {\em Nucl. Instrum. Methods A}, 240:410, 1985.

\bibitem{Noda09}
Y.~Aita et~al.
\newblock {\em 31th Intl. Cosmic Ray Conf. (Lodz), ID0313}, 2009.

\bibitem{Zas93}
E.~Zas et~al.
\newblock {\em Astropart.Phys.}, 1:297, 1993.

\bibitem{Enberg08}
R.~Enberg et~al.
\newblock {\em Physical Review D}, 78:043005, 2008.

\bibitem{Martin03}
A.~D. Martin et~al.
\newblock {\em Acta Physica B}, 34:3273, 2003.

\bibitem{Athar03}
H.~Athar et~al.
\newblock {\em Astropart.Phys.}, 18:581, 2003.

\end{thebibliography}
\end{document}